\documentclass{scrartcl}

\usepackage{
    adjustbox, 
	amsmath, 
    amsfonts,
    amssymb, 
	amsthm, 
	booktabs, 
	caption, 
	cite, 
	enumerate,
	float, 
	fullpage, 
	listings, 
	longtable, 
	mathabx, 
	multirow, 
	numprint, 
	subcaption, 
	tikz, 
	xcolor 
}

\usepackage[pagebackref]{hyperref}
\usepackage[LGR,T1]{fontenc}
\usepackage[utf8]{inputenc}
\usepackage[english]{babel} 

\usepackage{cleveref} 

\usetikzlibrary{decorations.markings, calc, patterns, arrows}
\usetikzlibrary{decorations.pathreplacing,calligraphy}
\tikzstyle{stuff_fill_green}=[circle,draw,fill=green!]
\tikzstyle{stuff_fill_red}=[circle,draw,fill=red!40]
\tikzstyle{stuff_fill_blue}=[circle,draw,fill=cyan!70]
\tikzstyle{stuff_fill_connect}=[circle,draw,fill=orange!70]
\tikzstyle{stuff_fill_high}=[circle,draw,fill=purple!70]

\theoremstyle{plain}
\newtheorem{thrm}{Theorem}[section]
\theoremstyle{definition}
\newtheorem{defn}[thrm]{Definition}
\newtheorem{exmp}[thrm]{Example}
\newtheorem{fig}[thrm]{Figure}
\newtheorem{cor}[thrm]{Corollary}
\newtheorem{lem}{Lemma}
\newtheorem{prop}[thrm]{Proposition}
\newtheorem*{note}{Note}
\newtheorem*{remark}{Remark}

\newtheorem{algo}{Algorithm}
\allowdisplaybreaks
\newtheoremstyle{break}  
{\topsep}   
{\topsep}   
{}  
{0pt}       
{\bfseries} 
{:}         
{\newline}  
{}          

\theoremstyle{break}

\makeatletter
\let\@addpunct\@gobble
\makeatother
\newenvironment{myproof}[1][\bfseries \proofname]{%
	\begin{proof}[#1]$ $\par\nobreak\ignorespaces
	}{%
	\end{proof}
}

\hypersetup{
	pdftitle={Improved statistics for F-theory standard models},
	pdfauthor={Martin Bies, Mirjam Cvetic, Ron Donagi, Marielle Ong},
	pdfsubject={Root bundles, Limit roots, Line bundles, Global sections, F-theory, MSSM, Standard Model, vector-like spectra, Brill-Noether theory}
}

\begin{document}

\vspace*{-2cm}
\begin{flushright}
    \texttt{UPR-1324-T}\\
\end{flushright}

\vspace*{0.8cm}
\begin{center}
    {\LARGE Improved statistics for F-theory standard models}
    
    \vspace*{1.8cm}
    {Martin Bies$^{1}$, Mirjam Cveti{\v c}$^{2,3,4}$, Ron Donagi$^{3,2}$, Marielle Ong$^{3}$}
    
    \vspace*{1cm}

     \bigskip
    {\textit{$^1$Department of Mathematics, RPTU Kaiserlautern-Landau, \\Kaiserslautern, Germany}}
    
     \bigskip
    {\textit{$^2$Department of Physics and Astronomy, University of Pennsylvania,  \\Philadelphia, PA 19104-6396, USA}}
    
    \bigskip
    {\textit{$^3$Department of Mathematics, University of Pennsylvania,  \\Philadelphia, PA 19104-6396, USA}}
    
    \bigskip
    {\textit{$^4$Center for Applied Mathematics and Theoretical Physics, University of Maribor, \\Maribor, Slovenia}}
        
\vspace*{0.8cm}
\end{center}
\noindent

Much of the analysis of F-theory-based Standard Models boils down to computing cohomologies of line bundles on matter curves. By varying parameters one can degenerate such matter curves to singular ones, typically with many nodes, where the computation is combinatorial and straightforward. The question remains to relate the (a priori possibly smaller) value on the original curve to the singular one. In this work, we introduce some elementary techniques (pruning trees and removing interior edges) for simplifying the resulting nodal curves to a small collection of terminal ones that can be handled directly. When applied to the QSMs, these techniques yield optimal results in the sense that obtaining more precise answers would require currently unavailable information about the QSM geometries. This provides us with an opportunity to enhance the statistical bounds established in earlier research regarding the absence of vector-like exotics on the quark-doublet curve.

\newpage

\tableofcontents

\newpage

\section{Introduction}

String Theory, a framework that seeks to unify quantum mechanics and general relativity, encompasses various regimes that offer valuable insights into the fundamental nature of the universe. Among these regimes, F-theory stands out as a non-perturbative approach that has garnered significant attention \cite{Vaf96, MV96, MV96*1}. In this paper, we aim to enhance our comprehension of a prominent category of F-theory Standard Models known as F-theory QSMs \cite{CHLLT19}, which represent the largest known class characterized by gauge coupling unification and the absence of chiral exotics. Our primary focus lies in the study of their vector-like spectra. Through this investigation, we make significant strides in refining our previous findings \cite{BCDLO21, BCL21, BCDO22}, which centered around the arithmetic aspects of Brill-Noether theory of root bundles on nodal curves.

F-theory is based on the powerful interplay between algebraic geometry and theoretical physics \cite{Vaf96, MV96, MV96*1}. Here, singular elliptic fibrations play a pivotal role and significant progress has been made in studying these geometries over the years. An overview of the developing connections between F-theory and the Standard Model can be found in \cite{CHST22}. For a broader and relatively recent overview, see \cite{Wei18} and the references therein.

This research paper focuses on 4-dimensional compactifications of F-theory with $\mathcal{N} = 1$ supersymmetry for the purpose of phenomenological investigations. Our goal is to construct of models of partial physics that closely align with the Standard Model of particle physics. We recall that 4-dimensional F-theory compactifications with $\mathcal{N} = 1$ supersymmetry exhibit chiral (super)fields in a specific representation $\mathbf{R}$ of the gauge group, as well as their charge conjugate counterparts in the representation $\mathbf{\overline{R}}$. The difference in the number of the chiral fields in representations $\mathbf{R}$ and $\mathbf{\overline{R}}$ is referred to as the chiral spectrum. Meanwhile, the separate numbers of fields in the representations $\mathbf{R}$ and $\mathbf{\overline{R}}$ provide the vector-like spectrum. Since the Higgs is a vector-like pair, the study of vector-like pairs is inevitable in the quest for F-theory Minimal Supersymmetric Standard Models (MSSMs). 

Significant efforts in perturbative String Theory have been dedicated to finding Standard model compactifications. This includes extensive work on the $E_8 \times E_8$ heterotic string \cite{CHSW85, GKMR86, BHOP05, BD06, BCD06, AGHL10, AGLP11, AGLP12} and intersecting branes models in type II String Theory \cite{BDL96, AFIRU01, AFIRU01*1, IMR01, BKLO01, CSU01, CSU01*1} (also refer to \cite{BCLS05} and references therein). While these perturbative String Theory compactifications realized the gauge sector of the Standard Model with its chiral spectrum, many constructions have shortcomings, such as the presence of exotic chiral or vector-like matter. The models that surpassed these obstacles and provided the first globally consistent construction of the MSSM in String Theory can be found in \cite{BD06, BCD06}. For further insight into the intricate global conditions relating to slope-stability of vector bundles, additional details are offered in \cite{GLS07, BD08}.

The chiral fermionic spectrum of 4d $\mathcal{N} = 1$ F-theory compactifications is well understood. The physics is encoded in the geometry of an elliptically fibered Calabi-Yau space $\pi \colon Y_{4} \twoheadrightarrow B_3$ and the chiral spectrum is fixed by the background gauge flux $G_4 = dC_3 \in H^{(2,2)} (Y_4)$ \cite{Hec10, DW11, MS11, BCV12, GH12, KMW12, KMW12*1, IJMMP13, DW14}, where $C_3$ refers to the internal $C_3$ profile in the dual M-theory geometry. Consequently, many tools were developed to construct and count the primary vertical subspace of $G_4$ configurations and to work out the physical quantities determined by such fluxes \cite{GH12, KMW12, BGK13, CGKP14, CKPOR15, LMTW16, LW16, JTT21, JT22}. Computationally, the most challenging condition is the $G_4$-flux quantization condition $G_4 + \frac{1}{2} c_2(Y_4) \in H^4_{\mathbb{Z}}(Y_4)$. One often examines necessary conditions for a specific $G_4$-flux to satisfy this quantization, and then proceeds with the resulting flux candidates under the assumption that the quantization condition is indeed satisfied. In applications of such toolkits, many globally consistent chiral F-theory models were presented \cite{KMW12*1, CKPOR15, LW16, CLLO18, JTT22}. This exploration culminated in the largest known group of explicit string vacua that realize the Standard Model gauge group with the exact chiral spectrum and gauge coupling unification. This class of F-theory models is known as the \emph{Quadrillion F-theory Standard Models} (QSMs) \cite{CHLLT19}. The $G_4$-flux candidate in the F-theory QSMs satisfies ``only'' necessary conditions for flux quantization. We proceed under the assumption that the quantization condition is satisfied. This $G_4$-flux candidate cancels the D3-tadpole and ensures that the $U(1)$ gauge boson remains massless.

The vector-like spectrum in F-theory depends not only on $G_4$ but also on $C_3$. We refer to the F-theory ``object'' dual to $C_3$ as an F-theory gauge potential. Such gauge potentials can be thought of as elements of the Deligne cohomology of $Y_4$ (see \cite{BMPW14} and references therein). Previous work, described in \cite{BMPW14, BMW17, Bie18}, utilizes the fact that certain subsets of the Deligne cohomology can be expressed in the Chow ring. This allowed us to compute line bundles $L_{\mathbf{R}}$ on the matter curves $C_\mathbf{R}$ in $B_3$. These line bundles can be interpreted as the localization of gauge flux on matter curves $C_\mathbf{R}$ in the dual IIB picture, resulting in massive vector-like pairs on these curves. The massless zero modes on the matter curves are given by the sheaf cohomology groups of the line bundle $L_{\mathbf{R}}$, where $h^{0}(C_{\mathbf{R}}, {L}_{\mathbf{R}})$ represents the number of massless chiral superfields in the representation $\mathbf{R}$ on $C_{\mathbf{R}}$, and $h^1 (C_{\mathbf{R}}, {L}_{\mathbf{R}})$ represents the number of massless chiral superfields in the representation $\mathbf{\overline{R}}$ on $C_{\mathbf{R}}$.

By construction, the line bundles ${L}_{\mathbf{R}}$ are twists of one of the many spin bundles on $C_{\mathbf{R}}$. We recall that there are $2^{2g}$ inequivalent spin bundles on a smooth, irreducible genus $g$ curve \cite{Ati71, Mum71}. Thus, one must investigate which spin bundles are compatible with the F-theory geometry, as well as their origin in the Deligne cohomology. To the best of the authors' knowledge, such a study has not been conducted and the answer is unknown. In similar spirit, one usually approaches vector-like spectra by lifting an existing $G_4$-flux to a gauge potential in the Deligne cohomology. Such lifts are not unique due to the intermediate Jacobian of $Y_4$. It can also be hard to find a single lift in many situations. For instance, this happens in the QSMs \cite{BCDLO21}, which is the origin of the root bundle program initiated in this work. Should we surpass these challenges, we still face a rather complicated interplay between the complex struture moduli of $Y_4$ and the line bundle cohomologies $h^i( C_{\mathbf{R}}, L_{\mathbf{R}} )$. To illustrate this challenge, let us mention that even with advanced algorithms \cite{GS23, Dev23, Tea23, **key*} and powerful supercomputers, the necessary computations could not be performed in realistic compactification geometries. So, the early works \cite{BMPW14, BMW17, Bie18} focused on models with computationally simple geometries even though the compactifications exhibited unrealistically large numbers of chiral fermions.

Machine learning techniques have and are being employed in the string theory context. Such works include but are not limited to \cite{CHKN17, HNR19, ACHL21, LLRS21, Abel:2023zwg}. For a broader overview, the interested reader may wish to consult \cite{Rue20} and the references therein. In this spirit, in \cite{BCDLLR21}, machine learning techniques were employed to systematically investigate the complex structure dependence of line bundle cohomologies. A large dataset \cite{BCDLLR20} of line bundle cohomologies for various complex structure moduli was generated using algorithms from \cite{aut19}. By combining data science techniques and well-known results about Brill-Noether theory \cite{BN74} \footnote{For a contemporary presentation of Brill-Noether theory, the interested reader is referred to \cite{EGH96}. An earlier use of Brill-Noether theory in F-theory can be found in \cite{Wat16}.}, a quantitative exploration of the jumps in charged matter vector pairs in relation to the complex structure moduli of the matter curve $C_\mathbf{R}$ was conducted.

Inspite of the challenges we face, we find inspiration in the attractive aspects of F-theory QSMs \cite{CHLLT19} and the goal of providing explicit realizations of F-theory MSSMs. Therefore, our focus is on exploring the vector-like spectra within F-theory QSMs, specifically those localized on their five matter curves. This investigation was initiated in a previous study \cite{BCDLO21} and subsequently expanded upon in later works \cite{BCL21, BCDO22}. This program was recently summarized in \cite{Bie23*1}. For yet more background information, the interested reader may consult \cite{L22}.

The first obstacle that we encounter is the lack of knowledge regarding the lift to the Deligne cohomology of the proposed $G_4$-flux candidate in the F-theory QSMs. The second challenge involves allowing only for twists of spin bundles that align with this F-theory compactification. As we do not have answers to either of these questions, we have established a necessary condition that all induced line bundles $L_{\mathbf{R}}$ must satisfy. The initial study \cite{BCDLO21} revealed that on three of the five matter curves in the QSM geometry, the line bundle ${L}_{\mathbf{R}}$ must necessarily be a fractional power $P{\mathbf{R}}$ of the canonical bundle on that particular matter curve $C_{\mathbf{R}}$. On the remaining two curves, the line bundles are, in addition, modified by contributions from Yukawa points. Such fractional power line bundles are known as root bundles, which are  generalizations of spin bundles. Like spin bundles, root bundles are generally not unique. The mathematics suggests that different root bundles may arise from non-equivalent F-theory gauge potentials in Deligne cohomology; all of which produce the same $G_4$-flux. However, not all root bundles that satisfy our constraints on $C_{\mathbf{R}}$ are necessarily induced by F-theory gauge potentials. In other words, just like the spin bundles, only specific root bundles on the matter curves are expected to be consistent with the F-theory geometry $Y_4$. Therefore, an F-theory gauge potential gives rise to a set $\{ P_{\mathbf{R}} \}$, which includes one root bundle for each matter curve $C_{\mathbf{R}}$. By repeating this process for all physically relevant F-theory gauge potentials, it is expected that only a subset of all root bundles on $C_{\mathbf{R}}$ will generally be obtained. Conversely, if we are given a set of root bundles $\{ P_{\mathbf{R}} \}$ (one on each matter curve), then this set may not originate from an F-theory gauge potential. Namely, it is possible for one of the root bundles to not be induced at all or not be induced in combination with one of the other roots.

Determining which root bundles are physical, i.e. induced from an F-theory gauge potential, is crucial for gaining a comprehensive understanding of the physics involved. However, this task presents a significant challenge. Instead of directly addressing this question, the study conducted in \cite{BCDLO21} took a systematic approach by examining all root and all spin bundles on each matter curve. By adopting this method, it was possible to assign a specific root bundle to each matter curve (except for the Higgs curve) in a particular QSM geometry \cite{CHLLT19} such that their cohomologies exhibit the desired MSSM vector-like spectra. These bundles are obtained by deforming the smooth and irreducible physical matter curve $C_{\mathbf{R}}$ into a nodal curve $C^\bullet_{\mathbf{R}}$ and describing the root bundles on the latter using a diagrammatic representation called \emph{limit roots} \cite{CCC07}. Counting the global sections of limit root bundles on the complete blow-up of $C^\bullet_{\mathbf{R}}$ can be accomplished using the techniques developed in \cite{BCDLO21}.

In the expanded study conducted in \cite{BCL21}, a computer algorithm was utilized to count all root bundles with three global sections on the complete blow-up of $C^\bullet_{\mathbf{R}}$ in 33 different families of QSM geometries. To accomplish this, families of toric spaces $B_3( \Delta^\circ )$ associated with polytopes $\Delta^\circ$ from the Kreuzer-Skarke list \cite{KS98} were examined (see also \cite{HT17} for a different study of these geometries in a string theory context). These toric spaces represent alternative desingularizations of toric K3-surfaces \cite{Bat94, PS97, CK99, Roh04,BLMSS17}. The study revealed that the structure of a canonical nodal curve $C^\bullet_{\mathbf{R}}$ and, consequently, the count of limit root bundles, solely depends on the polytope $\Delta^\circ$. These techniques were further refined in \cite{BCDO22} to handle limit roots on ``simple'' partial blow-ups. The outcomes of the computer scan strongly indicate that the absence of vector-like exotics in the $(\mathbf{3}, \mathbf{2})_{1/6}$ representation is a highly probable scenario within the F-theory QSMs. It is essential to emphasize that this finding carries a statistical meaning, as the precise origin of the limit roots from F-theory gauge potentials has yet to be fully understood.

\subsection*{Results}

This study builds upon previous works \cite{BCDLO21, BCL21, BCDO22} and takes their techniques further. Previous research revealed that counting global sections of root bundles leads to exploring line bundle cohomology on nodal curves. Specifically, \cite{BCL21} focused on line bundle cohomology for disconnected nodal curves while \cite{BCDO22} expanded the analysis to include tree-like nodal curves. In this study, we extend the investigation to line bundles on nodal curves with a non-zero first Betti number, i.e. circuits. To accomplish this, a two-step procedure is employed:
\begin{enumerate}
 \item A line bundle on a difficult nodal curve is replaced by a line bundle on a simpler nodal curve, which we refer to as a \emph{terminal circuit}.
 \item It is demonstrated that there are only a finite number of terminal circuits for a fixed Betti number. Consequently, the terminal circuits are classified and line bundle cohomologies on this nodal curves are examined through a case-by-case analysis.
\end{enumerate}
Based on the available data, we can confirm that our counts of global sections for root bundles are optimal. To achieve improved results, revisiting the QSM geometries and extracting more refined geometric information would be necessary. This includes acquiring information about the descent data of the root bundles and the precise line bundle divisor on elliptic components of the nodal curves. However, without this additional information, our computer scan \cite{Bie23} provides the best possible answer.

Having achieved the optimal outcome, we assess the probability of no vector-like exotics precisely on the quark-doublet curve. Across the union of all 33 distinct QSM families explored, this scenario is estimated to occur in at least $93.91\%$ of all these configurations. This result strengthens the earlier findings that statistically support the likelihood of this desired physical scenario. However, it is crucial to note that these conclusions rely on the top-down origin of the root bundles, which we anticipate future research will clarify.

Estimating the likelihood of one exotic vector-like pair on the Higgs curve would be desirable, but is infeasible with our current computational techniques. However, we can use the compute Brill-Noether numbers on the quark-doublet curve to again at least some indications. Namely, our results show that this undesirable phenomenon of exactly one vector-like exotic on the quark-doublet curve occurs with probability of at most $5.17\%$. We view this result as a positive indication towards exactly one vector-like pair on the Higgs curve.

\subsection*{Outline}

In \cref{sec:SummaryOfOldTechniques}, we provide a concise overview of the previous findings. For a more comprehensive understanding, we direct interested readers to the original works \cite{BCDLO21, BCL21, BCDO22}. This section also serves to justify the need for our more refined techniques, which expand upon the previous results. These advanced techniques are introduced in \cref{sec:SimplificationTechniques}. Of particular significance is the classification of the terminal graphs discussed in \cref{subsec:TerminalGraphs}. The investigation of jumps on these terminal graphs represents a technically challenging aspect of this research work. Further details and derivations can be found in \cref{Appendix:Derivations} and \cref{sec:AppendixEllipticCircuitJumps}. These techniques are then applied to the F-theory QSMs, and the implications are elaborated upon in \cref{sec:ImplicationsForFTheoryQSMs}. Finally, we conclude this work with a summary and outlook in \cref{sec:SummaryAndOutlook}.

\newpage

\section{Limit root bundles meet F-theory QSMs} \label{sec:SummaryOfOldTechniques}

We begin with a very brief revision on our root bundle program. The interested reader can find much more information in the original works \cite{BCDLO21, BCL21, BCDO22} and the recent summary on this program in \cite{Bie23*1}.

Vector-like spectra in F-theory are counted by the line bundle cohomologies of certain line bundles on the matter curves \cite{BMPW14, BMW17, Bie18}. For the Quadrillion F-theory standard models \cite{CHLLT19}, we could not determine the exact line bundle whose cohomologies count the vector-like spectra. However, it was possible to derive necessary and highly non-trivial constraints for these line bundles. Our initial study \cite{BCDLO21} showed that these line bundles $P_{\mathbf{R}}$ must be root bundles. To elaborate more on this finding, we first recall that the F-theory QSMs \cite{CHLLT19} are defined in terms of toric 3-folds with $\overline{K}^3_{B_3} \in \{ 6, 10, 18, 30 \}$. The five matter curves of these models are defined by sections $s_i \in H^0 \left( B_3, \overline{K}_{B_3} \right)$. In \cite{BCDLO21}, we argued that the line bundles $P_{\mathbf{R}}$, whose cohomologies count the vector-like spectra, must necessarily satisfy the following conditions:
\begin{align}
\begin{tabular}{cc}
\toprule
curve & root bundle constraint \\
\midrule
$C_{(\mathbf{3},\mathbf{2})_{1/6}} = V( s_3, s_9 )$ & $P_{(\mathbf{3},\mathbf{2})_{1/6}}^{\otimes 2 \overline{K}^3_{B_3}} = K_{{(\mathbf{3},\mathbf{2})_{1/6}}}^{\otimes \left( 6 + \overline{K}^3_{B_3} \right)}$ \\
$C_{(\mathbf{1},\mathbf{2})_{-1/2}}= V \left( s_3, P_H \right)$ & $P_{(\mathbf{1},\mathbf{2})_{-1/2}}^{\otimes 2 \overline{K}^3_{B_3}} = K_{{(\mathbf{1},\mathbf{2})_{-1/2}}}^{\otimes \left( 4 + \overline{K}^3_{B_3} \right)} \otimes \mathcal{O}_{(\mathbf{1},\mathbf{2})_{-1/2}} \left( - 30 Y_1 \right)$ \\
$C_{(\overline{\mathbf{3}},\mathbf{1})_{-2/3}} = V( s_5, s_9 )$ & $P_{(\overline{\mathbf{3}},\mathbf{1})_{-2/3}}^{\otimes 2 \overline{K}^3_{B_3}} = K_{{(\overline{\mathbf{3}},\mathbf{1})_{-2/3}}}^{\otimes \left( 6 + \overline{K}^3_{B_3} \right)}$ \\
$C_{(\overline{\mathbf{3}},\mathbf{1})_{1/3}} = V \left( s_9, P_R \right)$ & $P_{(\overline{\mathbf{3}},\mathbf{1})_{1/3}}^{\otimes 2 \overline{K}^3_{B_3}} = K_{{(\overline{\mathbf{3}},\mathbf{1})_{1/3}}}^{\otimes \left( 4 + \overline{K}^3_{B_3} \right)} \otimes \mathcal{O}_{(\overline{\mathbf{3}},\mathbf{1})_{1/3}} \left( - 30 Y_3 \right)$ \\
$C_{(\mathbf{1},\mathbf{1})_{1}} = V( s_1, s_5 )$ & $P_{(\mathbf{1},\mathbf{1})_{1}}^{\otimes 2 \overline{K}^3_{B_3}} = K_{{(\mathbf{1},\mathbf{1})_{1}}}^{\otimes \left( 6 + \overline{K}^3_{B_3} \right)}$ \\
\bottomrule
\end{tabular} \label{tab:RootBundles}
\end{align}
In this table, we make use of two polynomials: $P_H = s_2 s_5^2 + s_1 ( s_1 s_9 - s_5 s_6 )$ and $P_R = s_3 s_5^2 + s_6 ( s_1 s_6 - s_2 s_5 )$. It is important to note that the line bundles on both the Higgs curve $C_{(\mathbf{1},\mathbf{2}){-1/2}}$ and the curve $C_{(\overline{\mathbf{3}},\mathbf{1})_{1/3}}$ are dependent on the Yukawa points $Y_1 = V( s_3, s_5, s_9 )$ and $Y_3 = V( s_3, s_6, s_9 )$, respectively. Furthermore, it is important to keep in mind that if two divisors $D$ and $E$ are linearly equivalent, i.e., $D \sim E$, then $n \cdot D \sim n \cdot E$ for any integer $n$. However, the converse is not true, and that is why we do not cancel common factors.

For instance, $P_{(\mathbf{3},\mathbf{2})_{1/6}}$ must be a $2 \overline{K}^3_{B_3}$-th root of the $\left( 6 + \overline{K}^3_{B_3} \right)$-th power of $K_{{(\mathbf{3},\mathbf{2})_{1/6}}}$. Hence, the name root bundles. It can be argued that solutions to the root bundle constraints in \cref{tab:RootBundles} do exist, but are not unique. This is to be contrasted with the physics' expectation that only a subset of these roots is realized from F-theory gauge potentials in the Deligne cohomology $H^4_D( \widehat{Y}_4, \mathbb{Z}(2) )$). Such top-down consideration were and still are beyond our reach. In addition, there are computational limitations that stop us from enumerating the huge number of root bundles on the Higgs curve and then working out their number of global sections. To circumvent both of these shortcomings, starting from our analysis in \cite{BCL21}, we have therefore employed the following philosophy:
\begin{itemize}
 \item Focus on F-theory QSM geometries, such that there are at least as many different F-theory gauge potentials as there are solutions to the root bundle constraint on the quark-doublet curve $C_{(\mathbf{3},\mathbf{2})_{1/6}}$. Then, at least statistically speaking, we could hope that every root bundle on the quark-doublet curve is realized from an F-theory gauge potential. It turned out that this condition is equivalent to $h^{2,1}(\widehat{Y}_4) \geq g$ where $g$ is the genus of $C_{(\mathbf{3},\mathbf{2})_{1/6}}$ \cite{BCL21}.
 \item Subsequently, we analyze the line bundle cohomologies of all root bundles on the quark-doublet systematically in order to draw conclusions about the presence/absence of vector-like exotics.
\end{itemize}
We thus make a selection among the F-theory QSMs. To this end, we recall that these geometries arise from desingularizations of toric K3-surfaces. Such desingularizations were first studied in \cite{Bat94} and correspond to 3-dimensional, reflexive lattice polytopes $\Delta \subset M_{\mathbf{R}}$ and their polar duals $\Delta^\circ \subset N_{\mathbf{R}}$ defined by $\left\langle \Delta, \Delta^\circ \right\rangle \geq -1$. The complete list of such 3-dimensional polytopes is available in \cite{KS98}. Just as in \cite{BCDLO21, BCL21, BCDO22} and denote the i-th polytope in this list as $\Delta_i^\circ \subset N_{\mathbf{R}}$. We denote the family of F-theory QSM geometries associated to the i-th 3-dimensional, reflexive lattice polytope $\Delta_i^\circ$ in \cite{KS98} by $B_3( \Delta_i^\circ )$. The different geometries in $B_3( \Delta_i^\circ )$ are one-to-one to the fine regular star triangulations of $\Delta_i^\circ$. Some quantities are the same for a family $B_3( \Delta_i^\circ )$, for instance $\overline{K}_{B_3}^3$, $h^{2,1}(\widehat{Y}_4)$ and the genus of $C_{(\mathbf{3},\mathbf{2})_{1/6}}$. It turns out that there are exactly 37 F-theory QSM families $B_3( \Delta_i^\circ )$ which satisfy $h^{2,1}(\widehat{Y}_4) \geq g$.

It thus remains to construct solutions for the root bundle constraint in \cref{tab:RootBundles} on the matter curve $C^\bullet_{(\mathbf{3},\mathbf{2})_{1/6}}$ of these 37 QSM families and then to work out the line bundle cohomologies of those root bundles. Sadly, it turns out that this is a rather tough task for smooth, irreducible curves. Historic references indicating this obstacle include \cite{Ati71, Mum71}. Instead of working out the exact solutions, we have therefore opted for an approximation technique, which is based on the following two key insights:
\begin{itemize}
 \item Root bundles on nodal curves are well understood and follow a diagrammatic pattern \cite{CCC07}. We refer the interested reader to \cite{BCDLO21} and \cite{Bie23*1} for more details.
 \item There is a canonical deformation $C_{(\mathbf{3},\mathbf{2})_{1/6}} \to C^\bullet_{(\mathbf{3},\mathbf{2})_{1/6}}$, such that the nodal curve $C^\bullet_{(\mathbf{3},\mathbf{2})_{1/6}}$ is the same for all spaces in $B_3( \Delta_i^\circ )$, i.e. depends only on the polytope $\Delta_i^\circ$. This we explained in large detail in \cite{BCL21}.
\end{itemize}
It is worth noting that this is a rather impressive and stunning finding! Despite there being an enormous number of F-theory QSM geometries (around $10^{15}$), it is possible to estimate the vector-like spectra of the majority of these setups from analyzing root bundles on only 37 different nodal curves!

Despite these pleasing features of nodal curves, we eventually want to go back and determine the cohomologies of root bundles on the initial smooth, irreducible matter curve $C_{(\mathbf{3},\mathbf{2})_{1/6}}$. Hence, we should study how cohomologies of root bundles $P^\bullet_{\mathbf{R}}$ on nodal curves relate to the cohomologies of root bundles $P^\bullet_{\mathbf{R}}$ on the smooth irreducible curve. In principle, we can trace the roots $P^\bullet_{\mathbf{R}}$ along the deformation $C^\bullet_{\mathbf{R}} \to C_{\mathbf{R}}$ to find roots $P_{\mathbf{R}}$ on the original curve $C_{\mathbf{R}}$. While such a deformation can alter the number of global sections, it is known that it is known that the number of global sections is an upper semi-continuous function:
\begin{align}
h^0( C_{\mathbf{R}}, P_{\mathbf{R}} ) \leq h^0( C^{\bullet}_{\mathbf{R}}, P^{\bullet}_{\mathbf{R}} ) = \chi( P_{\mathbf{R}} ) + \delta \, , \qquad \delta \in \mathbb{Z}_{\geq 0} \, .
\end{align}
In order to find roots for which the number of sections remains constant upon deformation to the original curve, we can look for instances where equality holds. One such instance is the generic case $\delta=0$, in which $h^0( C^{\bullet}_{\mathbf{R}}, P^{\bullet}_{\mathbf{R}} ) = \chi( P_{\mathbf{R}} )$. This is because the number of sections is then already minimal on $C^{\bullet}_{\mathbf{R}}$ and therefore must remain constant throughout the deformation. As a result, the number of limit roots on the partial blow-ups $C^\circ_{\mathbf{R}}$ of the nodal curve $C^\bullet_{\mathbf{R}}$ that have the generic number of global sections provides a lower bound on the number of roots $P_{\mathbf{R}}$ without vector-like exotics. Indeed, this is exactly the situation that desribes/approximates root bundles on $C_{(\mathbf{3},\mathbf{2})_{1/6}}$ without vector-like exotics.

In reassessing the 37 F-theory QSM families $B_3( \Delta_i^\circ )$ with $h^{2,1}(\widehat{Y}_4) \geq g$, we realized that for four of them, the canonical nodal quark-doublet curve $C^\bullet_{(\mathbf{3},\mathbf{2})_{1/6}}$ has a smooth, irreducible component whose genus is greater than one. This makes the global section counting for limit roots a lot harder, and so we decided to ignored those four QSM families. Consequently, we focused on the following 33 families of QSM geometries. Note that the numbers of fine regular star triangulations were worked out/estimated in \cite{HT17}.
\begin{equation}
\begin{tabular}{cc|c|cc}
\toprule
Polytope & $\overline{K}_{B_3}^3$ & $N_{\text{roots}}$ & $\widecheck{N}_{\text{FRST}}$ & $\widehat{N}_{\text{FRST}}$ \\
\midrule
$\Delta_8^\circ$ & $6$ & $12^8$ & $3.867 \times 10^{13}$ & $2.828 \times 10^{16}$ \\
$\Delta_4^\circ$ & $6$ & $12^8$ & $3.188 \times 10^{11}$ & \\
$\Delta_{134}^\circ$ & $6$ & $12^8$ & $7.538 \times 10^{10}$ & \\
$\Delta_{128}^\circ$, $\Delta_{130}^\circ$, $\Delta_{136}^\circ$, $\Delta_{236}^\circ$ & $6$ & $12^8$ & $3.217 \times 10^{11}$ \\
\midrule \midrule
$\Delta_{88}^\circ$ & $10$ & $20^{12}$ & $5.231 \times 10^{10}$ & $1.246 \times 10^{14}$ \\
$\Delta_{110}^\circ$ & $10$ & $20^{12}$ & $5.239 \times 10^8$ \\
$\Delta_{272}^\circ$, $\Delta_{274}^\circ$ & $10$ & $20^{12}$ & $3.212 \times 10^{12}$ & $2.481 \times 10^{15}$ \\
$\Delta_{387}^\circ$ & $10$ & $20^{12}$ & $6.322 \times 10^{10}$ & $6.790 \times 10^{12}$ \\
$\Delta_{798}^\circ$, $\Delta_{808}^\circ$, $\Delta_{810}^\circ$, $\Delta_{812}^\circ$ & $10$ & $20^{12}$ & $1.672 \times 10^{10}$ & $2.515 \times 10^{13}$ \\
\midrule
$\Delta_{254}^\circ$ & $10$ & $20^{12}$ & $1.568 \times 10^{10}$ \\
$\Delta_{52}^\circ$ & $10$ & $20^{12}$ & $1.248 \times 10^8$ \\
$\Delta_{302}^\circ$ & $10$ & $20^{12}$ & $5.750 \times 10^7$ \\
$\Delta_{786}^\circ$ & $10$ & $20^{12}$ & $9.810 \times 10^8$ \\
$\Delta_{762}^\circ$ & $10$ & $20^{12}$ & $1.087 \times 10^{11}$ & $2.854 \times 10^{13}$ \\
\midrule
$\Delta_{417}^\circ$ & $10$ & $20^{12}$ & $1.603 \times 10^9$ & \\
$\Delta_{838}^\circ$ & $10$ & $20^{12}$ & $4.461 \times 10^9$ & \\
$\Delta_{782}^\circ$ & $10$ & $20^{12}$ & $3.684 \times 10^9$ & \\
$\Delta_{377}^\circ$, $\Delta_{499}^\circ$, $\Delta_{503}^\circ$ & $10$ & $20^{12}$ & $4.461 \times 10^9$ & \\
$\Delta_{1348}^\circ$ & $10$ & $20^{12}$ & $4.285 \times 10^9$ & \\
\midrule
$\Delta_{882}^\circ$, $\Delta_{856}^\circ$ & $10$ & $20^{12}$ & $3.180 \times 10^9$ & \\
$\Delta_{1340}^\circ$ & $10$ & $20^{12}$ & $4.496 \times 10^9$ & \\
$\Delta_{1879}^\circ$ & $10$ & $20^{12}$ & $4.461 \times 10^9$ & \\
$\Delta_{1384}^\circ$ & $10$ & $20^{12}$ & $7.040 \times 10^9$ & \\
\bottomrule
\end{tabular}
\end{equation}

We conclude this section with an illustrative example of constructing limit roots and determining their number of global sections. This example is chosen to illustrate our key advances. The interested reader may consider this example as a motivation for the more involved study of these techniques in \cref{sec:SimplificationTechniques}. For this example, let us look at a nodal curve with three irreducible, smooth, rational components, whose dual graph $\Gamma$ is as follows:
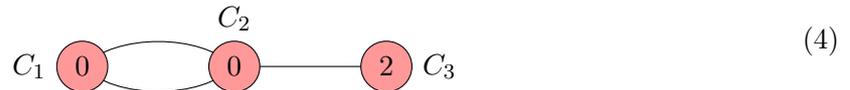
\begin{equation}
\begin{tikzpicture}[baseline=(current  bounding  box.center)]
\def\s{1};
\path[-, out = 45, in = 135, looseness = 0.8] (-\s,0) edge (\s,0);
\path[-, out = -45, in = -135, looseness = 0.8] (-\s,0) edge (\s,0);
\path[-, out = 0, in = -180, looseness = 0] (\s,0) edge (3*\s,0);
\node at (-\s,0) [stuff_fill_red, circle, draw, label=left:$C_1$] {$0$};
\node at (\s,0)  [stuff_fill_red, circle, draw, label=above:$C_2$] {$0$};
\node at (3*\s,0)  [stuff_fill_red, circle, draw,  label=right:$C_3$] {$2$};
\end{tikzpicture}
\end{equation}
This means that the vertices represent the irreducible components and the edges correspond to the nodal singularities. On this curve, we consider a line bundle $L$ with $\mathrm{deg} \left( \left. L \right|_{C_i} \right) = ( 0,0,2 )$, i.e. the degree of $L$ restricted to the i-th component is exactly the number that we display in the above graph inside of the i-th vertex. It is worth recalling that this information does not fix $L$ uniquely. In order to uniquely specify $L$, we would have to specify the so-called descent data, which in this case, boils down to fixing one parameter $\lambda \in \mathbb{C}^\ast$ (see \cite{HM06} for details). This parameter specifies that the sections on $C_1$ and $C_2$ agree up to $\lambda$ at one of the two nodes in which they intersect, while the sections agree in all remaining nodes.

Let us now construct limit 2nd roots of this line bundle $L$. Note that $\beta_1( \Gamma ) = 1$, so on a smooth, irreducible curve, there are $2^2 = 4$ roots. Here is how to construct the corresponding limit root. First, make a binary decision for each node. Namely, we decide whether to blow this node up or not. If we do, then this amounts to putting a bi-weight $1-1$ on the corresponding edge in the above graph and subtracting $1$ from each of the adjacent vertices. Then, one removes the corresponding edge and tries to divide the resulting degrees by $2$. If doing so leads to integer degrees on all vertices, we have encoded a limit root. In this case, this procedure leads to the following two admissible configurations:
\begin{equation}
\begin{tikzpicture}[baseline=(current  bounding  box.center)]
\def\s{1};
\def\d{6};
\path[-, out = 45, in = 135, looseness = 0.8] (-\s,0) edge (\s,0);
\path[-, out = -45, in = -135, looseness = 0.8] (-\s,0) edge (\s,0);
\path[-, out = 0, in = -180, looseness = 0] (\s,0) edge (3*\s,0);
\node at (-\s,0) [stuff_fill_red, circle, draw, label=above:$C_1$] {$0$};
\node at (\s,0)  [stuff_fill_red, circle, draw, label=above:$C_2$] {$0$};
\node at (3*\s,0)  [stuff_fill_red, circle, draw,  label=above:$C_3$] {$1$};
\draw [thick,decoration={brace,mirror,raise=0.5cm},decorate] (-1.5*\s,0) -- (3.5*\s,0) node [pos=0.5,anchor=north,yshift=-0.55cm] {Limit root 1};
\path[-, out = 0, in = -180, looseness = 0] (\s+\d,0) edge (3*\s+\d,0);
\node at (-\s+\d,0) [stuff_fill_red, circle, draw,  label=above:$C_1$] {$-1$};
\node at (\s+\d,0)  [stuff_fill_red, circle, draw, label=above:$C_2$] {$-1$};
\node at (3*\s+\d,0)  [stuff_fill_red, circle, draw,  label=above:$C_3$] {$1$};
\draw [thick,decoration={brace,mirror,raise=0.5cm},decorate] (-1.5*\s+\d,0) -- (3.5*\s+\d,0) node [pos=0.5,anchor=north,yshift=-0.55cm] {Limit root 2};
\end{tikzpicture}
\end{equation}
Limit root 1 is obtained by not blowing up any node, whereas limit root 2 is obtained by blowing up both nodes between $C_1$ and $C_2$. Each of these graphs correspond according to the geometric multiplicity to $\mu = 2^{\beta_1(\Gamma)} = 2^1 = 2$ roots on a corresponding smooth, irreducible curve.

Now that we have enumerated the limit roots, it remains to compute the global sections. This is exactly where our new techniques appear. We begin by looking at limit root 1. Our simplification techniques for this line bundle on this nodal curve proceed as follows:
\begin{align}
h^0 \left(
\begin{tikzpicture}[baseline=(current  bounding  box.center)]
\def\s{1};
\path[-, out = 45, in = 135, looseness = 0.8] (-\s,0) edge (\s,0);
\path[-, out = -45, in = -135, looseness = 0.8] (-\s,0) edge (\s,0);
\path[-, out = 0, in = -180, looseness = 0] (\s,0) edge (3*\s,0);
\node at (-\s,0) [stuff_fill_red, circle, draw] {$0$};
\node at (\s,0)  [stuff_fill_red, circle, draw] {$0$};
\node at (3*\s,0)  [stuff_fill_red, circle, draw] {$1$};
\end{tikzpicture}
\right) &\stackrel{\text{T1}}{=} h^0 \left( 
\begin{tikzpicture}[baseline=(current  bounding  box.center)]
\def\s{1};
\path[-, out = 45, in = 135, looseness = 0.8] (-\s,0) edge (\s,0);
\path[-, out = -45, in = -135, looseness = 0.8] (-\s,0) edge (\s,0);
\node at (-\s,0) [stuff_fill_red, circle, draw] {$0$};
\node at (\s,0)  [stuff_fill_red, circle, draw] {$0$};
\end{tikzpicture}
\right) + 1 \stackrel{\text{T2}}{=} h^0 \left(
\begin{tikzpicture}[baseline=(current  bounding  box.center)]
\def\s{1};
\path[-,out = -30, in = 30, looseness = 6] (-\s,-0.2*\s) edge (-\s,+0.2*\s);
\node at (-\s,0) [stuff_fill_red, circle, draw] {$0$};
\end{tikzpicture}
\right) + 1 \, .
\end{align}
The step $T1$ prunes a leaf while $T2$ removes an interior edge. We will give more details on these techniques in sections \ref{subsec:PruningLeaves} and \ref{subsec:RemovingInteriorEdges}. By following these steps, we were able to reduce the original problem to a line bundle on an arguably simpler nodal curve:
\begin{equation}
\begin{tikzpicture}[baseline=(current  bounding  box.center)]
\def\s{1};
\path[-,out = -30, in = 30, looseness = 6] (-\s,-0.2*\s) edge (-\s,+0.2*\s);
\node at (-\s,0) [stuff_fill_red, circle, draw] {$0$};
\end{tikzpicture}
\end{equation}
This last graph cannot be simplified further, which is why we call such a configuration a \emph{terminal graph}. We classify these terminal graphs in \cref{subsec:TerminalGraphs} and analyze their number of global sections. In this case, it is not hard to see that there is exactly one global section provided that the descent data satisfies $\lambda = 1$. This corresponds to the canonical bundle on this nodal curve. In all other cases, we have no global sections. This is why we say that this terminal graph is jumping. 

We now return to the second limit root. For this limit root, we obtain along the same lines:
\begin{align}
h^0 \left(
\begin{tikzpicture}[baseline=(current  bounding  box.center)]
\def\s{1};
\path[-, out = 0, in = -180, looseness = 0] (\s,0) edge (3*\s,0);
\node at (-\s,0) [stuff_fill_red, circle, draw] {$-1$};
\node at (\s,0)  [stuff_fill_red, circle, draw] {$-1$};
\node at (3*\s,0)  [stuff_fill_red, circle, draw] {$1$};
\end{tikzpicture}
\right) &\stackrel{\text{T1}}{=} h^0 \left( 
\begin{tikzpicture}[baseline=(current  bounding  box.center)]
\def\s{1};
\node at (-\s,0) [stuff_fill_red, circle, draw] {$-1$};
\node at (\s,0)  [stuff_fill_red, circle, draw] {$-1$};
\end{tikzpicture}
\right) + 1 \, .
\end{align}
For this limit root, there are always no global sections, irrespective of the descent data. We summarize this information as follows:
\begin{equation}
\begin{tabular}{cc}
\toprule
$h^0 = 1$ & $h^0 \geq 1$ \\
\midrule
2 & 2 \\
(Limit root 2) & (Limit root 1) \\
\bottomrule
\end{tabular}
\end{equation}
Certainly, the nodal curves encountered for the F-theory QSMs are more involved. But still, the steps to be taken are analogous to the above example. In the following section, we provide more details on the leaf pruning and removal of interior edges that eventually lead to the classification of terminal graphs in \cref{subsec:TerminalGraphs}. Subsequently, we return to the F-theory QSMs in \cref{sec:ImplicationsForFTheoryQSMs}.

\section{Simplification techniques} \label{sec:SimplificationTechniques}

\subsection{Pruning leaves} \label{subsec:PruningLeaves}

Let $(C, L)$ be a pair consisting of a nodal curve $C$ and line bundle $L$. Suppose that the dual graph $\Gamma$ of $C$ contains a tree (consisting of only rational components) that is attached to exactly one component (of genus 0 or 1) at exactly one node. We put forward an algorithm that prunes leaves repeatedly until the entire tree is removed. 

\begin{algo}[Algorithmic pruning of leafs] \label{prune}
Suppose that $\Gamma$ contains a leaf $T$, i.e. $T$ is a vertex that is attached to exactly one vertex $U$ of $\Gamma$ at a nodal singularity $p$. If $d_T = \deg(L|_T) \geq 0$, then let $C^\prime = C \setminus T$ and define a line bundle $L' = L|_{C'}$ on $C'$. Then,
\begin{equation}
h^0(C, L) = h^0(C^\prime, L') + d_T \, .
\end{equation}
If $d_T < 0$, then let $C^\prime = C \setminus T$ and define a new line bundle $L^\prime = L \otimes O_{C^\prime}(-p)$ on $C^\prime$. In particular, it satisfies $L^\prime|_U = L|_U \otimes \mathcal{O}_U(-p)$ and $L^{\prime}|_{C^\prime \setminus U} = L|_{C^\prime \setminus U}$
Then, \begin{equation}
h^0(C, L) = h^0(C^\prime, L').
\end{equation}
\end{algo}
 By repeating this procedure and pruning leaves repeatedly, we can remove subtrees of $\Gamma$. Let us explain the case where $d_T < 0$ further. 
 If $U$ has genus 0, then replacing $L$ with $L'$ amounts to replacing $d_U$ with $d_U-1$. 
 This is because line bundles on rational curves are uniquely specified by their degree. However, this is no longer true when $U$ has genus 1. That is, non-isomorphic bundles on an elliptic curve can have the same degree and this adds ambiguity when calculating $h^0$. Indeed, the Riemann-Roch Theorem states that for a degree $d$ line bundle $L$ on a elliptic curve $E$,
\begin{equation}
h^0(E, L) = \begin{cases}
d, &\text{if } d \geq 1, \\
0 \text{ or } 1, & \text{if } d = 0, \\
0, &\text{if } d < 0. \\
\end{cases}
\end{equation}
So if the pruning leads to a degree 0 line bundle on the elliptic component, then we can only find a lower bound for $h^0$. For example, let $d < 0$ and consider the curve $C$ and line bundle $L$: 
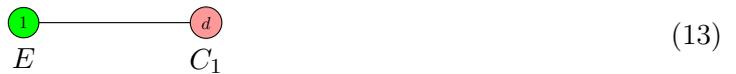
\begin{equation}
\begin{tikzpicture}[scale=0.6, baseline=(current  bounding  box.center)]
      \def\s{4.0};
      \draw (0,0) -- (\s,0);
      \node at (0,0) [stuff_fill_green, scale=0.6, label=below:$E$]{$1$};
      \node at (\s,0) [stuff_fill_red, scale=0.6, label=below:$C_1$]{$d$};
\end{tikzpicture}
\label{equ:SpecialConfigurationForEllipticTree}
\end{equation}
The degree 1 line bundle on $E$ is either $\mathcal{O}_E(p)$, where $p$ is the node on $E$, or $\mathcal{O}_E(q)$ for some other point $q \in E$. Applying the algorithm involves pruning $C_1$ and demanding that sections of the line bundle on $E$ have a zero at $p$. This leads to the weighted dual graph $\Gamma^\prime$:
\begin{equation}
\begin{tikzpicture}[scale=0.6, baseline=(current  bounding  box.center)]
      \node at (0,0) [stuff_fill_green, scale=0.6, label=below:$E$]{$0$};
\end{tikzpicture}
\end{equation}
At this stage, the line bundle on $E$ is either $\mathcal{O}_E(p-p) = \mathcal{O}_E$, which has $h^0=1$, or $\mathcal{O}_E(q-p)$, which has $h^0 = 0$.  

We illustrate this algorithm with a few examples. Here, genus 0 components are marked in pink while genus 1 components are marked in green. When it is possible for two vertices to be connected to each other by more than one edge, we draw two dashed edges between the two vertices. 

\begin{exmp}[When the tree is based at a rational component] Let $(C, L)$ be depicted by the following graph $\Gamma$. Suppose that $\Gamma$ contains a tree, consisting of rational components $C_1$ to $C_4$, that is connected to a rational component $R$. Let $S$ be the complement of the tree in $\Gamma$. 
\begin{equation}
\begin{tikzpicture}[scale=0.6, baseline=(current  bounding  box.center)]
      \def\s{4.0};
      \path[-,out = 30, in = 150, dashed] (-2*\s,0) edge (-1*\s,0);
      \path[-,out = -30, in = -150, dashed] (-2*\s,0) edge (-1*\s,0);
      \draw (-1*\s,0) -- (3*\s,0);
      \node at (-2*\s,0) [stuff_fill_connect, scale=0.6, label=below:$S \setminus R$]{$d$};
      \node at (-1*\s,0) [stuff_fill_red, scale=0.6, label=below:$R$]{$d_R = -2$};
      \node at (0,0) [stuff_fill_red, scale=0.6, label=below:$C_1$]{$d_1 = 0$};
      \node at (1*\s,0) [stuff_fill_red, scale=0.6, label=below:$C_2$]{$d_2 = 3$};
      \node at (2*\s,0) [stuff_fill_red, scale=0.6, label=below:$C_3$]{$d_3 = -2$};
      \node at (3*\s,0) [stuff_fill_red, scale=0.6, label=below:$C_4$]{$d_4 = 1$};
\end{tikzpicture} 
\end{equation}
First, prune the leaf $C_4$ from $C_3$. This leads to the following pair $(C', L')$: 
\begin{equation}
\begin{tikzpicture}[scale=0.6, baseline=(current  bounding  box.center)]
      \def\s{4.0};
      \path[-,out = 30, in = 150, dashed] (-2*\s,0) edge (-1*\s,0);
      \path[-,out = -30, in = -150, dashed] (-2*\s,0) edge (-1*\s,0);
      \draw (-1*\s,0) -- (2*\s,0);
      \node at (-2*\s,0) [stuff_fill_connect, scale=0.6, label=below:$S \setminus R$]{$d$};
      \node at (-1*\s,0) [stuff_fill_red, scale=0.6, label=below:$R$]{$d_R = -2$};
      \node at (0,0) [stuff_fill_red, scale=0.6, label=below:$C_1$]{$d_1 = 0$};
      \node at (1*\s,0) [stuff_fill_red, scale=0.6, label=below:$C_2$]{$d_2 = 3$};
      \node at (2*\s,0) [stuff_fill_red, scale=0.6, label=below:$C_3$]{$d_3 = -2$};
\end{tikzpicture} 
\end{equation}
Since $d_4 \geq 0$, we have that $h^0( C, L ) = h^0( C', L' ) + 1$. Next, prune $C_3$ from $C_2$. This leads to the following pair $(C'', L'')$\begin{equation}
\begin{tikzpicture}[scale=0.6, baseline=(current  bounding  box.center)]
      \def\s{4.0};
      \path[-,out = 30, in = 150, dashed] (-2*\s,0) edge (-1*\s,0);
      \path[-,out = -30, in = -150, dashed] (-2*\s,0) edge (-1*\s,0);
      \draw (-1*\s,0) -- (1*\s,0);
      \node at (-2*\s,0) [stuff_fill_connect, scale=0.6, label=below:$S \setminus R$]{$d$};
      \node at (-1*\s,0) [stuff_fill_red, scale=0.6, label=below:$R$]{$d_R = -2$};
      \node at (0,0) [stuff_fill_red, scale=0.6, label=below:$C_1$]{$d_1 = 0$};
      \node at (1*\s,0) [stuff_fill_red, scale=0.6, label=below:$C_2$]{$d_2 = 2$};
\end{tikzpicture} 
\end{equation}
Note that $d_2$ was reduced by $1$ since $d_3 = -2 < 0$. We have that $h^0( C, L) = h^0( C'', L'') + 1$. Next, prune $C_2$ from $C_1$, which leads to the following pair $(C''', L''')$:
\begin{equation}
\begin{tikzpicture}[scale=0.6, baseline=(current  bounding  box.center)]
      \def\s{4.0};
      \path[-,out = 30, in = 150, dashed] (-2*\s,0) edge (-1*\s,0);
      \path[-,out = -30, in = -150, dashed] (-2*\s,0) edge (-1*\s,0);
      \draw (-1*\s,0) -- (0,0);
      \node at (-2*\s,0) [stuff_fill_connect, scale=0.6, label=below:$S \setminus R$]{$d$};
      \node at (-1*\s,0) [stuff_fill_red, scale=0.6, label=below:$R$]{$d_R = -2$};
      \node at (0,0) [stuff_fill_red, scale=0.6, label=below:$C_1$]{$d_1 = 0$};
\end{tikzpicture} 
\end{equation}
Since $d_2 = 2$, the algorithm tells us to increase the offset by $2$. Thus, $h^0( C, L ) = h^0( C''', L''') + 3$. Next, prune $C_1$ from $R$ and obtain the pair $(C'''' = S, L'''')$.
\begin{equation}
\begin{tikzpicture}[scale=0.6, baseline=(current  bounding  box.center)]
      \def\s{4.0};
      \path[-,out = 30, in = 150, dashed] (-2*\s,0) edge (-1*\s,0);
      \path[-,out = -30, in = -150, dashed] (-2*\s,0) edge (-1*\s,0);
      \node at (-2*\s,0) [stuff_fill_connect, scale=0.6, label=below:$S \setminus R$]{$d$};
      \node at (-1*\s,0) [stuff_fill_red, scale=0.6, label=below:$R$]{$d_R = -2$};
\end{tikzpicture} 
\end{equation}
Since $d_1 = 0$, the offset is 0 and $h^0( C, L ) = h^0( C^{\prime\prime\prime\prime}, L'''') + 3$. \end{exmp}

\begin{exmp}[When the tree is based at an elliptic component] Suppose that $\Gamma$ contains a tree, consisting of rational components $C_1$ to $C_4$, that is connected to an elliptic component $E$.
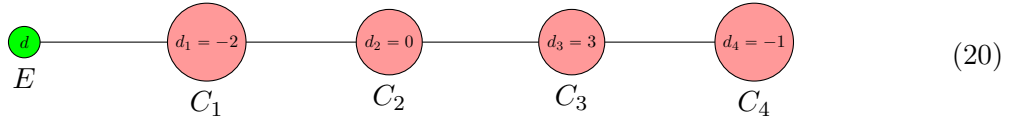
\begin{equation}
\begin{tikzpicture}[scale=0.6, baseline=(current  bounding  box.center)]

      \def\s{4.0};

      \draw (-2*\s,0) -- (2*\s,0);

      \node at (-2*\s,0) [stuff_fill_green, scale=0.6, label=below:$E$]{$d$};
      \node at (-1*\s,0) [stuff_fill_red, scale=0.6, label=below:$C_1$]{$d_1 = -2$};
      \node at (0,0) [stuff_fill_red, scale=0.6, label=below:$C_2$]{$d_2 = 0$};
      \node at (1*\s,0) [stuff_fill_red, scale=0.6, label=below:$C_3$]{$d_3 = 3$};
      \node at (2*\s,0) [stuff_fill_red, scale=0.6, label=below:$C_4$]{$d_4 = -1$};
\end{tikzpicture} 
\end{equation}
First, prune $C_4$ from $C_3$. Since $d_4 < 0$, we replace $d_3$ with $d_3-1$. This leads to the pair $(C', L')$:
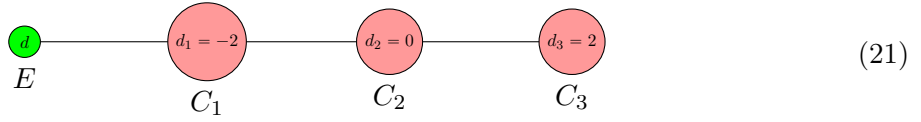
\begin{equation}
\begin{tikzpicture}[scale=0.6, baseline=(current  bounding  box.center)]

      \def\s{4.0};

      \draw (-2*\s,0) -- (1*\s,0);

      \node at (-2*\s,0) [stuff_fill_green, scale=0.6, label=below:$E$]{$d$};
      \node at (-1*\s,0) [stuff_fill_red, scale=0.6, label=below:$C_1$]{$d_1 = -2$};
      \node at (0,0) [stuff_fill_red, scale=0.6, label=below:$C_2$]{$d_2 = 0$};
      \node at (1*\s,0) [stuff_fill_red, scale=0.6, label=below:$C_3$]{$d_3 = 2$};
\end{tikzpicture} 
\end{equation}
We have that $h^0(C, L) = h^0(C', L')$. Next, prune $C_3$ from $C_2$. This leads to the pair $(C'', L'')$:
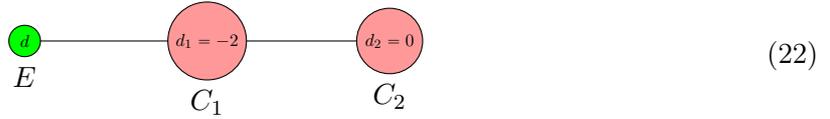
\begin{equation}
\begin{tikzpicture}[scale=0.6, baseline=(current  bounding  box.center)]

      \def\s{4.0};

      \draw (-2*\s,0) -- (0,0);

      \node at (-2*\s,0) [stuff_fill_green, scale=0.6, label=below:$E$]{$d$};
      \node at (-1*\s,0) [stuff_fill_red, scale=0.6, label=below:$C_1$]{$d_1 = -2$};
      \node at (0,0) [stuff_fill_red, scale=0.6, label=below:$C_2$]{$d_2 = 0$};
\end{tikzpicture} 
\end{equation}
We have that $h^0(C, L) = h^0(C'', L'')+2$. Pruning $C_2$ from $C_1$ leads to the pair $(C''', L''')$:
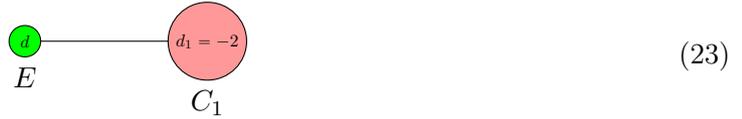
\begin{equation}
\begin{tikzpicture}[scale=0.6, baseline=(current  bounding  box.center)]

      \def\s{4.0};

      \draw (-2*\s,0) -- (-1*\s,0);

      \node at (-2*\s,0) [stuff_fill_green, scale=0.6, label=below:$E$]{$d$};
      \node at (-1*\s,0) [stuff_fill_red, scale=0.6, label=below:$C_1$]{$d_1 = -2$};
     \end{tikzpicture} 
     \end{equation}
We have that $h^0(C, L) = h^0(C''', L''')+2$. Next, prune the leaf $C_1$ from $E$. Since $d_1 < 0$, we replace $d$ with $d-1$. This leaves us with the pair $(C'''' = E, L'''')$:
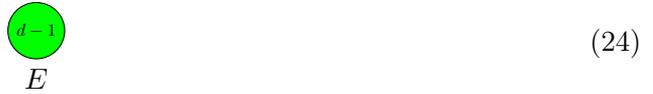
\begin{equation}
\begin{tikzpicture}[scale=0.6, baseline=(current  bounding  box.center)]

      \def\s{4.0};


      \node at (-2*\s,0) [stuff_fill_green, scale=0.6, label=below:$E$]{$d-1$};
     \end{tikzpicture} 
     \end{equation}  
The final formula for $h^0(C'''', L'''')$ is 
$$h^0(C'''', L'''') = h^0(E, L'''')+2 = 2 +  \begin{cases}
d, &\text{if } d \geq 2, \\
0 \text{ or } 1, & \text{if } d = 1, \\
0, &\text{if } d < 1. \\
\end{cases}$$ 
\end{exmp}

\begin{exmp}[When the tree is based at a elliptic component] Let $(C, L)$ be depicted by the following graph $\Gamma$. Suppose that $\Gamma$ contains a tree, consisting of rational components $C_1$ to $C_4$, that is connected to an elliptic component $E$. Let $S$ be the complement of the tree in $\Gamma$. 
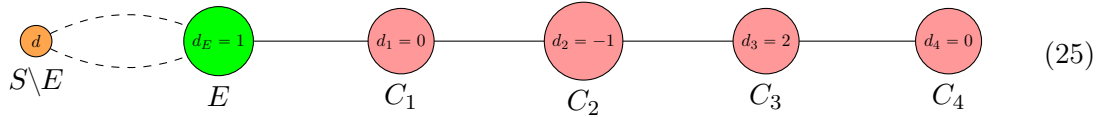
\begin{equation}
\begin{tikzpicture}[scale=0.6, baseline=(current  bounding  box.center)]
      \def\s{4.0};
      \path[-,out = 30, in = 150, dashed] (-2*\s,0) edge (-1*\s,0);
      \path[-,out = -30, in = -150, dashed] (-2*\s,0) edge (-1*\s,0);
      \draw (-1*\s,0) -- (3*\s,0);
      \node at (-2*\s,0) [stuff_fill_connect, scale=0.6, label=below:$S \setminus E$]{$d$};
      \node at (-1*\s,0) [stuff_fill_green, scale=0.6, label=below:$E$]{$d_E = 1$};
      \node at (0,0) [stuff_fill_red, scale=0.6, label=below:$C_1$]{$d_1 = 0$};
      \node at (1*\s,0) [stuff_fill_red, scale=0.6, label=below:$C_2$]{$d_2 = -1$};
      \node at (2*\s,0) [stuff_fill_red, scale=0.6, label=below:$C_3$]{$d_3 = 2$};
      \node at (3*\s,0) [stuff_fill_red, scale=0.6, label=below:$C_4$]{$d_4 = 0$};
\end{tikzpicture} 
\end{equation}
First, prune the leaf $C_4$ from $C_3$. This leads to the following pair $(C', L')$: 
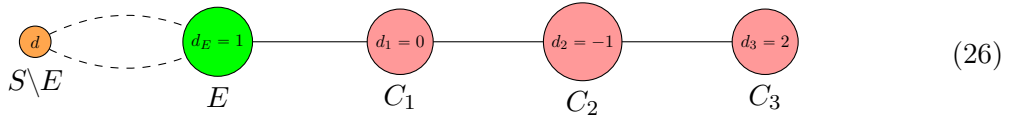
\begin{equation}
\begin{tikzpicture}[scale=0.6, baseline=(current  bounding  box.center)]
      \def\s{4.0};
      \path[-,out = 30, in = 150, dashed] (-2*\s,0) edge (-1*\s,0);
      \path[-,out = -30, in = -150, dashed] (-2*\s,0) edge (-1*\s,0);
      \draw (-1*\s,0) -- (2*\s,0);
      \node at (-2*\s,0) [stuff_fill_connect, scale=0.6, label=below:$S \setminus E$]{$d$};
      \node at (-1*\s,0) [stuff_fill_green, scale=0.6, label=below:$E$]{$d_E = 1$};
      \node at (0,0) [stuff_fill_red, scale=0.6, label=below:$C_1$]{$d_1 = 0$};
      \node at (1*\s,0) [stuff_fill_red, scale=0.6, label=below:$C_2$]{$d_2 = -1$};
      \node at (2*\s,0) [stuff_fill_red, scale=0.6, label=below:$C_3$]{$d_3 = 2$};
\end{tikzpicture} 
\end{equation}
We have that $h^0( C, L ) = h^0( C', L' )$. Next, prune $C_3$ from $C_2$. This leads to the pair $(C'', L''):$
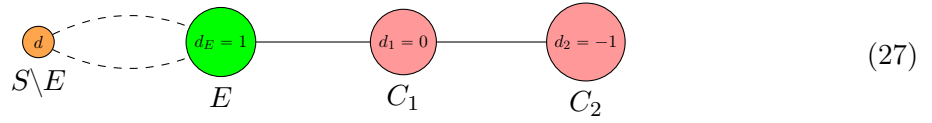
\begin{equation}
\begin{tikzpicture}[scale=0.6, baseline=(current  bounding  box.center)]
      \def\s{4.0};
      \path[-,out = 30, in = 150, dashed] (-2*\s,0) edge (-1*\s,0);
      \path[-,out = -30, in = -150, dashed] (-2*\s,0) edge (-1*\s,0);
      \draw (-1*\s,0) -- (1*\s,0);
      \node at (-2*\s,0) [stuff_fill_connect, scale=0.6, label=below:$S \setminus E$]{$d$};
      \node at (-1*\s,0) [stuff_fill_green, scale=0.6, label=below:$E$]{$d_E = 1$};
      \node at (0,0) [stuff_fill_red, scale=0.6, label=below:$C_1$]{$d_1 = 0$};
      \node at (1*\s,0) [stuff_fill_red, scale=0.6, label=below:$C_2$]{$d_2 = -1$};
\end{tikzpicture} 
\end{equation}
Since $d_3 \geq 0$, we have that $h^0( C, L) = h^0( C'', L'') + 2$. Next, prune $C_2$ from $C_1$, which leads to the following pair $(C''', L''')$:
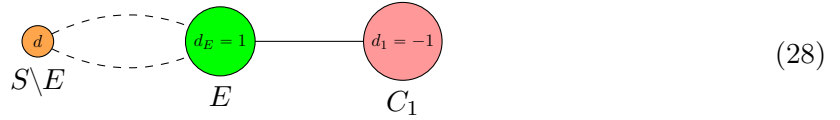
\begin{equation}
\begin{tikzpicture}[scale=0.6, baseline=(current  bounding  box.center)]
      \def\s{4.0};
      \path[-,out = 30, in = 150, dashed] (-2*\s,0) edge (-1*\s,0);
      \path[-,out = -30, in = -150, dashed] (-2*\s,0) edge (-1*\s,0);
      \draw (-1*\s,0) -- (0,0);
      \node at (-2*\s,0) [stuff_fill_connect, scale=0.6, label=below:$S \setminus E$]{$d$};
      \node at (-1*\s,0) [stuff_fill_green, scale=0.6, label=below:$E$]{$d_E = 1$};
      \node at (0,0) [stuff_fill_red, scale=0.6, label=below:$C_1$]{$d_1 = -1$};
\end{tikzpicture} 
\end{equation}
Since $d_2 <0$, we reduce $d_1$ by 1 and we have $h^0( C, L ) = h^0( C''', L''') + 2$. Next, prune $C_1$ from $E$ and obtain the pair $(C'''' = S, L'''')$.
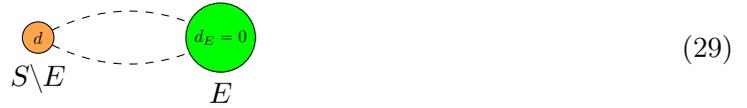
\begin{equation}
\begin{tikzpicture}[scale=0.6, baseline=(current  bounding  box.center)]
      \def\s{4.0};
      \path[-,out = 30, in = 150, dashed] (-2*\s,0) edge (-1*\s,0);
      \path[-,out = -30, in = -150, dashed] (-2*\s,0) edge (-1*\s,0);
      \node at (-2*\s,0) [stuff_fill_connect, scale=0.6, label=below:$S \setminus E$]{$d$};
      \node at (-1*\s,0) [stuff_fill_green, scale=0.6, label=below:$E$]{$d_E = 0$};
\end{tikzpicture} 
\end{equation}
Since $d_1 < 0$, we reduce $d_E$ by 1 and we have $h^0( C, L ) = h^0( C^{\prime\prime\prime\prime}, L'''') + 2$. 
\end{exmp}

\subsection{Removing interior edges} \label{subsec:RemovingInteriorEdges}

We present our second major technique that simplifies the pair $(C, L)$.
Suppose that a rational component $C_{\mathrm{mid}}$ of $C$ is connected to two other components $C_1$ and $C_2$ (not necessarily distinct) via a single edge each. 
We call $C_{\mathrm{mid}}$ and the two half-edges emanating from it an \emph{interior edge}. 

\begin{algo} \label{algo2}
Suppose that $C$ contains an interior edge $C_{\mathrm{mid}}$ that meets $C_1$ at node $p_1$ and $C_2$ at node $p_2$. Let $d = \deg(L|_{C_{\mathrm{mid}}})$.
\begin{enumerate}
\item If $d > 0$, let $C' = C \setminus C_{\mathrm{mid}}$ and $L' = L|_{C'}$. Then, $h^0(C, L) = h^0(C', L') + d-1$.
\item If $d < 0$, let $C' = C \setminus C_{\mathrm{mid}}$ and define a new line bundle $L' = L \otimes O_{C^\prime}(-p_1-p_2)$ on $C'$. Then, $h^0(C, L) = h^0(C', L')$.
\item If $d =0$, form a new curve $C^\prime$ by removing $C_{\mathrm{mid}}$ from $C$ and joining the remaining half-edges to form an edge. This gives rise to a map $\pi: C \rightarrow C^\prime$ that collapses the interior edge to the new node. Define the sheaf $L' = \pi_*(L)$ on $C'$, which is a line bundle because $d=0$. Then, $h^0(C, L) = h^0(C', L')$.
\end{enumerate}
\end{algo}
Again, if $C_i$ is rational, then twisting $L_{C_i}$ by $\mathcal{O}_{C_i}( - p_i )$ in the algorithm amounts to decreasing $\deg(L|_{C_i})$ by 1. We illustrate the different cases of the algorithm in the following figures. We mark the vertex orange if its genus is either 0 or 1. 

\begin{fig}[Case 1: $d > 0$]
Consider the pair $(C, L)$ given by the following graph and let $S = C \setminus (C_1 \cup C_{\mathrm{mid}} \cup C_2)$. Denoting $d_i = \deg(L|_{C_i})$, we have that
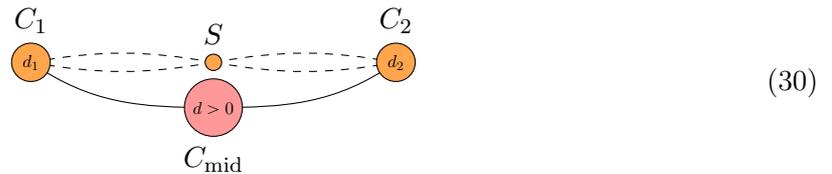
\begin{equation}
\begin{tikzpicture}[scale=0.6, baseline=(current  bounding  box.center)]

      \def\s{1.0};

      \path[-,out = -35, in = 180] (-2*\s,0) edge (2*\s,-1*\s);
      \path[-,out = -145, in = 0] (6*\s,0) edge (2*\s,-1*\s);
       \path[-,out = 10, in = 170, dashed] (-2*\s,0) edge (2*\s,0);
      \path[-,out = -10, in = -170, dashed] (-2*\s,0) edge (2*\s,0);
       \path[-,out = 10, in = 170, dashed] (2*\s,0) edge (6*\s,0);
      \path[-,out = -10, in = -170, dashed] (2*\s,0) edge (6*\s,0);
      \node at (-2*\s,0) [stuff_fill_connect, scale=0.6, label=above:$C_1$]{$d_1$};
      \node at (2*\s,0) [stuff_fill_connect, scale=0.6, label=above:$S$]{};
      \node at (6*\s,0) [stuff_fill_connect, scale=0.6, label=above:$C_2$]{$d_2$};
      \node at (2*\s,-1*\s) [stuff_fill_red, scale=0.6, label=below:$C_{\mathrm{mid}}$]{$d > 0$};
\end{tikzpicture}
\end{equation}
By the algorithm, we define a new pair $(C' = C \setminus C_{\mathrm{mid}}, L' = L|_{C'})$, which is given by the following graph:
\begin{equation}
\begin{tikzpicture}[scale=0.6, baseline=(current  bounding  box.center)]

      \def\s{1.0};

     \path[-,out = 10, in = 170, dashed] (-2*\s,0) edge (2*\s,0);
      \path[-,out = -10, in = -170, dashed] (-2*\s,0) edge (2*\s,0);
       \path[-,out = 10, in = 170, dashed] (2*\s,0) edge (6*\s,0);
      \path[-,out = -10, in = -170, dashed] (2*\s,0) edge (6*\s,0);      
      \node at (-2*\s,0) [stuff_fill_connect, scale=0.6, label=above:$C_1$]{$d_1$};
      \node at (2*\s,0) [stuff_fill_connect, scale=0.6, label=above:$S$]{};
      \node at (6*\s,0) [stuff_fill_connect, scale=0.6, label=above:$C_2$]{$d_2$};
\end{tikzpicture} 
\end{equation}
We have that $h^0(C, L) = h^0(C', L') + d-1$.
\end{fig}
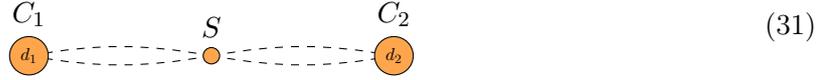
\begin{fig}[Case 2: $d < 0$] Consider the pair $(C, L)$ given by the following graph and let $S = C \setminus (C_1 \cup C_{\mathrm{mid}} \cup C_2)$. Denoting $d_i = \deg(L|_{C_i})$, we have that
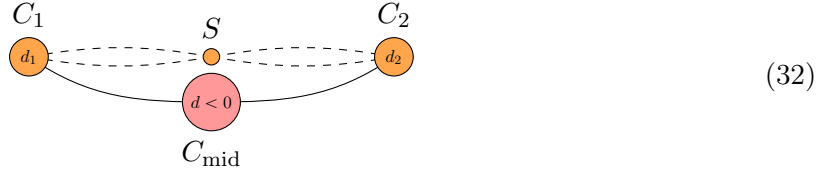
\begin{equation}
\begin{tikzpicture}[scale=0.6, baseline=(current  bounding  box.center)]

      \def\s{1.0};

      \path[-,out = -35, in = 180] (-2*\s,0) edge (2*\s,-1*\s);
      \path[-,out = -145, in = 0] (6*\s,0) edge (2*\s,-1*\s);
      \path[-,out = 10, in = 170, dashed] (-2*\s,0) edge (2*\s,0);
      \path[-,out = -10, in = -170, dashed] (-2*\s,0) edge (2*\s,0);
       \path[-,out = 10, in = 170, dashed] (2*\s,0) edge (6*\s,0);
      \path[-,out = -10, in = -170, dashed] (2*\s,0) edge (6*\s,0);
      
      \node at (-2*\s,0) [stuff_fill_connect, scale=0.6, label=above:$C_1$]{$d_1$};
      \node at (2*\s,0) [stuff_fill_connect, scale=0.6, label=above:$S$]{};
      \node at (6*\s,0) [stuff_fill_connect, scale=0.6, label=above:$C_2$]{$d_2$};
      \node at (2*\s,-1*\s) [stuff_fill_red, scale=0.6, label=below:$C_{\mathrm{mid}}$]{$d < 0$};
\end{tikzpicture} 
\end{equation}
By the above algorithm, we define a new pair $(C', L')$ by the following graph:
\begin{equation}
\begin{tikzpicture}[scale=0.6, baseline=(current  bounding  box.center)]

      \def\s{1.0};

      \path[-,out = 10, in = 170, dashed] (-2*\s,0) edge (2*\s,0);
      \path[-,out = -10, in = -170, dashed] (-2*\s,0) edge (2*\s,0);
       \path[-,out = 10, in = 170, dashed] (2*\s,0) edge (6*\s,0);
      \path[-,out = -10, in = -170, dashed] (2*\s,0) edge (6*\s,0);
      
      \node at (-2*\s,0) [stuff_fill_connect, scale=0.6, label=above:$C_1$]{$d_1-1$};
      \node at (2*\s,0) [stuff_fill_connect, scale=0.6, label=above:$S$]{};
      \node at (6*\s,0) [stuff_fill_connect, scale=0.6, label=above:$C_2$]{$d_2-1$};
\end{tikzpicture} 
\end{equation}
Then, $h^0(C, L) = h^0(C', L')$. 
\end{fig}
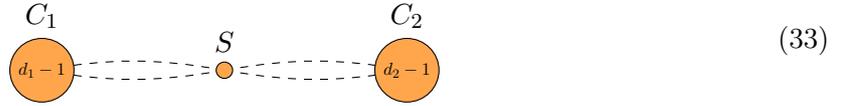
\begin{fig}[Case 3: $d = 0$]
Consider the pair $(C, L)$ given by the following graph and let $S = C \setminus (C_1 \cup C_{\mathrm{mid}} \cup C_2)$. Denoting $d_i = \deg(L|_{C_i})$, we have that
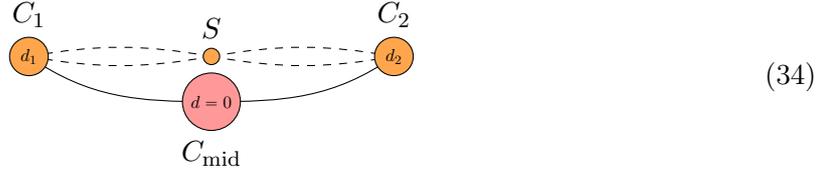
\begin{equation}
\begin{tikzpicture}[scale=0.6, baseline=(current  bounding  box.center)]

      \def\s{1.0};

      \path[-,out = -35, in = 180] (-2*\s,0) edge (2*\s,-1*\s);
      \path[-,out = -145, in = 0] (6*\s,0) edge (2*\s,-1*\s);
     \path[-,out = 10, in = 170, dashed] (-2*\s,0) edge (2*\s,0);
      \path[-,out = -10, in = -170, dashed] (-2*\s,0) edge (2*\s,0);
       \path[-,out = 10, in = 170, dashed] (2*\s,0) edge (6*\s,0);
      \path[-,out = -10, in = -170, dashed] (2*\s,0) edge (6*\s,0);
      
      \node at (-2*\s,0) [stuff_fill_connect, scale=0.6, label=above:$C_1$]{$d_1$};
      \node at (2*\s,0) [stuff_fill_connect, scale=0.6, label=above:$S$]{};
      \node at (6*\s,0) [stuff_fill_connect, scale=0.6, label=above:$C_2$]{$d_2$};
      \node at (2*\s,-1*\s) [stuff_fill_red, scale=0.6, label=below:$C_{\mathrm{mid}}$]{$d =0$};
\end{tikzpicture} 
\end{equation}
By the algorithm, define a new pair $(C', L')$ given by the following graph:
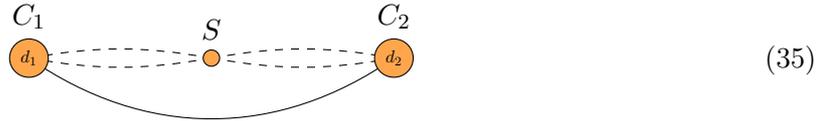
\begin{equation}
\begin{tikzpicture}[scale=0.6, baseline=(current  bounding  box.center)]

      \def\s{1.0};

      \path[-,out = 10, in = 170, dashed] (-2*\s,0) edge (2*\s,0);
      \path[-,out = -10, in = -170, dashed] (-2*\s,0) edge (2*\s,0);
       \path[-,out = 10, in = 170, dashed] (2*\s,0) edge (6*\s,0);
      \path[-,out = -10, in = -170, dashed] (2*\s,0) edge (6*\s,0);      \path[-,out = -35, in = -145] (-2*\s,0) edge (6*\s,0);
      
      \node at (-2*\s,0) [stuff_fill_connect, scale=0.6, label=above:$C_1$]{$d_1$};
      \node at (2*\s,0) [stuff_fill_connect, scale=0.6, label=above:$S$]{};
      \node at (6*\s,0) [stuff_fill_connect, scale=0.6, label=above:$C_2$]{$d_2$};
\end{tikzpicture}
\end{equation}
Then, $h^0(C, L) = h^0(C', L')$, 
\end{fig}
We will now apply the algorithm to a few examples.
\begin{exmp}
Consider the pair $(C, L)$ given by the following graph:
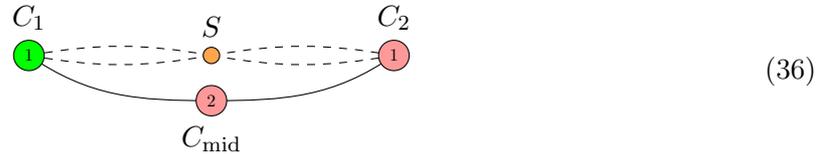
\begin{equation}
\begin{tikzpicture}[scale=0.6, baseline=(current  bounding  box.center)]
      \def\s{1.0};

      \path[-,out = -35, in = 180] (-2*\s,0) edge (2*\s,-1*\s);
      \path[-,out = -145, in = 0] (6*\s,0) edge (2*\s,-1*\s);
      \path[-,out = 10, in = 170, dashed] (-2*\s,0) edge (2*\s,0);
      \path[-,out = -10, in = -170, dashed] (-2*\s,0) edge (2*\s,0);
       \path[-,out = 10, in = 170, dashed] (2*\s,0) edge (6*\s,0);
      \path[-,out = -10, in = -170, dashed] (2*\s,0) edge (6*\s,0);
      
      \node at (-2*\s,0) [stuff_fill_green, scale=0.6, label=above:$C_1$]{$1$};
      \node at (2*\s,0) [stuff_fill_connect, scale=0.6, label=above:$S$]{};
      \node at (6*\s,0) [stuff_fill_red, scale=0.6, label=above:$C_2$]{$1$};
      \node at (2*\s,-1*\s) [stuff_fill_red, scale=0.6, label=below:$C_{\mathrm{mid}}$]{$2$};
\end{tikzpicture}
\end{equation}
Since $\deg(L|_{C_{\mathrm{mid}}}) > 0$, the algorithm tells us to form the new pair $(C', L')$ given by the graph:
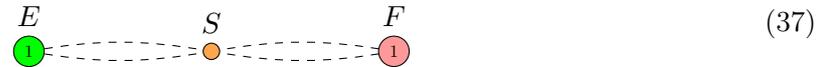
\begin{equation}
\begin{tikzpicture}[scale=0.6, baseline=(current  bounding  box.center)]
      \def\s{1.0};

      \path[-,out = 10, in = 170, dashed] (-2*\s,0) edge (2*\s,0);
      \path[-,out = -10, in = -170, dashed] (-2*\s,0) edge (2*\s,0);
       \path[-,out = 10, in = 170, dashed] (2*\s,0) edge (6*\s,0);
      \path[-,out = -10, in = -170, dashed] (2*\s,0) edge (6*\s,0);
      
      \node at (-2*\s,0) [stuff_fill_green, scale=0.6, label=above:$E$]{$1$};
      \node at (2*\s,0) [stuff_fill_connect, scale=0.6, label=above:$S$]{};
      \node at (6*\s,0) [stuff_fill_red, scale=0.6, label=above:$F$]{$1$};
\end{tikzpicture}
\end{equation}
Then, $h^0(C, L) = h^0(C', L') + 2-1= h^0(C', L') + 1$.
\end{exmp}
\begin{exmp}
Consider the pair $(C, L)$ given by the following graph:
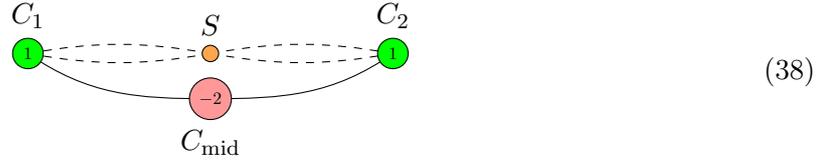
\begin{equation}
\begin{tikzpicture}[scale=0.6, baseline=(current  bounding  box.center)]
      \def\s{1.0};

      \path[-,out = -35, in = 180] (-2*\s,0) edge (2*\s,-1*\s);
      \path[-,out = -145, in = 0] (6*\s,0) edge (2*\s,-1*\s);
      \path[-,out = 10, in = 170, dashed] (-2*\s,0) edge (2*\s,0);
      \path[-,out = -10, in = -170, dashed] (-2*\s,0) edge (2*\s,0);
       \path[-,out = 10, in = 170, dashed] (2*\s,0) edge (6*\s,0);
      \path[-,out = -10, in = -170, dashed] (2*\s,0) edge (6*\s,0);
      
      \node at (-2*\s,0) [stuff_fill_green, scale=0.6, label=above:$C_1$]{$1$};
      \node at (2*\s,0) [stuff_fill_connect, scale=0.6, label=above:$S$]{};
      \node at (6*\s,0) [stuff_fill_green, scale=0.6, label=above:$C_2$]{$1$};
      \node at (2*\s,-1*\s) [stuff_fill_red, scale=0.6, label=below:$C_{\mathrm{mid}}$]{$-2$};
\end{tikzpicture}
\end{equation}
Since $\deg(C_{\mathrm{mid}}) < 0$, the algorithm tells us to remove $C_{\mathrm{mid}}$, adjust the degrees and form the new pair $(C', L')$:
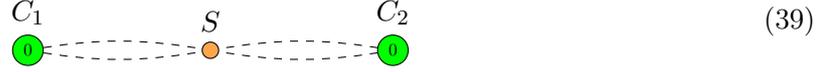
\begin{equation}
\begin{tikzpicture}[scale=0.6, baseline=(current  bounding  box.center)]
      \def\s{1.0};

    \path[-,out = 10, in = 170, dashed] (-2*\s,0) edge (2*\s,0);
      \path[-,out = -10, in = -170, dashed] (-2*\s,0) edge (2*\s,0);
       \path[-,out = 10, in = 170, dashed] (2*\s,0) edge (6*\s,0);
      \path[-,out = -10, in = -170, dashed] (2*\s,0) edge (6*\s,0);
      
      \node at (-2*\s,0) [stuff_fill_green, scale=0.6, label=above:$C_1$]{$0$};
      \node at (2*\s,0) [stuff_fill_connect, scale=0.6, label=above:$S$]{};
      \node at (6*\s,0) [stuff_fill_green, scale=0.6, label=above:$C_2$]{$0$};
\end{tikzpicture}
\end{equation}
Then, $h^0(C, L) = h^0(C', L').$ Since there are degree 0 bundles on $C_1$ and $C_2$, we can only compute the lower bound of $h^0$. \end{exmp}
\begin{exmp}
Consider the pair $(C, L)$ given by the following graph:
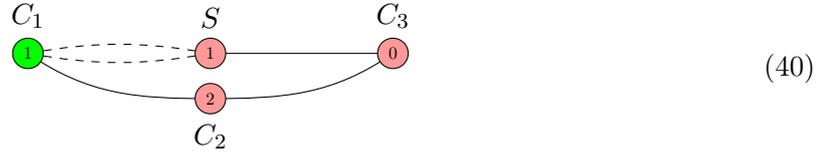
\begin{equation}
\begin{tikzpicture}[scale=0.6, baseline=(current  bounding  box.center)]
      \def\s{1.0};

      \path[-,out = -35, in = 180] (-2*\s,0) edge (2*\s,-1*\s);
      \path[-,out = -145, in = 0] (6*\s,0) edge (2*\s,-1*\s);
      \path[-,out = 10, in = 170, dashed] (-2*\s,0) edge (2*\s,0);
      \path[-,out = -10, in = -170, dashed] (-2*\s,0) edge (2*\s,0);
      \draw (2*\s,0) -- (6*\s,0);
      \node at (-2*\s,0) [stuff_fill_green, scale=0.6, label=above:$C_1$]{$1$};
      \node at (2*\s,0) [stuff_fill_red, scale=0.6, label=above:$S$]{$1$};
      \node at (6*\s,0) [stuff_fill_red, scale=0.6, label=above:$C_3$]{$0$};
      \node at (2*\s,-1*\s) [stuff_fill_red, scale=0.6, label=below:$C_2$]{$2$};
\end{tikzpicture}
\end{equation}
We can first remove $C_3$ using the algorithm. Since $\deg(L|_{C_3})= 0$, we remove $C_2$ and join the half-edges to produce a new pair $(C', L')$:
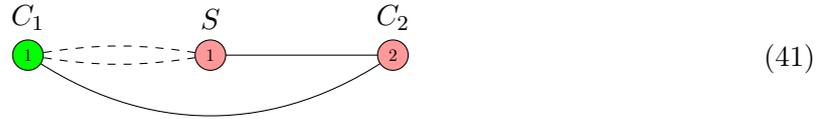
\begin{equation}
\begin{tikzpicture}[scale=0.6, baseline=(current  bounding  box.center)]
      \def\s{1.0};

      \path[-,out = 10, in = 170, dashed] (-2*\s,0) edge (2*\s,0);
      \path[-,out = -10, in = -170, dashed] (-2*\s,0) edge (2*\s,0);
     \draw (2*\s, 0) -- (6*\s, 0);
      \path[-,out = -35, in = -145] (-2*\s,0) edge (6*\s,0);

      \node at (-2*\s,0) [stuff_fill_green, scale=0.6, label=above:$C_1$]{$1$};
      \node at (2*\s,0) [stuff_fill_red, scale=0.6, label=above:$S$]{$1$};
      \node at (6*\s,0) [stuff_fill_red, scale=0.6, label=above:$C_2$]{$2$};
\end{tikzpicture}
\end{equation}
Then, $h^0(C, L) = h^0(C', L')$. To remove $C_2$, we apply the algorithm to form a new pair $(C'', L'')$: 
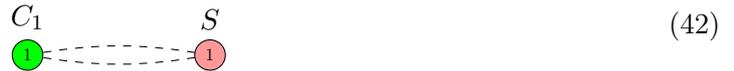
\begin{equation}
\begin{tikzpicture}[scale=0.6, baseline=(current  bounding  box.center)]
      \def\s{1.0};

      \path[-,out = 10, in = 170, dashed] (-2*\s,0) edge (2*\s,0);
      \path[-,out = -10, in = -170, dashed] (-2*\s,0) edge (2*\s,0);

      \node at (-2*\s,0) [stuff_fill_green, scale=0.6, label=above:$C_1$]{$1$};
      \node at (2*\s,0) [stuff_fill_red, scale=0.6, label=above:$S$]{$1$};
\end{tikzpicture}
\end{equation}
Then, $h^0(C, L) = h^0(C'', L'') + 1$.

Alternatively, we could remove $C_2$ first. This involves forming a new curve $(\widetilde{C}', \widetilde{L}')$:
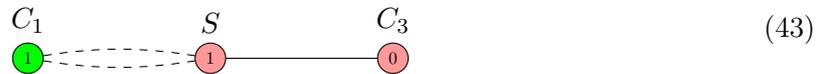
\begin{equation}
\begin{tikzpicture}[scale=0.6, baseline=(current  bounding  box.center)]
      \def\s{1.0};

      \path[-,out = 10, in = 170, dashed] (-2*\s,0) edge (2*\s,0);
      \path[-,out = -10, in = -170, dashed] (-2*\s,0) edge (2*\s,0);
     \draw (2*\s, 0) -- (6*\s, 0);

      \node at (-2*\s,0) [stuff_fill_green, scale=0.6, label=above:$C_1$]{$1$};
      \node at (2*\s,0) [stuff_fill_red, scale=0.6, label=above:$S$]{$1$};
      \node at (6*\s,0) [stuff_fill_red, scale=0.6, label=above:$C_3$]{$0$};
\end{tikzpicture}
\end{equation}
Then, $h^0(C, L) = h^0(\widetilde{C}', \widetilde{L}') + 1$. Next, we view $C_3$ as a leaf and prune it by forming the new pair $(\widetilde{C}'', \widetilde{L}'')$:
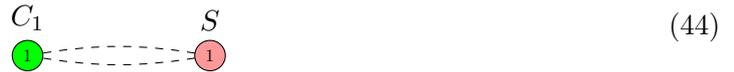
\begin{equation}
\begin{tikzpicture}[scale=0.6, baseline=(current  bounding  box.center)]
      \def\s{1.0};

      \path[-,out = 10, in = 170, dashed] (-2*\s,0) edge (2*\s,0);
      \path[-,out = -10, in = -170, dashed] (-2*\s,0) edge (2*\s,0);

      \node at (-2*\s,0) [stuff_fill_green, scale=0.6, label=above:$C_1$]{$1$};
      \node at (2*\s,0) [stuff_fill_red, scale=0.6, label=above:$S$]{$1$};
\end{tikzpicture}
\end{equation}
Then, $h^0(C, L) = h^0(\widetilde{C}'', \widetilde{L}'') + 1$. Note that $(\widetilde{C}'', \widetilde{L}'') = (C'', L'')$. So, both procedures conclude with the same result.
\end{exmp}
We can also apply this to remove interior edges connected to the same curve component.
\begin{exmp}[When $C_1=C_2$]
Consider the pair $(C, L)$ given by the following graph:
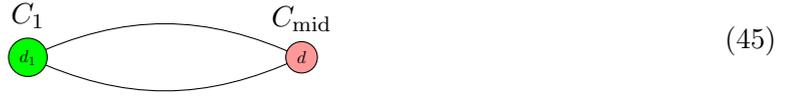
\begin{equation}
\begin{tikzpicture}[scale=0.6, baseline=(current  bounding  box.center)]

      \def\s{1.0};

      \path[-,out = -25, in = -155] (-2*\s,0) edge (4*\s,0);
      \path[-,out = 25, in = 155] (-2*\s,0) edge (4*\s,0);
      
      \node at (-2*\s,0) [stuff_fill_green, scale=0.6, label=above:$C_1$]{$d_1$};
      \node at (4*\s,0) [stuff_fill_red, scale=0.6,  label=above:$C_{\mathrm{mid}}$]{$d$};
\end{tikzpicture}
\end{equation}
If $d = 0$, the algorithm tells us to remove $C_{\mathrm{mid}}$ and join the half-edges to form the new pair $(C', L')$:
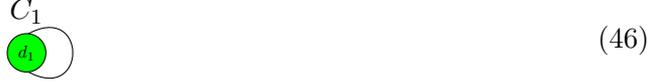
\begin{equation}
\begin{tikzpicture}[scale=0.6, baseline=(current  bounding  box.center)]
      \def\s{2.0};
      \path[-,out = -30, in = 30, looseness = 5] (-2*\s,-0.2*\s) edge (-2*\s,+0.2*\s);
      \node at (-2*\s,0) [stuff_fill_green, scale=0.6, label=above:$C_1$]{$d_1$};
    \end{tikzpicture}
\end{equation}
Then $h^0(C, L) = h^0(C', L')$. 
If $d < 0$, the algorithm tells us to remove $C_{\mathrm{mid}}$ and adjust the degrees on $C_1$ in the following way to obtain the new pair $(C', L')$:
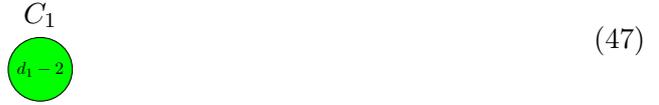
\begin{equation}
\begin{tikzpicture}[scale=0.6, baseline=(current  bounding  box.center)]
      \def\s{2.0};
      \node at (-2*\s,0) [stuff_fill_green, scale=0.6, label=above:$C_1$]{$d_1-2$};
    \end{tikzpicture}
\end{equation}
By Riemann-Roch we have that:
\begin{equation}
h^0(C, L) = h^0(C', L') = \begin{cases}
 	d_1-2, & \text{if } d_1 > 2, \\
 	0 \text{ or } 1, & \text{if } d_1 = 2,\\
 	0, & \text{if } d_1 < 2.
 \end{cases}
 \end{equation}
To explain the $d_1 = 2$ case further, let $p$ and $q$ be the nodes on $C_1$. 
Applying the algorithm involves demanding that the line bundle sections on $C_1$ vanish at $p$ and $q$. 
If $L|_{C_1} \cong \mathcal{O}_{C_1}(p+q)$, then the algorithm results in the line bundle $L' = \mathcal{O}_{C_1}(p+q-p-q) = \mathcal{O}_{C_1}$, which has $h^0(C', L')=1$. Otherwise, it is a non-trivial, degree 0 line bundle on $C_1$ and $h^0(C', L') = 0$. 

If $d > 0$, we remove $C_{\mathrm{mid}}$ and obtain the new pair $(C', L')$. The algorithm states that $h^0(C, L) = h^0(C', L') + d-1$.
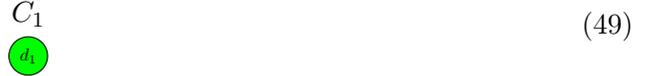
\begin{equation}
\begin{tikzpicture}[scale=0.6, baseline=(current  bounding  box.center)]
      \def\s{2.0};
      \node at (-2*\s,0) [stuff_fill_green, scale=0.6, label=above:$C_1$]{$d_1$};
    \end{tikzpicture}
\end{equation}
\end{exmp}

\subsubsection{Proofs of interior edge removal algorithm}

In this section, we present a case-by-case proof of the interior-edge-removal techniques presented earlier. Some of these results could be proved more efficiently using techniques from algebraic geometry but we choose to leave the elementary proofs for accessibility. Suppose that a pair $(C, L)$ contains an interior edge $C_{\mathrm{mid}}$ between two components $C_1$ and $C_2$. Denote $d = \deg(L|_{C_{\mathrm{mid}}})$ and $d_i = \deg(L|_{C_i})$. Since $\mathrm{PGL}_2$ acts on $\mathbb{P}^1$ transitively, we may assume that $C_{\mathrm{mid}}$ meets $C_1$ and $C_2$ at the points $[1: 0]$ and $[0: 1]$  on $C_{\mathrm{mid}}$ respectively.

\subsubsection*{Case 1: \texorpdfstring{$d > 0$.}{d > 0.}}
\begin{lem} If $C' = C \setminus C_{\mathrm{mid}}$ and $L' = L|_{C'}$, then $h^0(C, L) = h^0(C', L') + d-1$. \end{lem}
\begin{myproof}
Let $n_1, n_2$ be the nodes at which $C_{\mathrm{mid}}$ is attached to $C_1$ and $C_2$. Observe that we have the following:
\begin{align}
h^0(C, L) &= h^0(C' \cup C_{\mathrm{mid}}, L) \\
&\geq h^0(C', L') + h^0(C_{\mathrm{mid}}, L|_{C_{\mathrm{mid}}}) - h^0(\{n_\alpha, n_\beta\}, L|_{\{n_\alpha, n_\beta\}}) \\
&= h^0(C', L') + d- 1. 
 \end{align}
To prove that equality is achieved, we need to show that the nodes impose exactly two independent conditions. Let $[z: w]$ be the coordinates on $C_{\mathrm{mid}} \cong \mathbb{P}^1$. Then, any section $s \in H^0(C_{\mathrm{mid}}, L)$ is of the form 
\begin{equation}
s([z:w]) = \sum^d_{i=0}\gamma_i z^iw^{d-i}, \quad \gamma_i \in \mathbb{C}.	
\end{equation}
Let $s_1 \in H^0(C_1, L|_{C_1})$ and $s_2 \in H^0(C_2, L|_{C_2})$ be arbitrary sections. Then, $s$ coincides with $s_1$ at the node $[1:0]$ and $s$ coincies with $s_2$ at the node $[0:1]$ up to descent data, i.e.  
\begin{equation}
\gamma_0 = s([1:0]) = \lambda_1 s_1([1:0]), \quad \gamma_d = r=s([0:1]) = \lambda_2 s_2([0:1]) \, ,
\end{equation}
where $\lambda_1, \lambda_2 \in \mathbb{C}^*$ are descent data parameters assigned at the nodes. These equations are well-defined in the sense that scaling the intersection points on $\mathbb{P}^1$ does not change $h^0(C, L)$. Indeed, the basis sections of $H^0(C, L)$ will differ by a $\mathrm{PGL}_2(\mathbb{C})$-transformation under such scaling. Hence, these euqations are two independent conditions imposed onto $H^0(C_{\mathrm{mid}}, L|_{C_\mathrm{mid}})$. So, the result follows.\end{myproof}

\subsubsection*{Case 2: \texorpdfstring{$d < 0$.}{d < 0.}}
\begin{lem} Let $(C', L')$ be the pair defined as in Algorithm \ref{algo2}. Then, $h^0(C, L) = h^0(C', L')$. \end{lem} 

\begin{myproof}
Suppose $d_i = \deg(L|_{C_i}) > 0$ and $p_i$ be the nodes on $C_i$ for $i = 1,2$. Then, sections $s_1 \in H^0(C_1, L|_{C_1})$ and $s_2 \in H^0(C_2, L|_{C_2})$ satisfy the following gluing conditions at the nodes regardless of the descent data:
\begin{equation}
s_1(p_1) = 0 \, , \quad  s_2(p_2) = 0 \, .
\end{equation}
Demanding $s_i$ to vanish at $p_i$ amounts to twisting the line bundle on $C_i$ by $O_{C_i}(-p_i)$. Hence, the result $h^0(C, L) = h^0(C', L')$ follows for $d_i > 0$. 
 
 If one of the $d_i$ is negative, then the zero section on the corresponding component $C_i$ glues to the zero section on $C_{\mathrm{mid}}$ uniquely. Since $d_i < 0$, twisting the line bundle on $C_i$ still results in a negative degree $d_i - 1 < 0$. Hence, $h^0(C_i, L'|_{C_i}) = h^0(C_i, L|_{C_i}) = 0$ and the result follows. \end{myproof}

\subsubsection*{Case 3: \texorpdfstring{$d=0$.}{d = 0.}}
\begin{lem} Suppose that $C_{\mathrm{mid}}$ met $C_1$ at the node $p_1 \in C_1$ and $C_2$ at the node $p_2 \in C_2$. Remove $C_{\mathrm{mid}}$ and join the half-edges together so that $C_1$ meets $C_2$ at $p_1 \in C_1$ and $p_2 \in C_2$. is gives rise to a map $\pi: C \rightarrow C^\prime$ that collapses the interior edge to the new node. Define the sheaf $L' = \pi_*(L)$ on $C'$, which is a line bundle because $d=0$. Then, $h^0(C, L) = h^0(C', L')$. \end{lem}

\begin{myproof}
Since $d$ is zero, $H^0(C_{\mathrm{mid}}, L|_{C_{\mathrm{mid}}})$ is spanned by one constant section $\gamma$. 
Suppose that sections over $C_1$ (resp. $C_2$) is glued to sections over $C_{\mathrm{mid}}$ with descent data $1 \in \mathbb{C}^*$ (resp. $\lambda$). 
Any section $s_1 \in H^0(C_1, L|_{C_1})$ and $s_2 \in H^0(C_2, L|_{C_2})$ satisfy $s_1(p_1)= \gamma$ and $s_2(p_2) = \lambda \gamma = \lambda s_1(p_1)$. This shows that $h^0(C, L) = h^0(C', L')$.
\end{myproof}

\subsection{Terminal graphs} \label{subsec:TerminalGraphs}

\begin{defn}
We call a connected graph $\Gamma$ that cannot be reduced further by our techniques (pruning external trees and removing internal edges) terminal.
\end{defn}

\subsubsection{Rational circuits}

\begin{defn}
A rational circuit is a terminal graph whose components are all rational.
\end{defn}

\begin{exmp}
The only rational circuit $\Gamma$ with $\beta_1( \Gamma ) = 0$ consists of a single vertex and no edges, i.e. is a rational curve.
\end{exmp}

\begin{lem}
If $\Gamma$ is a rational circuit with $1 < V$ vertices and first Betti number $\beta_1( \Gamma )$. Then
\begin{align}
V \leq 2 \beta_1( \Gamma ) - 2 \, .
\end{align}
\end{lem}

\begin{myproof}
Recall that the first Betti number of a graph is given by
\begin{align}
\beta_1( \Gamma ) = E + C - V \, ,
\end{align}
where $E$ is the number of edges, $C$ the number of connected components and $V$ the number of vertices. We assume that $\Gamma$ is a rational circuit. So, by definition, $\Gamma$ is connected and $C = 1$. We see that
\begin{align}
E = \beta_1( \Gamma ) + V - 1 \, .
\end{align}
Assume that there are at least two (rational) vertices $V$ in $\Gamma$. Such a vertex can be removed by our techniques if there are less than three edges connecting to it. Otherwise, this vertex either belongs to a tree or is an internal edge. So in order for the graph $\Gamma$ to be terminal, we must have at least 3 edges connecting to each vertex:
\begin{align}
\frac{3V}{2} \leq E = \beta_1( \Gamma ) + V - 1 \, .
\end{align}
Consequently, we conclude that for a fixed first Betti number $\beta_1( \Gamma )$, a rational circuit has a number of vertices $V$ subject to the condition $V \leq 2 \beta_1( \Gamma ) - 2$.
\end{myproof}

\begin{exmp}
Let us consider rational circuits $\Gamma$ with $\beta_1( \Gamma ) = 1$ and $1 < V$. Then, the above bound states:
\begin{align}
V \leq 2 \beta_1( \Gamma ) - 2 = 0 \, .
\end{align}
Consequently, there are no such rational circuits. Rather, the only rational circuits $\Gamma$ with $\beta_1( \Gamma ) = 1$ satisfy $V = 1$ and $E = 1$. Consequently, the only rational circuit with $\beta_1( \Gamma ) = 1$ is as follows:
\begin{align}
\begin{tikzpicture}[scale=0.6, baseline=(current  bounding  box.center)]
      \def\s{2.0};
      \path[-,out = -30, in = 30, looseness = 5] (-2*\s,-0.2*\s) edge (-2*\s,+0.2*\s);
      \node at (-2*\s,0) [stuff_fill_red, scale=0.6]{$d$};
\end{tikzpicture}
\end{align}
\end{exmp}

Observe that each rational circuit $\Gamma$ describes, and is determined by, a family of stable curves of arithmetic genus $\beta_1 \left( \Gamma \right)$. Since the Deligne-Mumford compactified moduli space classifying stable curves of fixed genus has finitely many strata, we can conclude the following.

\begin{cor}
The number of rational circuits with fixed first Betti number $\beta_1( \Gamma )$ is finite.
\end{cor}

\begin{exmp}
For $\beta_1( \Gamma ) \in \{ 0, 1, 2, 3 \}$, the rational circuits are as follows:
\begin{itemize}
\item $\beta_1( \Gamma ) = 0$:\\
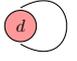
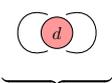
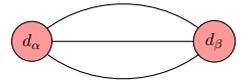
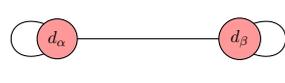
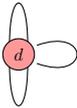
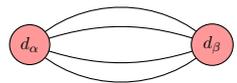
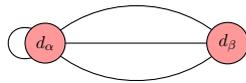
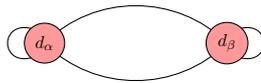
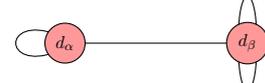
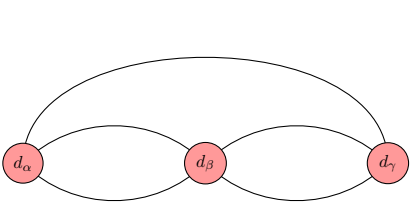
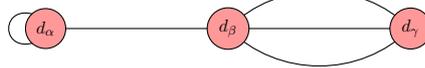
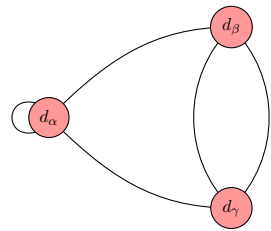
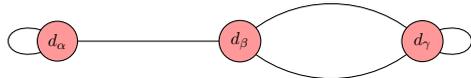
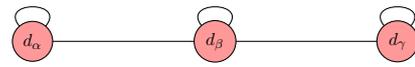
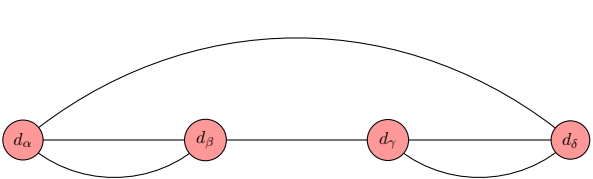
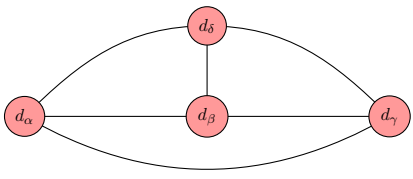
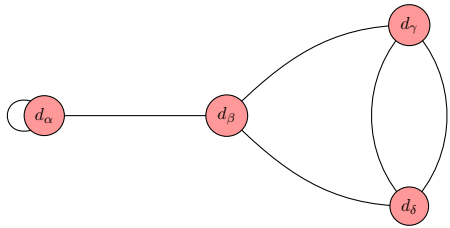
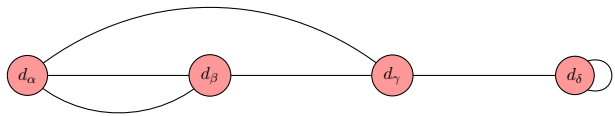
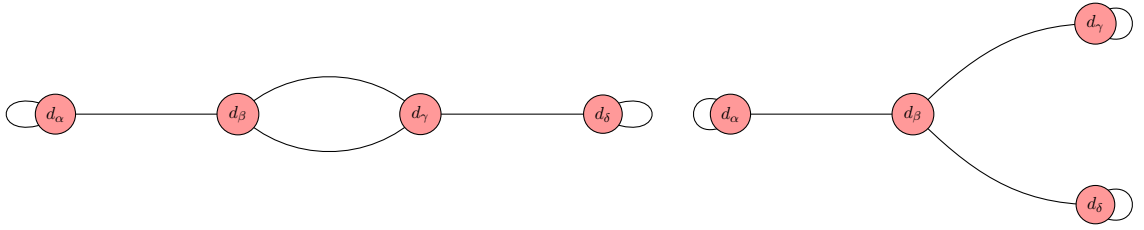
\begin{tikzpicture}[scale=0.6, baseline=(current  bounding  box.center)]
      \def\s{2.0};
      \node at (-2*\s,0) [stuff_fill_red, scale=0.6]{$d$};
\end{tikzpicture}

 \item $\beta_1( \Gamma ) = 1$:\\
\begin{tikzpicture}[scale=0.6, baseline=(current  bounding  box.center)]
      \def\s{2.0};
      \path[-,out = -30, in = 30, looseness = 5] (-2*\s,-0.2*\s) edge (-2*\s,+0.2*\s);
      \node at (-2*\s,0) [stuff_fill_red, scale=0.6]{$d$};
      \draw [thick, decorate, decoration = {calligraphic brace}] (-1.4*\s,-1) --  (-2.2*\s,-1) node[pos=0.25*\s,below=6pt]{$G_1$};
\end{tikzpicture}

 \item $\beta_1( \Gamma ) = 2$:\\
\begin{tikzpicture}[scale=0.6, baseline=(current  bounding  box.center)]
      \def\s{2.0};
      \path[-,out = -20, in = 20, looseness = 3] (-1.9*\s,-0.2*\s) edge (-1.9*\s,+0.2*\s);
      \path[-,out = -160, in = 160, looseness = 3] (-2.1*\s,-0.2*\s) edge (-2.1*\s,+0.2*\s);
      \node at (-2*\s,0) [stuff_fill_red, scale=0.6]{$d$};
      \draw [thick, decorate, decoration = {calligraphic brace}] (-1.4*\s,-1) --  (-2.6*\s,-1) node[pos=0.25*\s,below=6pt]{$G_2$};
\end{tikzpicture}\;
\begin{tikzpicture}[scale=0.6, baseline=(current  bounding  box.center)]
      \def\s{1.0};
      \path[-,out = 45, in = 135] (-2*\s,0) edge (+2*\s,0);
      \path[-,out = 0, in = 180] (-2*\s,0) edge (+2*\s,0);
      \path[-,out = -45, in = -135] (-2*\s,0) edge (+2*\s,0);
      \node at (-2*\s,0) [stuff_fill_red, scale=0.6]{$d_\alpha$};
      \node at (+2*\s,0) [stuff_fill_red, scale=0.6]{$d_\beta$};
      \draw [thick, decorate, decoration = {calligraphic brace}] (2.6*\s,-1) --  (-2.6*\s,-1) node[pos=0.5*\s,below=6pt]{$G_3$};
\end{tikzpicture}\;
\begin{tikzpicture}[scale=0.6, baseline=(current  bounding  box.center)]
      \def\s{1.0};
      \path[-,out = -150, in = 150, looseness = 10] (-2*\s,-0.2*\s) edge (-2*\s,+0.2*\s);
      \path[-,out = -30, in = 30, looseness = 10] (2*\s,-0.2*\s) edge (2*\s,+0.2*\s);
      \path[-,out = 180, in = 0] (-2*\s,0) edge (2*\s,+0);
      \node at (-2*\s,0) [stuff_fill_red, scale=0.6]{$d_\alpha$};
      \node at (2*\s,0) [stuff_fill_red, scale=0.6]{$d_\beta$};
      \draw [thick, decorate, decoration = {calligraphic brace}] (2.6*\s,-1) --  (-2.6*\s,-1) node[pos=0.5*\s,below=6pt]{$G_4$};
\end{tikzpicture}

 \item $\beta_1( \Gamma ) = 3$:\\
\begin{tikzpicture}[scale=0.6, baseline=(current  bounding  box.center)]
      \def\s{2.0};
      \path[-,out = -20, in = 20, looseness = 8] (0.2*\s,-0.1*\s) edge (0.2*\s,+0.1*\s);
      \path[-,out = -90, in = -90, looseness = 8] (-0.1*\s,-0.1*\s) edge (0.1*\s,-0.1*\s);
      \path[-,out = 90, in =90, looseness = 8] (-0.1*\s,0.1*\s) edge (0.1*\s,0.1*\s);
      \node at (0,0) [stuff_fill_red, scale=0.6]{$d$};
\end{tikzpicture}

\begin{tikzpicture}[scale=0.6, baseline=(current  bounding  box.center)]
      \def\s{1.0};
      \path[-,out = 45, in = 135] (-2*\s,0) edge (+2*\s,0);
      \path[-,out = 20, in = 160] (-2*\s,0) edge (+2*\s,0);
      \path[-,out = -20, in = -160] (-2*\s,0) edge (+2*\s,0);
      \path[-,out = -45, in = -135] (-2*\s,0) edge (+2*\s,0);
      \node at (-2*\s,0) [stuff_fill_red, scale=0.6]{$d_\alpha$};
      \node at (+2*\s,0) [stuff_fill_red, scale=0.6]{$d_\beta$};
      \draw [thick, decorate, decoration = {calligraphic brace}] (2.6*\s,-1) --  (-2.6*\s,-1) node[pos=0.5*\s,below=6pt]{$G_5$};
\end{tikzpicture}\;
\begin{tikzpicture}[scale=0.6, baseline=(current  bounding  box.center)]
      \def\s{1.0};
      \path[-,out = -150, in = 150, looseness = 8] (-2*\s,-0.2*\s) edge (-2*\s,+0.2*\s);
      \path[-,out = 45, in = 135] (-2*\s,0) edge (+2*\s,0);
      \path[-,out = 0, in = 180] (-2*\s,0) edge (+2*\s,0);
      \path[-,out = -45, in = -135] (-2*\s,0) edge (+2*\s,0);
      \node at (-2*\s,0) [stuff_fill_red, scale=0.6]{$d_\alpha$};
      \node at (+2*\s,0) [stuff_fill_red, scale=0.6]{$d_\beta$};
\end{tikzpicture}\;
\begin{tikzpicture}[scale=0.6, baseline=(current  bounding  box.center)]
      \def\s{1.0};
      \path[-,out = -150, in = 150, looseness = 8] (-2*\s,-0.2*\s) edge (-2*\s,+0.2*\s);
      \path[-,out = 45, in = 135] (-2*\s,0) edge (+2*\s,0);
      \path[-,out = -45, in = -135] (-2*\s,0) edge (+2*\s,0);
      \path[-,out = -30, in = 30, looseness = 8] (2*\s,-0.2*\s) edge (2*\s,+0.2*\s);
      \node at (-2*\s,0) [stuff_fill_red, scale=0.6]{$d_\alpha$};
      \node at (+2*\s,0) [stuff_fill_red, scale=0.6]{$d_\beta$};
\end{tikzpicture}\;
\begin{tikzpicture}[scale=0.6, baseline=(current  bounding  box.center)]
      \def\s{1.0};
      \path[-,out = -160, in = 160, looseness = 8] (1.8*\s,-0.2*\s) edge (1.8*\s,+0.2*\s);
      \path[-,out = 0, in = 180] (2*\s,0) edge (6*\s,0);
      \path[-,out = -90, in = -90, looseness = 8] (5.8*\s,-0.1*\s) edge (6.2*\s,-0.1*\s);
      \path[-,out = 90, in =90, looseness = 8] (5.8*\s,0.1*\s) edge (6.2*\s,0.1*\s);
      \node at (+2*\s,0) [stuff_fill_red, scale=0.6]{$d_\alpha$};
      \node at (+6*\s,0) [stuff_fill_red, scale=0.6]{$d_\beta$};
\end{tikzpicture}

\begin{tikzpicture}[scale=0.6, baseline=(current  bounding  box.center)]
      \def\s{1.0};
      \path[-,out = 45, in = 135] (-2*\s,0) edge (2*\s,0);
      \path[-,out = -45, in = -135] (-2*\s,0) edge (2*\s,0);
      \path[-,out = 45, in = 135] (2*\s,0) edge (6*\s,0);
      \path[-,out = -45, in = -135] (2*\s,0) edge (6*\s,0);
      \path[-,out = 90, in = 90] (-2*\s,0) edge (6*\s,0);
      \node at (-2*\s,0) [stuff_fill_red, scale=0.6]{$d_\alpha$};
      \node at (+2*\s,0) [stuff_fill_red, scale=0.6]{$d_\beta$};
      \node at (+6*\s,0) [stuff_fill_red, scale=0.6]{$d_\gamma$};
\end{tikzpicture}\;
\begin{tikzpicture}[scale=0.6, baseline=(current  bounding  box.center)]
      \def\s{1.0};
      \path[-,out = -150, in = 150, looseness = 8] (-6*\s,-0.2*\s) edge (-6*\s,+0.2*\s);
      \path[-,out = 0, in = 180] (-6*\s,0) edge (-2*\s,0);
      \path[-,out = 45, in = 135] (-2*\s,0) edge (+2*\s,0);
      \path[-,out = 0, in = 180] (-2*\s,0) edge (+2*\s,0);
      \path[-,out = -45, in = -135] (-2*\s,0) edge (+2*\s,0);
      \node at (-6*\s,0) [stuff_fill_red, scale=0.6]{$d_\alpha$};
      \node at (-2*\s,0) [stuff_fill_red, scale=0.6]{$d_\beta$};
      \node at (+2*\s,0) [stuff_fill_red, scale=0.6]{$d_\gamma$};
\end{tikzpicture}\;
\begin{tikzpicture}[scale=0.6, baseline=(current  bounding  box.center)]
      \def\s{1.0};
      \path[-,out = -150, in = 150, looseness = 8] (-2*\s,-0.2*\s) edge (-2*\s,+0.2*\s);
      \path[-,out = 45, in = 180] (-2*\s,0) edge (+2*\s,+2*\s);
      \path[-,out = -45, in = 180] (-2*\s,0) edge (+2*\s,-2*\s);
      \path[-,out = -45, in = 45] (+2*\s,+2*\s) edge (+2*\s,-2*\s);
      \path[-,out = -135, in = 135] (+2*\s,+2*\s) edge (+2*\s,-2*\s);
      \node at (-2*\s,0) [stuff_fill_red, scale=0.6]{$d_\alpha$};
      \node at (+2*\s,+2*\s) [stuff_fill_red, scale=0.6]{$d_\beta$};
      \node at (+2*\s,-2*\s) [stuff_fill_red, scale=0.6]{$d_\gamma$};
\end{tikzpicture}

\begin{tikzpicture}[scale=0.6, baseline=(current  bounding  box.center)]
      \def\s{1.0};
      \path[-,out = -160, in = 160, looseness = 8] (1.8*\s,-0.2*\s) edge (1.8*\s,+0.2*\s);
      \path[-,out = 0, in = 180] (2*\s,0) edge (6*\s,0);
      \path[-,out = 45, in = 135] (6*\s,0) edge (10*\s,0);
      \path[-,out = -45, in = -135] (6*\s,0) edge (10*\s,0);
      \path[-,out = -20, in = 20, looseness = 8] (10.2*\s,-0.2*\s) edge (10.2*\s,+0.2*\s);
      \node at (+2*\s,0) [stuff_fill_red, scale=0.6]{$d_\alpha$};
      \node at (+6*\s,0) [stuff_fill_red, scale=0.6]{$d_\beta$};
      \node at (+10*\s,0) [stuff_fill_red, scale=0.6]{$d_\gamma$};
\end{tikzpicture}\;
\begin{tikzpicture}[scale=0.6, baseline=(current  bounding  box.center)]
      \def\s{1.0};
      \path[-,out = 40, in = 140, looseness = 10] (-1.85*\s,0.2*\s) edge (-2.15*\s,+0.2*\s);
      \path[-,out = 40, in = 140, looseness = 10] (2.15*\s,0.2*\s) edge (1.85*\s,+0.2*\s);
      \path[-,out = 40, in = 140, looseness = 10] (6.15*\s,0.2*\s) edge (5.85*\s,+0.2*\s);
      \path[-,out = 0, in = 180] (-2*\s,0) edge (6*\s,+0);
      \node at (-2*\s,0) [stuff_fill_red, scale=0.6]{$d_\alpha$};
      \node at (2*\s,0) [stuff_fill_red, scale=0.6]{$d_\beta$};
      \node at (6*\s,0) [stuff_fill_red, scale=0.6]{$d_\gamma$};
\end{tikzpicture}

\begin{tikzpicture}[scale=0.6, baseline=(current  bounding  box.center)]
      \def\s{1.0};
      \path[-,out = 0, in = 180] (-2*\s,0) edge (2*\s,0);
      \path[-,out = -45, in = -135] (-2*\s,0) edge (2*\s,0);
      \path[-,out = 0, in = 180] (2*\s,0) edge (6*\s,0);
      \path[-,out = 0, in = 180] (6*\s,0) edge (10,0);
      \path[-,out = -45, in = -135] (6*\s,0) edge (10,0);
      \path[-,out = 40, in = 140] (-2*\s,0) edge (10,0);
      \node at (-2*\s,0) [stuff_fill_red, scale=0.6]{$d_\alpha$};
      \node at (+2*\s,0) [stuff_fill_red, scale=0.6]{$d_\beta$};
      \node at (+6*\s,0) [stuff_fill_red, scale=0.6]{$d_\gamma$};
      \node at (+10,0) [stuff_fill_red, scale=0.6]{$d_\delta$};
\end{tikzpicture}\;
\begin{tikzpicture}[scale=0.6, baseline=(current  bounding  box.center)]
      \def\s{1.0};
      \path[-,out = 0, in = 180] (-4*\s,0) edge (4*\s,0);
      \path[-,out = 45, in = -180] (-4*\s,0) edge (0*\s,2*\s);
      \path[-,out = 90, in = -90] (0*\s,0) edge (0*\s,2*\s);
      \path[-,out = 135, in = 0] (4*\s,0) edge (0*\s,2*\s);
      \path[-,out = -30, in = -150] (-4*\s,0*\s) edge (4*\s,0*\s);
      \node at (-4*\s,0) [stuff_fill_red, scale=0.6]{$d_\alpha$};
      \node at (0*\s,0) [stuff_fill_red, scale=0.6]{$d_\beta$};
      \node at (+4*\s,0) [stuff_fill_red, scale=0.6]{$d_\gamma$};
      \node at (0*\s,2*\s) [stuff_fill_red, scale=0.6]{$d_\delta$};
\end{tikzpicture}

\begin{tikzpicture}[scale=0.6, baseline=(current  bounding  box.center)]
      \def\s{1.0};
      \path[-,out = -150, in = 150, looseness = 8] (-6*\s,-0.2*\s) edge (-6*\s,+0.2*\s);
      \path[-,out = 0, in = 180] (-6*\s,0) edge (-2*\s,0);
      \path[-,out = 45, in = 180] (-2*\s,0) edge (+2*\s,+2*\s);
      \path[-,out = -45, in = 180] (-2*\s,0) edge (+2*\s,-2*\s);
      \path[-,out = -45, in = 45] (+2*\s,+2*\s) edge (+2*\s,-2*\s);
      \path[-,out = -135, in = 135] (+2*\s,+2*\s) edge (+2*\s,-2*\s);
      \node at (-6*\s,0) [stuff_fill_red, scale=0.6]{$d_\alpha$};
      \node at (-2*\s,0) [stuff_fill_red, scale=0.6]{$d_\beta$};
      \node at (+2*\s,+2*\s) [stuff_fill_red, scale=0.6]{$d_\gamma$};
      \node at (+2*\s,-2*\s) [stuff_fill_red, scale=0.6]{$d_\delta$};
\end{tikzpicture}\;
\begin{tikzpicture}[scale=0.6, baseline=(current  bounding  box.center)]
      \def\s{1.0};
      \path[-,out = 0, in = 180] (-2*\s,0) edge (2*\s,0);
      \path[-,out = -45, in = -135] (-2*\s,0) edge (2*\s,0);
      \path[-,out = 0, in = 180] (2*\s,0) edge (6*\s,0);
      \path[-,out = 0, in = 180] (6*\s,0) edge (10*\s,0);
      \path[-,out = 40, in = 140] (-2*\s,0) edge (6*\s,0);
      \path[-,out = -30, in = 30, looseness = 8] (10*\s,-0.2*\s) edge (10*\s,0.2*\s);
      \node at (-2*\s,0) [stuff_fill_red, scale=0.6]{$d_\alpha$};
      \node at (+2*\s,0) [stuff_fill_red, scale=0.6]{$d_\beta$};
      \node at (+6*\s,0) [stuff_fill_red, scale=0.6]{$d_\gamma$};
      \node at (+10,0) [stuff_fill_red, scale=0.6]{$d_\delta$};
\end{tikzpicture}

\begin{tikzpicture}[scale=0.6, baseline=(current  bounding  box.center)]
      \def\s{1.0};
      \path[-,out = -160, in = 160, looseness = 8] (1.8*\s,-0.2*\s) edge (1.8*\s,+0.2*\s);
      \path[-,out = 0, in = 180] (2*\s,0) edge (6*\s,0);
      \path[-,out = 45, in = 135] (6*\s,0) edge (10*\s,0);
      \path[-,out = -45, in = -135] (6*\s,0) edge (10*\s,0);
      \path[-,out = 0, in = 180] (10*\s,0) edge (14*\s,0);
      \path[-,out = -20, in = 20, looseness = 8] (14.2*\s,-0.2*\s) edge (14.2*\s,+0.2*\s);
      \node at (+2*\s,0) [stuff_fill_red, scale=0.6]{$d_\alpha$};
      \node at (+6*\s,0) [stuff_fill_red, scale=0.6]{$d_\beta$};
      \node at (+10*\s,0) [stuff_fill_red, scale=0.6]{$d_\gamma$};
      \node at (+14*\s,0) [stuff_fill_red, scale=0.6]{$d_\delta$};
\end{tikzpicture}\;
\begin{tikzpicture}[scale=0.6, baseline=(current  bounding  box.center)]
      \def\s{1.0};
      \path[-,out = -150, in = 150, looseness = 8] (-6*\s,-0.2*\s) edge (-6*\s,+0.2*\s);
      \path[-,out = 0, in = 180] (-6*\s,0) edge (-2*\s,0);
      \path[-,out = 45, in = 180] (-2*\s,0) edge (+2*\s,+2*\s);
      \path[-,out = -45, in = 180] (-2*\s,0) edge (+2*\s,-2*\s);
      \path[-,out = -30, in = 30, looseness = 8] (2*\s,1.8*\s) edge (2*\s,+2.2*\s);
      \path[-,out = -30, in = 30, looseness = 8] (2*\s,-2.2*\s) edge (2*\s,-1.8*\s);
      \node at (-6*\s,0) [stuff_fill_red, scale=0.6]{$d_\alpha$};
      \node at (-2*\s,0) [stuff_fill_red, scale=0.6]{$d_\beta$};
      \node at (+2*\s,+2*\s) [stuff_fill_red, scale=0.6]{$d_\gamma$};
      \node at (+2*\s,-2*\s) [stuff_fill_red, scale=0.6]{$d_\delta$};
\end{tikzpicture}
\end{itemize}
\end{exmp}

\begin{note}
A simple computer scan shows that for the QSM applications in this paper, only the graphs $G_1$, $G_2$, $G_3$, $G_4$, $G_5$ appear. We may thus wonder if all line bundles on these curves with the prescribed degrees have the same number of global sections, or if there are exceptions (a.k.a. jumps). For $G_1$, we work out the answer below. The remaining graphs are addressed in \cref{Appendix:Derivations}.
\end{note}

\begin{defn}
Consider a rational circuit $\Gamma$. By definition, this is a terminal graph whose components are all rational. In addition, by our prescription, for each rational component/vertex $V_i$ of $\Gamma$, there is an integer $d_i$ which encodes a line bundle $L_i = \mathcal{O}_{V_i}( d_i ) \cong \mathcal{O}_{\mathbb{P}^1}( d_i )$. Altogether, this defines a family $\mathcal{F}$ of line bundles on the nodal curve $C^\bullet$ whose dual graph is $\Gamma$, such that for every $L \in \mathcal{F}$ and every vertex $V_i$ of $\Gamma$:
\begin{align}
\left. L \right|_{V_i} = L_i = \mathcal{O}_{V_i}( d_i ) \cong \mathcal{O}_{\mathbb{P}^1}( d_i ) \, .
\end{align}
We say that $\Gamma$ is non-jumping if $h^0( C^\bullet, L )$ is the same for all $L \in \mathcal{F}$. Otherwise, we say that $\Gamma$ is jumping.
\end{defn}

\begin{remark}
For our applications to the Quadrillion Standard models, we must tell under what conditions on the degree $d_i$, the rational circuits $G_1$, $G_2$, $G_3$, $G_4$ and $G_5$ are jumping.
\end{remark}

\begin{prop}
The rational circuit $G_1$ is jumping iff $d = 0$:
\begin{align}
\begin{tikzpicture}[scale=0.6, baseline=(current  bounding  box.center)]
\def\s{2.0};
\path[-,out = -30, in = 30, looseness = 5] (-2*\s,-0.2*\s) edge (-2*\s,+0.2*\s);
\node at (-2*\s,0) [stuff_fill_red, scale=0.6]{$d$};
\end{tikzpicture}
\end{align}
\end{prop}

\begin{myproof}
If $d < 0$, then $h^0( G_1, L ) = 0$ for all $L \in \mathcal{F}$. Next, consider $d = 0$. Then the sections on the $\mathbb{P}^1$ are constant. Provided that the descent data is $1$, then any constant section glues across the node and we have $h^0( G_1, L ) = 1$. However, if the descent data satisfies $\lambda \neq 1$, then only the section identically zero glues across the node an $h^0( G_1, L ) = 0$. So for $d = 0$, $G_1$ is jumping. Finally, consider $d > 0$. Pick $[x:y]$ as homogeneous coordinates of $V_1 = \mathbb{P}^1$ and position the nodes (my means of an $\mathrm{SL}(2, \mathbb{C})$ transformation) at $[1:0]$ and $0:1]$. Then the most general section $\varphi \in H^0( V_1, \left. L \right|_{V_1} )$ is of the form
\begin{align}
\varphi = \sum_{i = 0}^{d}{\alpha_i x^i y^{d-i}} \, .
\end{align}
This section is subject to the gluing condition $\alpha_0 = \lambda \alpha_d$. Hence, $h^0( G_1, L ) = d$ for all $L \in \mathcal{F}$.
\end{myproof}

\begin{note}
By similar arguments, one can identify exactly for what line bundle degreees $G_2$, $G_3$, $G_4$ and $G_5$ are jumping. Specifically, one finds the following:
\begin{itemize}
 \item $G_2$ is jumping iff $0 \leq d \leq 2$.
 \item $G_3$ is jumping iff $0 \leq d_\alpha, d_\beta \leq 1$.
 \item $G_4$ is jumping iff one of the following applies:
\begin{itemize}
    \item $d_\alpha < 0$ and $d_\beta = 1$,
    \item $d_\alpha = 0$ and $d_\beta \geq 0$,
    \item $d_\alpha = 1$ and $d_\beta < 0$,
    \item $d_\alpha \geq 0$ and $d_\beta = 0$,
    \item $d_\alpha = d_\beta = 1$.
\end{itemize}
 \item $G_5$ is jumping iff $0 \leq d_\alpha, d_\beta \leq 2$.
\end{itemize}
We provide the details in \cref{Appendix:Derivations}.
\end{note}

\begin{remark}
We notice an interesting pattern. The canonical degree typically appears to be a sort of ``tipping'' points beyond which there cannot be any jumps. For a smooth curve, this is indeed the known pattern by Serre duality. However, we also see exceptions to this rule. For instance, for the graph $G_4$, there is a jump for multidegree $(d,0)$ for all $d \geq 0$. Hence, in particular for any $d$ larger than $2$, i.e. larger than the degree of the canonical bundle of a smooth curve with genus $g = 2$.
\end{remark}

\subsubsection{Elliptic circuits}

\begin{defn}
We call a terminal graph with one elliptic component and otherwise only rational components, an elliptic circuit.
\end{defn}

\begin{note}
The key difference to the previous section is that our simplification steps never remove an elliptic curve. So there is no requirement for the elliptic curve to have at least three edges leading off.
\end{note}

\begin{cor}
The number of connected elliptic circuits is finite.
\end{cor}

\begin{exmp}
For $\beta_1( \Gamma ) \in \{ 0, 1, 2 \}$, the connected elliptic circuits are as follows:
\begin{itemize}
\item $\beta_1( \Gamma ) = 0$:\\
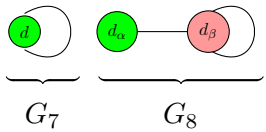
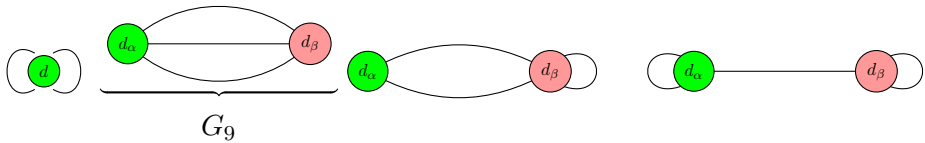
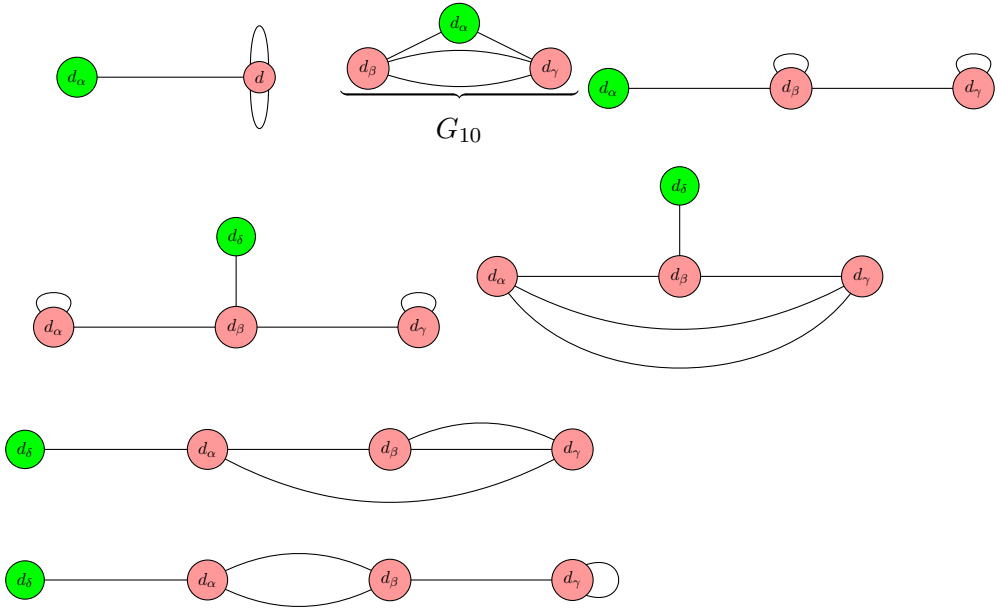
\begin{tikzpicture}[scale=0.6, baseline=(current  bounding  box.center)]
      \def\s{2.0};
      \node at (-2*\s,0) [stuff_fill_green, scale=0.6]{$d$};
      \draw [thick, decorate, decoration = {calligraphic brace}] (-1.8*\s,-0.6) --  (-2.2*\s,-0.6) node[pos=0.25*\s,below=6pt]{$G_6$};
\end{tikzpicture}

\item $\beta_1( \Gamma ) = 1$:\\
\begin{tikzpicture}[scale=0.6, baseline=(current  bounding  box.center)]
      \def\s{2.0};
      \path[-,out = -30, in = 30, looseness = 5] (-2*\s,-0.2*\s) edge (-2*\s,+0.2*\s);
      \node at (-2*\s,0) [stuff_fill_green, scale=0.6]{$d$};
      \draw [thick, decorate, decoration = {calligraphic brace}] (-1.4*\s,-1) --  (-2.2*\s,-1) node[pos=0.25*\s,below=6pt]{$G_7$};
\end{tikzpicture}
\begin{tikzpicture}[scale=0.6, baseline=(current  bounding  box.center)]
      \def\s{2.0};
      \path[-,out = 0, in = 180] (-3*\s,0) edge (-2*\s,0);
      \path[-,out = -30, in = 30, looseness = 5] (-2*\s,-0.2*\s) edge (-2*\s,+0.2*\s);
      \node at (-3*\s,0) [stuff_fill_green, scale=0.6]{$d_\alpha$};
      \node at (-2*\s,0) [stuff_fill_red, scale=0.6]{$d_\beta$};
      \draw [thick, decorate, decoration = {calligraphic brace}] (-1.4*\s,-1) --  (-3.2*\s,-1) node[pos=0.25*\s,below=6pt]{$G_8$};
\end{tikzpicture}

\item $\beta_1( \Gamma ) = 2$:\\
\begin{tikzpicture}[scale=0.6, baseline=(current  bounding  box.center)]
      \def\s{2.0};
      \path[-,out = -30, in = 30, looseness = 3] (-1.9*\s,-0.2*\s) edge (-1.9*\s,+0.2*\s);
      \path[-,out = -150, in = 150, looseness = 3] (-2.1*\s,-0.2*\s) edge (-2.1*\s,+0.2*\s);
      \node at (-2*\s,0) [stuff_fill_green, scale=0.6]{$d$};
\end{tikzpicture}
\begin{tikzpicture}[scale=0.6, baseline=(current  bounding  box.center)]
      \def\s{1.0};
      \path[-,out = 45, in = 135] (-2*\s,0) edge (+2*\s,0);
      \path[-,out = 0, in = 180] (-2*\s,0) edge (+2*\s,0);
      \path[-,out = -45, in = -135] (-2*\s,0) edge (+2*\s,0);
      \node at (-2*\s,0) [stuff_fill_green, scale=0.6]{$d_\alpha$};
      \node at (+2*\s,0) [stuff_fill_red, scale=0.6]{$d_\beta$};
      \draw [thick, decorate, decoration = {calligraphic brace}] (2.6*\s,-1) --  (-2.6*\s,-1) node[pos=0.5*\s,below=6pt]{$G_9$};
\end{tikzpicture}
\begin{tikzpicture}[scale=0.6, baseline=(current  bounding  box.center)]
      \def\s{1.0};
      \path[-,out = -30, in = 30, looseness = 10] (2*\s,-0.2*\s) edge (2*\s,+0.2*\s);
      \path[-,out = -30, in = -150] (-2*\s,0) edge (2*\s,+0);
      \path[-,out = 30, in = 150] (-2*\s,0) edge (2*\s,+0);
      \node at (-2*\s,0) [stuff_fill_green, scale=0.6]{$d_\alpha$};
      \node at (2*\s,0) [stuff_fill_red, scale=0.6]{$d_\beta$};
\end{tikzpicture}
\begin{tikzpicture}[scale=0.6, baseline=(current  bounding  box.center)]
      \def\s{1.0};
      \path[-,out = -150, in = 150, looseness = 10] (-2*\s,-0.2*\s) edge (-2*\s,+0.2*\s);
      \path[-,out = -30, in = 30, looseness = 10] (2*\s,-0.2*\s) edge (2*\s,+0.2*\s);
      \path[-,out = 180, in = 0] (-2*\s,0) edge (2*\s,+0);
      \node at (-2*\s,0) [stuff_fill_green, scale=0.6]{$d_\alpha$};
      \node at (2*\s,0) [stuff_fill_red, scale=0.6]{$d_\beta$};
\end{tikzpicture}

\begin{tikzpicture}[scale=0.6, baseline=(current  bounding  box.center)]
      \def\s{2.0};
      \path[-,out = -90, in = -90, looseness = 8] (-0.1*\s,-0.1*\s) edge (0.1*\s,-0.1*\s);
      \path[-,out = 90, in =90, looseness = 8] (-0.1*\s,0.1*\s) edge (0.1*\s,0.1*\s);
      \path[-,out = 180, in = 0] (-2*\s,0) edge (0,+0);
      \node at (-2*\s,0) [stuff_fill_green, scale=0.6]{$d_\alpha$};
      \node at (0,0) [stuff_fill_red, scale=0.6]{$d$};
\end{tikzpicture}
\begin{tikzpicture}[scale=0.6, baseline=(current  bounding  box.center)]
      \def\s{1.0};
      \draw (0,0) -- (-2*\s,-\s);
      \draw (0,0) -- (2*\s,-\s);
      \path[-,out = 20, in = 160] (-2*\s,-\s) edge (2*\s,-\s);
      \path[-,out = -20, in = -160] (-2*\s,-\s) edge (2*\s,-\s);      
      \node at (0,0) [stuff_fill_green, scale=0.6]{$d_\alpha$};
      \node at (-2*\s,-\s) [stuff_fill_red, scale=0.6]{$d_\beta$};
      \node at (+2*\s,-\s) [stuff_fill_red, scale=0.6]{$d_\gamma$};
      \draw [thick, decorate, decoration = {calligraphic brace}] (2.6*\s,-0.5-\s) --  (-2.6*\s,-0.5-\s) node[pos=0.5*\s,below=6pt]{$G_{10}$};
\end{tikzpicture}
\begin{tikzpicture}[scale=0.6, baseline=(current  bounding  box.center)]
      \def\s{1.0};
      \path[-,out = 40, in = 140, looseness = 10] (2.15*\s,0.2*\s) edge (1.85*\s,+0.2*\s);
      \path[-,out = 40, in = 140, looseness = 10] (6.15*\s,0.2*\s) edge (5.85*\s,+0.2*\s);
      \path[-,out = 0, in = 180] (-2*\s,0) edge (6*\s,+0);
      \node at (-2*\s,0) [stuff_fill_green, scale=0.6]{$d_\alpha$};
      \node at (2*\s,0) [stuff_fill_red, scale=0.6]{$d_\beta$};
      \node at (6*\s,0) [stuff_fill_red, scale=0.6]{$d_\gamma$};
\end{tikzpicture}

\begin{tikzpicture}[scale=0.6, baseline=(current  bounding  box.center)]
      \def\s{1.0};
      \path[-,out = 40, in = 140, looseness = 10] (-1.85*\s,0.2*\s) edge (-2.15*\s,+0.2*\s);
      \path[-,out = 40, in = 140, looseness = 10] (6.15*\s,0.2*\s) edge (5.85*\s,+0.2*\s);
      \path[-,out = 0, in = 180] (-2*\s,0) edge (6*\s,+0);
      \path[-,out = 90, in = -90] (2*\s,0) edge (2*\s,2*\s);
      \node at (-2*\s,0) [stuff_fill_red, scale=0.6]{$d_\alpha$};
      \node at (2*\s,0) [stuff_fill_red, scale=0.6]{$d_\beta$};
      \node at (2*\s,2*\s) [stuff_fill_green, scale=0.6]{$d_\delta$};
      \node at (6*\s,0) [stuff_fill_red, scale=0.6]{$d_\gamma$};
\end{tikzpicture}
\begin{tikzpicture}[scale=0.6, baseline=(current  bounding  box.center)]
      \def\s{1.0};
      \path[-,out = -30, in = -150] (-2*\s,0) edge (6*\s,0);
      \path[-,out = -60, in = -120] (-2*\s,0) edge (6*\s,0);
      \path[-,out = 0, in = 180] (-2*\s,0) edge (6*\s,+0);
      \path[-,out = 90, in = -90] (2*\s,0) edge (2*\s,2*\s);
      \node at (-2*\s,0) [stuff_fill_red, scale=0.6]{$d_\alpha$};
      \node at (2*\s,0) [stuff_fill_red, scale=0.6]{$d_\beta$};
      \node at (2*\s,2*\s) [stuff_fill_green, scale=0.6]{$d_\delta$};
      \node at (6*\s,0) [stuff_fill_red, scale=0.6]{$d_\gamma$};
\end{tikzpicture}

\begin{tikzpicture}[scale=0.6, baseline=(current  bounding  box.center)]
      \def\s{1.0};
      \path[-,out = -30, in = -150] (-2*\s,0) edge (6*\s,0);
      \path[-,out = 30, in = 150] (2*\s,0) edge (6*\s,0);
      \path[-,out = 0, in = 180] (-6*\s,0) edge (6*\s,+0);
      \node at (-2*\s,0) [stuff_fill_red, scale=0.6]{$d_\alpha$};
      \node at (2*\s,0) [stuff_fill_red, scale=0.6]{$d_\beta$};
      \node at (-6*\s,0) [stuff_fill_green, scale=0.6]{$d_\delta$};
      \node at (6*\s,0) [stuff_fill_red, scale=0.6]{$d_\gamma$};
\end{tikzpicture}

\begin{tikzpicture}[scale=0.6, baseline=(current  bounding  box.center)]
      \def\s{1.0};
      \path[-,out = -30, in = -150] (-2*\s,0) edge (2*\s,0);
      \path[-,out = 30, in = 150] (-2*\s,0) edge (2*\s,0);
      \path[-,out = 0, in = 180] (-6*\s,0) edge (-2*\s,+0);
      \path[-,out = 0, in = 180] (2*\s,0) edge (6*\s,+0);
      \path[-,out = -30, in = 30, looseness = 10] (6*\s,-0.2*\s) edge (6*\s,+0.2*\s);
      \node at (-2*\s,0) [stuff_fill_red, scale=0.6]{$d_\alpha$};
      \node at (2*\s,0) [stuff_fill_red, scale=0.6]{$d_\beta$};
      \node at (-6*\s,0) [stuff_fill_green, scale=0.6]{$d_\delta$};
      \node at (6*\s,0) [stuff_fill_red, scale=0.6]{$d_\gamma$};
\end{tikzpicture}
\end{itemize}
\end{exmp}

\begin{note}
A simple computer scan shows that only $G_6$, $G_7$, $G_8$, $G_9$, $G_{10}$ appear in our QSM applications. We focus on the degrees that matter for the QSM applications and show that each configuration is jumping. For this, we construct a line bundle on the terminal graph whose number of global sections is strictly larger than the lower bound given the sum of the number of local sections minus the number of edges. The detailed arguments are given in \cref{sec:AppendixEllipticCircuitJumps}.
\end{note}

\section{Implications for F-theory QSMs} \label{sec:ImplicationsForFTheoryQSMs}

\subsection{Improved statistics for ``rational'' QSMs}

\begin{table}[!htbp]
\begin{center}
\begin{tabular}{cc|cc|cc|cc|c}
\toprule
\multicolumn{9}{c}{Results based on \cite{BCDO22}}\\
Polytope & $\overline{K}_{B_3}^3$ & $= 3$ & $\geq 3$ & $= 4$ & $\geq 4$ & $= 5$ & $\geq 5$ & $= 6$ \\
\midrule
$\Delta_4^\circ$ & $6$ & $99.9952$ & & $0.0048$ & & & \\
$\Delta_{134}^\circ$ & $6$ & $99.7830$ & $0.2170$ & & & & \\
$\Delta_{128}^\circ$, $\Delta_{130}^\circ$, $\Delta_{136}^\circ$, $\Delta_{236}^\circ$ & $6$ & $99.8891$ & $0.1109$ & & & & \\
\midrule
$\Delta_{254}^\circ$ & $10$ & $95.9365$ & $0.5264$ & $3.5065$ & $0.0131$ & $0.0175$ & $0.0000$ \\
$\Delta_{52}^\circ$ & $10$ & $95.3269$ & $0.7489$ & $3.8812$ & $0.0254$ & $0.0175$ & $0.0001$ \\
$\Delta_{302}^\circ$ & $10$ & $95.9225 $ & $ 0.5372 $ & $ 3.5090 $ & $ 0.0133 $ & $ 0.0181$ & \\
$\Delta_{786}^\circ$ & $10$ & $94.8418 $ & $ 0.2689 $ & $ 4.8368 $ & $ 0.0100 $ & $ 0.0425 $ & $ 0.0000$ \\
$\Delta_{762}^\circ$ & $10$ & $94.7752 $ & $ 0.2785 $ & $ 4.8948 $ & $ 0.0110 $ & $ 0.0405 $ & $ 0.0000$ \\
\midrule
$\Delta_{417}^\circ$ & $10$ & $94.8380 $ & $ 0.2799 $ & $ 4.8290 $ & $ 0.0102 $ & $ 0.0428 $ & $ 0.0000 $ & $ 0.0001$ \\
$\Delta_{838}^\circ$ & $10$ & $94.6510 $ & $ 0.2798 $ & $ 5.0106 $ & $ 0.0109 $ & $ 0.0476 $ & $ 0.0001$ \\
$\Delta_{782}^\circ$ & $10$ & $94.6400 $ & $ 0.2780 $ & $ 5.0244 $ & $ 0.0115 $ & $ 0.0460 $ & $ 0.0001$ \\
$\Delta_{377}^\circ$, $\Delta_{499}^\circ$, $\Delta_{503}^\circ$ & $10$ & $93.4431 $ & $ 0.2416 $ & $ 6.2185 $ & $ 0.0132 $ & $ 0.0835 $ & $ 0.0001$ \\
$\Delta_{1348}^\circ$ & $10$ & $93.6799 $ & $ 0.0388 $ & $ 6.1970 $ & $ 0.0009 $ & $ 0.0833 $ & & $ 0.0001$ \\
\midrule
$\Delta_{882}^\circ$, $\Delta_{856}^\circ$ & $10$ & $93.4356 $ & $ 0.2615 $ & $ 6.2051 $ & $ 0.0148 $ & $ 0.0824 $ & $ 0.0001 $ & $ 0.0005$ \\
$\Delta_{1340}^\circ$ & $10$ & $92.2900 $ & $ 0.0153 $ & $ 7.5511 $ & $ 0.0005 $ & $ 0.1427 $ & & $ 0.0004$ \\
$\Delta_{1879}^\circ$ & $10$ & $92.2864 $ & $ 0.0260 $ & $ 7.5439 $ & $ 0.0010 $ & $ 0.1420 $ & & $ 0.0007$ \\
$\Delta_{1384}^\circ$ & $10$ & $90.8524 $ & $ 0.0031 $ & $ 8.9219 $ & $ 0.0001 $ & $ 0.2213 $ & & $ 0.0012$ \\
\midrule \midrule
\multicolumn{9}{c}{Results based on the current work}\\
Polytope & $\overline{K}_{B_3}^3$ & $= 3$ & $\geq 3$ & $= 4$ & $\geq 4$ & $= 5$ & $\geq 5$ & $= 6$ \\
\midrule
$\Delta_4^\circ$ & $6$ & $99.9952$ & & $0.0048$ & & & & \\
$\Delta_{134}^\circ$ & $6$ & $99.9952$ & & $0.0048$ & & & & \\
$\Delta_{128}^\circ$, $\Delta_{130}^\circ$, $\Delta_{136}^\circ$, $\Delta_{236}^\circ$ & $6$ & $99.9952$ & & $0.0048$ & & & & \\
\midrule
$\Delta_{254}^\circ$ & $10$ & $96.3942$ & $0.0687$ & $3.5193$ & $0.0003$ & $0.0175$ & & \\
$\Delta_{52}^\circ$ & $10$ & $96.0587$ & $0.0171$ & $3.9066$ & $0.0000$ & $0.0176$ & & \\
$\Delta_{302}^\circ$ & $10$ & $96.3960$ & $0.0636$ & $3.5222$ & $0.0001$ & $0.0181$ & & \\
$\Delta_{786}^\circ$ & $10$ & $95.0714$ & $0.0393$ & $4.8466$ & $0.0002$ & $0.0425$ & & \\
$\Delta_{762}^\circ$ & $10$ & $95.0167$ & $0.0369$ & $4.9052$ & $0.0005$ & $0.0407$ & & \\
\midrule
$\Delta_{417}^\circ$ & $10$ & $95.0745$ & $0.0433$ & $4.8389$ & $0.0003$ & $0.0429$ & & $0.0001$ \\
$\Delta_{838}^\circ$ & $10$ & $94.9092$ & $0.0215$ & $5.0216$ & $0.0000$ & $0.0477$ & & \\
$\Delta_{782}^\circ$ & $10$ &  $94.9019$ & $0.0161$ & $5.0359$ & $0.0000$ & $0.0461$ & & \\
$\Delta_{377}^\circ$, $\Delta_{499}^\circ$, $\Delta_{503}^\circ$ & $10$ & $93.6500$ & $0.0347$ & $6.2312$ & $0.0005$ & $0.0836$ & \\
$\Delta_{1348}^\circ$ & $10$ & $93.7075$ & $0.0112$ & $6.1978$ & $0.0001$ & $0.0833$ & & $0.0001$ \\
\midrule
$\Delta_{882}^\circ$, $\Delta_{856}^\circ$ & $10$ & $93.6546$ & $0.0425$ & $6.2190$ & $0.0009$	& $0.0825$ & & $0.0005$ \\
$\Delta_{1340}^\circ$ & $10$ & $92.2989$ & $0.0064$ & $7.5515$ & $0.0001$ & $0.1427$ & & $0.0004$ \\
$\Delta_{1879}^\circ$ & $10$ & $92.3015$ & $0.0108$ & $7.5447$ & $0.0002$ & $0.1421$ & & $0.0007$ \\
$\Delta_{1384}^\circ$ & $10$ & $90.8524$ & $0.0031$ & $8.9219$ & $0.0001$ & $0.2213$ & & $0.0012$ \\
\bottomrule
\end{tabular}
\end{center}
\caption{We compare the Brill-Noether numbers computed in \cite{BCDO22} with the refined results of the current work. The results are rounded to four decimal places.}
\label{tab:Results1}
\end{table}

We now employ the refined simplification techniques to recompute the number of limit root bundles for the 23 QSMs whose canonical nodal quark-doublet curve consists of rational curves only. In order to fully appreciate the changes, we compare the refined results computed in this work with the results obtained with \emph{the old techniques} \cite{BCDO22}. We list the corresponding numbers in \cref{tab:Results1}.

We must first recall how to read these tables. First, look at the line concerning $\Delta_{134}^\circ$, which lists the information for the F-theory QSM family $B_3( \Delta_4^\circ )$. On the smooth, irreducible quark-doublet curve $C_{(\mathbf{3},\mathbf{2})_{1/6}}$ in this family, there are $12^8$ root bundles of interest. We estimated their number of global sections from those of the $12^8$ corresponding limit roots on the canonical nodal quark-doublet curve $C^\bullet_{(\mathbf{3},\mathbf{2})_{1/6}}$ in this family. In \cite{BCDO22}, we found that (roughly) $99.9952\%$ of these $12^8$ limit root bundles have exactly three global sections. Consequently, there are (roughly) $0.0048\%$ of the limit roots left. Our techniques in \cite{BCDO22} would only bound the number of global sections of these $0.0048\%$ of the limit root bundles from below by three. Consequently, this fraction of the limit roots is listed in the column labeled as $\geq 3$.

At the bottom of \cref{tab:Results1}, we list the refined results computed in this work. These numbers are rounded to four decimal places. Notice that the error margins are always smaller than $0.23\%$, which is a stark improvement in comparison to the results in \cite{BCDO22} or even \cite{BCL21}.

By repeating the analysis in \cite{BCDO22}, one finds that for $\Delta_{134}^\circ$, there is a unique setup that could not be sorted algorithmically. This configuration has the degree of the canonical bundle. By observing that the canonical bundle solves the root bundle constraint in question,
\cite{BCDO22} concludes that this configuration does indeed correspond to the canonical bundle and admits four global sections. With the exception of $\Delta_8^\circ$, whose canonical nodal quark-doublet curve admits an elliptic component, one can repeat this argument for all remaining setups with $\overline{K}_{B_3}^3 = 6$. This leads to the results listed in \cref{tab:Results1}.

Crucially, note that this argument cannot be repeated for setups with $\overline{K}_{B_3}^3 = 10$; essentially because we then have to solve a root bundle constraint which does not admit the canonical bundle as solution. Still, our results are optimal in a certain sense. The list of terminal graphs for which our improved computer scan \cite{Bie23} could only compute a lower bound on the number of global sections is as follows:
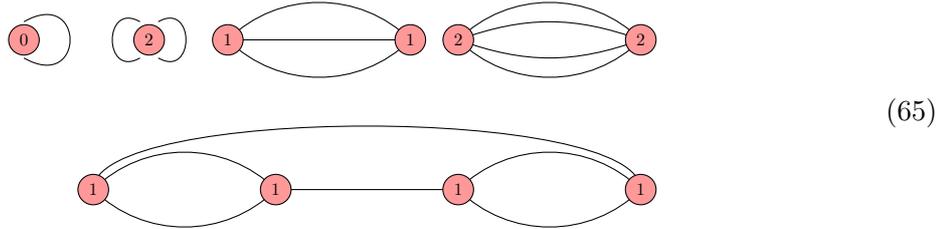
\begin{align}
\begin{split}
\begin{tikzpicture}[scale=0.6, baseline=(current  bounding  box.center)]
      \def\s{2.0};
      \path[-,out = -30, in = 30, looseness = 5] (-2*\s,-0.2*\s) edge (-2*\s,+0.2*\s);
      \node at (-2*\s,0) [stuff_fill_red, scale=0.6]{$0$};
\end{tikzpicture}\; \;
\begin{tikzpicture}[scale=0.6, baseline=(current  bounding  box.center)]
      \def\s{2.0};
      \path[-,out = -30, in = 30, looseness = 3] (-1.9*\s,-0.2*\s) edge (-1.9*\s,+0.2*\s);
      \path[-,out = -150, in = 150, looseness = 3] (-2.1*\s,-0.2*\s) edge (-2.1*\s,+0.2*\s);
      \node at (-2*\s,0) [stuff_fill_red, scale=0.6]{$2$};
\end{tikzpicture}\; \;
\begin{tikzpicture}[scale=0.6, baseline=(current  bounding  box.center)]
      \def\s{1.0};
      \path[-,out = 45, in = 135] (-2*\s,0) edge (+2*\s,0);
      \path[-,out = 0, in = 180] (-2*\s,0) edge (+2*\s,0);
      \path[-,out = -45, in = -135] (-2*\s,0) edge (+2*\s,0);
      \node at (-2*\s,0) [stuff_fill_red, scale=0.6]{$1$};
      \node at (+2*\s,0) [stuff_fill_red, scale=0.6]{$1$};
\end{tikzpicture}\; \;
\begin{tikzpicture}[scale=0.6, baseline=(current  bounding  box.center)]
      \def\s{1.0};
      \path[-,out = 45, in = 135] (-2*\s,0) edge (+2*\s,0);
      \path[-,out = 20, in = 160] (-2*\s,0) edge (+2*\s,0);
      \path[-,out = -20, in = -160] (-2*\s,0) edge (+2*\s,0);
      \path[-,out = -45, in = -135] (-2*\s,0) edge (+2*\s,0);
      \node at (-2*\s,0) [stuff_fill_red, scale=0.6]{$2$};
      \node at (+2*\s,0) [stuff_fill_red, scale=0.6]{$2$};
\end{tikzpicture}\\
\begin{tikzpicture}[scale=0.6, baseline=(current  bounding  box.center)]
      \def\s{1.0};
      \path[-,out = 45, in = 135] (-2*\s,0) edge (2*\s,0);
      \path[-,out = -45, in = -135] (-2*\s,0) edge (2*\s,0);
      \path[-,out = 0, in = 180] (2*\s,0) edge (6*\s,0);
      \path[-,out = 45, in = 135] (6*\s,0) edge (10,0);
      \path[-,out = -45, in = -135] (6*\s,0) edge (10,0);
      \path[-,out = 90, in = 90, looseness = 0.4] (-2*\s,0) edge (10,0);
      \node at (-2*\s,0) [stuff_fill_red, scale=0.6]{1};
      \node at (+2*\s,0) [stuff_fill_red, scale=0.6]{1};
      \node at (+6*\s,0) [stuff_fill_red, scale=0.6]{1};
      \node at (+10,0) [stuff_fill_red, scale=0.6]{1};
\end{tikzpicture}
\end{split}
\end{align}
We argued in \cref{Appendix:Derivations}, that all of these terminal circuits are jumping. In other words, the number of global sections depends on the descent data. As our computer scan \cite{Bie23} is (at least of now) not being provided with that information, it derives the best possible answer based on the combinatoric data provided.

\subsection{Improved statistics for ``elliptic'' QSMs}

\begin{table}[!htb]
\begin{center}
\begin{tabular}{c|cc|cc|cc}
\toprule
\multicolumn{7}{c}{Results based on \cite{BCDO22}} \\
$N$ & $= 3$ & $\geq 3$ & $= 4$ & $\geq 4$ & $= 5$ & $\geq 5$ \\
\midrule
0 & \numprint{781680888} & \numprint{62712} & \numprint{25196800} & 0 & \numprint{106800} & 0 \\
1 & \numprint{163221088} & \numprint{206886912} & \numprint{5967200} & \numprint{5200000} & 0 & \numprint{7200} \\
2 & \numprint{13270504} & \numprint{66014696} & \numprint{399200} & \numprint{1355600} & 0 & 0 \\
3 & \numprint{504800} & \numprint{9318400} & 0 & \numprint{88800} & 0 & 0 \\
4 & \numprint{4400} & \numprint{687600} & 0 & 0 & 0 & 0 \\
5 & 0 & \numprint{24800} & 0 & 0 & 0 & 0 \\
6 & 0 & \numprint{1600} & 0 & 0 & 0 & 0 \\
\midrule
$\Sigma$ & \numprint{958681680} & \numprint{282996720} & \numprint{31563200} & \numprint{6644400} & \numprint{106800} & \numprint{7200} \\
\midrule \midrule
\multicolumn{7}{c}{Results based on the current work} \\
N & $= 3$ & $\geq 3$ & $= 4$ & $\geq 4$ & $= 5$ & $\geq 5$ \\
\midrule
0 & \numprint{781680888} & \numprint{62712} & \numprint{25196800} & 0 & \numprint{106800} & 0 \\
1 & \numprint{366819888} & \numprint{3288112} & \numprint{11167200} & 0 & \numprint{7200} & 0 \\
2 & \numprint{76851304} & \numprint{2433896} & \numprint{1694800} & \numprint{60000} & 0 & 0 \\
3 & \numprint{8250400} & \numprint{1572800} & \numprint{64800} & \numprint{24000} & 0 & 0 \\
4 & \numprint{362000} & \numprint{330000} & 0 & 0 & 0 & 0 \\
5 & 0 & \numprint{24800} & 0 & 0 & 0 & 0 \\
6 & 0 & \numprint{1600} & 0 & 0 & 0 & 0 \\
\midrule
$\Sigma$ & \numprint{1233964480} & \numprint{7713920} & \numprint{38123600} & \numprint{84000} & \numprint{114000} & 0 \\
\bottomrule
\end{tabular}
\end{center}
\caption{The Brill-Noether table computed for $\Delta_{88}^\circ$ in \cite{BCDO22} (top) and its refined counterpart based on the current work (bottom). Those numbers are listed as multiples of the geometric multiplicity $\mu$, which happens to be $20^5$ for $\Delta_{88}^\circ$.}
\label{tab:Results2}
\end{table}

\begin{table}[!htb]
\begin{center}
\begin{tabular}{cc|cc|cc|cc}
\toprule
\multicolumn{8}{c}{Results based on \cite{BCDO22}} \\
Polytope & $\overline{K}_{B_3}^3$ & $= 3$ & $\geq 3$ & $= 4$ & $\geq 4$ & $= 5$ & $\geq 5$ \\
\midrule
$\Delta_8^\circ$ & 6 & $76.3889$ & $23.6111$ & & \\
$\Delta_{88}^\circ$ & 10 & $74.8970$ & $22.1091$ & $2.4659$ & $0.5191$ & $0.0083$ & $0.0006$ \\
$\Delta_{110}^\circ$ & 10 & $82.3757 $ & $ 14.1284 $ & $ 3.100 $ & $ 0.3827 $ & $ 0.0132$ \\
$\Delta_{272}^\circ$, $\Delta_{274}^\circ$ &10 & $78.0509 $ & $ 18.0470 $ & $ 3.3587 $ & $ 0.5255 $ & $ 0.0155 $ & $ 0.0024$ \\
$\Delta_{387}^\circ$ & 10 & $73.8375 $ & $ 21.9358 $ & $ 3.5044 $ & $ 0.6900 $ & $ 0.0295 $ & $ 0.0028$ \\
$\Delta_{798}^\circ$, $\Delta_{808}^\circ$, $\Delta_{810}^\circ$, $\Delta_{812}^\circ$ & 10 & $76.9517 $ & $ 17.8638 $ & $ 4.4362 $ & $ 0.6972 $ & $ 0.0468$ & $ 0.0043$ \\
\midrule \midrule
\multicolumn{8}{c}{Results based on the current work}\\
Polytope & $\overline{K}_{B_3}^3$ & $3$ & $\geq 3$ & $4$ & $\geq 4$ & $5$ & $\geq 5$ \\
\midrule
$\Delta_8^\circ$ & $6$ & $99.9421$ & & $0.0579$ & & & \\
$\Delta_{88}^\circ$ & $10$ & $96.4034$ & $0.6027$ & $2.9784$ & $0.0066$ & $0.0089$ & \\
$\Delta_{110}^\circ$ & $10$ & $95.5794$ & $0.8846$ & $3.5170$ & $0.0059$ & $0.0131$ & \\
$\Delta_{272}^\circ$, $\Delta_{274}^\circ$ & $10$ & $95.3336$ & $0.6916$ & $3.9450$ & $0.0118$ & $0.0180$ & \\
$\Delta_{387}^\circ$ & $10$ & $95.1545$ & $0.5358$ & $4.2769$ & $0.0005$ & $0.0323$ & \\
$\Delta_{798}^\circ$, $\Delta_{808}^\circ$, $\Delta_{810}^\circ$, $\Delta_{812}^\circ$ & $10$ & $93.8212$ & $0.8852$ & $5.2384$ & $0.0034$ & $0.0518$ & \\
\bottomrule
\end{tabular}
\end{center}
\caption{We compare the Brill-Noether numbers computed in \cite{BCDO22} with the refined results of the current work. The results are rounded to four decimal places.}
\label{tab:Results3}
\end{table}

Our advanced techniques help to improve the results for the QSM with a single elliptic curve. Since most of the remaining ignorance in the previous work \cite{BCDO22} originates from trees with an elliptic component. So we expect a massive improvement. As an example, let us look at $\Delta^\circ_{88}$. The Brill-Noether table obtained from application of our \emph{old techniques} in \cite{BCDO22} and its refined counterpart based on the techniques presented in this work are displayed in \cref{tab:Results2}.

We first recall the notation in this table. We have analyzed $20^{12}$ limit root bundles on the canonical quark-doublet curve $C^\bullet_{(\mathbf{3},\mathbf{2})_{1/6}}$ in the F-theory QSM-family $B_3( \Delta_{88}^\circ )$. Those limit roots are described by line bundles on a (possibly nodal) curve that is obtained from blowing up $0 \leq N \leq 6$ nodes of $C^\bullet_{(\mathbf{3},\mathbf{2})_{1/6}}$. For fixed $N$, only a fraction of the $20^{12}$ limit roots are obtained. We have enumerated them all and made an effort to compute their number of global sections. For instance, in \cite{BCDO22}, we found that for $N = 0$, there are $\numprint{781680888}$ limit root bundles with exactly three global sections.\footnote{More precisely, there are actually $\numprint{781680888} \times 20^5$ such limit roots. However, all counts in this table are multiples of the geometric multiplicity $\mu = 20^5$. For ease of presentation and in agreement with \cite{BCDO22}, we omit the factor $20^5$.} You find this number at the top-left of \cref{tab:Results2}. Similarly, we encountered $\numprint{62712}$ limit root bundles for $N = 0$, for which we could only compute a lower bound on the number of global sections in \cite{BCDO22}. Namely, we found that the number of global sections is bounded below by three. Hence, you find this number $\numprint{62712}$ at the top row with $N = 0$ and column with $\geq 3$. We regard this number of limit root bundles for which we could not uniquely determine the number of global sections as an ignorance to our analysis. Finally, the row beginning by $\Sigma$ lists the sum of all limit root bundles with $0 \leq N \leq 6$ and certain (lower bound on the) number of global sections. For instance, there is a total of $\numprint{31563200}$ limit root bundles with exactly four global sections in \cite{BCDO22}.

Let us now compare this Brill-Noether table from \cite{BCDO22} with the results obtained from our \emph{refined techniques}. This improved answer is listed at the bottom of \cref{tab:Results2}. Hence, with the techniques in \cite{BCDO22} we found the lower bound three to the number of global sections of $\numprint{282996720}$ limit root bundles, whereas this happened for merely $\numprint{7713920}$ limit roots once our advanced techniques were employed. This is a drastic reduction of ignorance! It is enlightening to express the two $\Sigma$-rows in percentages of the $20^{12}$ limit root bundles:
\begin{equation}
\begin{tabular}{c|cc|cc|cc}
\toprule
 & $= 3$ & $\geq 3$ & $= 4$ & $\geq 4$ & $= 5$ & $\geq 5$ \\
\midrule
$\Sigma$ in \cite{BCDO22} & $74.8970$ & $22.1091$ & $2.4659$ & $0.5191$ & $0.0083$ & $0.0006$ \\
$\Sigma$ in current work & $96.4034$ & $0.6027$ & $2.9784$ & $0.0066$ & $0.0089$ & \\
\bottomrule
\end{tabular}
\end{equation}
So the total ignorance dropped from about $22.7\%$ to about $0.7\%$. Clearly, this is a stark improvement! Indeed, this finding applies more generally to the 10 families of elliptic F-theory QSMs that we analyze in this work. We provide the comparison among the results from \cite{BCDO22} and the refined answers computed in this work in \cref{tab:Results3}. While we definitely reduced the ignorances significantly, let us emphasize that these results are optimal in a certain sense . Namely, we can easily list all setups for which our computer algorithm could only compute a lower bound on the number of global sections. Then, we can analyze what it would take to turn these lower bounds into exact counts of global sections. For instance, the remaining computation for $\Delta_8^\circ$ concerns (a family of) line bundles of the canonical degree. In fact, we could argue implicitly that these must be the canonical bundle \cite{BCDO22}. However, purely based on the combinatorics data that the computer uses, it is impossible to arrive at that conclusion. Rather, the descent data of the limit root bundles is needed. Hence, based on the information provided by the current computer scan, this result cannot be improved. Likewise, for the remaining QSM geometries in the above table, the remaining computations concern one or multiple of the following configurations:
\begin{align}
\begin{split}
\begin{tikzpicture}[scale=0.6, baseline=(current  bounding  box.center)]
      \def\s{2.0};
      \node at (-2*\s,0) [stuff_fill_green, scale=0.6]{$d = 0$};
\end{tikzpicture}
\qquad
\begin{tikzpicture}[scale=0.6, baseline=(current  bounding  box.center)]
      \def\s{2.0};
      \path[-,out = -30, in = 30, looseness = 5] (-2*\s,-0.2*\s) edge (-2*\s,+0.2*\s);
      \node at (-2*\s,0) [stuff_fill_red, scale=0.6]{$d = 0$};
\end{tikzpicture}
\qquad
\begin{tikzpicture}[scale=0.6, baseline=(current  bounding  box.center)]
      \def\s{2.0};
      \path[-,out = -30, in = 30, looseness = 5] (-2*\s,-0.2*\s) edge (-2*\s,+0.2*\s);
      \node at (-2*\s,0) [stuff_fill_green, scale=0.6]{$d$};
\end{tikzpicture}
\; (d \in \{2,3,4\})
\qquad
\begin{tikzpicture}[scale=0.6, baseline=(current  bounding  box.center)]
      \def\s{2.0};
      \path[-,out = 0, in = 180] (-3*\s,0) edge (-2*\s,0);
      \path[-,out = -30, in = 30, looseness = 5] (-2*\s,-0.2*\s) edge (-2*\s,+0.2*\s);
      \node at (-3*\s,0) [stuff_fill_green, scale=0.6]{$d$};
      \node at (-2*\s,0) [stuff_fill_red, scale=0.6]{$1$};
\end{tikzpicture}
\; (d \in \{1,2,3\})\\
\begin{tikzpicture}[scale=0.6, baseline=(current  bounding  box.center)]
      \def\s{1.0};
      \path[-,out = 45, in = 135] (-2*\s,0) edge (+2*\s,0);
      \path[-,out = 0, in = 180] (-2*\s,0) edge (+2*\s,0);
      \path[-,out = -45, in = -135] (-2*\s,0) edge (+2*\s,0);
      \node at (-2*\s,0) [stuff_fill_red, scale=0.6]{$1$};
      \node at (+2*\s,0) [stuff_fill_red, scale=0.6]{$1$};
\end{tikzpicture}
\qquad
\begin{tikzpicture}[scale=0.6, baseline=(current  bounding  box.center)]
      \def\s{1.0};
      \path[-,out = 45, in = 135] (-2*\s,0) edge (+2*\s,0);
      \path[-,out = 0, in = 180] (-2*\s,0) edge (+2*\s,0);
      \path[-,out = -45, in = -135] (-2*\s,0) edge (+2*\s,0);
      \node at (-2*\s,0) [stuff_fill_green, scale=0.6]{$d$};
      \node at (+2*\s,0) [stuff_fill_red, scale=0.6]{$1$};
\end{tikzpicture}
\; (d \in \{3,4\})
\qquad
\begin{tikzpicture}[scale=0.6, baseline=(current  bounding  box.center)]
      \def\s{1.0};
      \draw (0,0) -- (-2*\s,-2*\s);
      \draw (0,0) -- (2*\s,-2*\s);
      \path[-,out = 20, in = 160] (-2*\s,-2*\s) edge (2*\s,-2*\s);
      \path[-,out = -20, in = -160] (-2*\s,-2*\s) edge (2*\s,-2*\s);      
      \node at (0,0) [stuff_fill_green, scale=0.6]{$2$};
      \node at (-2*\s,-2*\s) [stuff_fill_red, scale=0.6]{$1$};
      \node at (+2*\s,-2*\s) [stuff_fill_red, scale=0.6]{$1$};
\end{tikzpicture}
\end{split}
\end{align}
All of these are jumping terminal graphs, which were derived in \cref{Appendix:Derivations} and \cref{sec:AppendixEllipticCircuitJumps}. To tell the number of global sections precisely, one needs the descent data and/or the exact line bundle divisor on the elliptic curve. So based on the currently available combinatoric data for our computer scan, the above results are fully optimized.

\subsection{Absence of vector-like quark-doublets}
\label{subsec:Absenceofvector-likequark-doublets}
Let us summarize our findings regarding the analyzed 33 families of F-theory QSMs:
\begin{equation}
\begin{tabular}{cc|cc|cc}
\toprule
Polytope & $\overline{K}_{B_3}^3$ & $P_{(3)}$ & $N_{\text{roots}}$ & $\widecheck{N}_{\text{FRST}}$ & $\widehat{N}_{\text{FRST}}$ \\
\midrule
$\Delta_8^\circ$ & $6$ & $99.9421$ & $12^8$ & $3.867 \times 10^{13}$ & $2.828 \times 10^{16}$ \\
$\Delta_4^\circ$ & $6$ & $99.9952$ & $12^8$ & $3.188 \times 10^{11}$ & \\
$\Delta_{134}^\circ$ & $6$ & $99.9952$ & $12^8$ & $7.538 \times 10^{10}$ & \\
$\Delta_{128}^\circ$, $\Delta_{130}^\circ$, $\Delta_{136}^\circ$, $\Delta_{236}^\circ$ & $6$ & $99.9952$ & $12^8$ & $3.217 \times 10^{11}$ \\
\midrule \midrule
$\Delta_{88}^\circ$ & $10$ & $96.4034$ & $20^{12}$ & $5.231 \times 10^{10}$ & $1.246 \times 10^{14}$ \\
$\Delta_{110}^\circ$ & $10$ & $95.5794$ & $20^{12}$ & $5.239 \times 10^8$ \\
$\Delta_{272}^\circ$, $\Delta_{274}^\circ$ & $10$ & $95.3336$ & $20^{12}$ & $3.212 \times 10^{12}$ & $2.481 \times 10^{15}$ \\
$\Delta_{387}^\circ$ & $10$ & $95.1545$ & $20^{12}$ & $6.322 \times 10^{10}$ & $6.790 \times 10^{12}$ \\
$\Delta_{798}^\circ$, $\Delta_{808}^\circ$, $\Delta_{810}^\circ$, $\Delta_{812}^\circ$ & $10$ & $93.8212$ & $20^{12}$ & $1.672 \times 10^{10}$ & $2.515 \times 10^{13}$ \\
\midrule
$\Delta_{254}^\circ$ & $10$ & $96.3942$ & $20^{12}$ & $1.568 \times 10^{10}$ \\
$\Delta_{52}^\circ$ & $10$ & $96.0587$ & $20^{12}$ & $1.248 \times 10^8$ \\
$\Delta_{302}^\circ$ & $10$ & $96.3960$ & $20^{12}$ & $5.750 \times 10^7$ \\
$\Delta_{786}^\circ$ & $10$ & $95.0714$ & $20^{12}$ & $9.810 \times 10^8$ \\
$\Delta_{762}^\circ$ & $10$ & $95.0167$ & $20^{12}$ & $1.087 \times 10^{11}$ & $2.854 \times 10^{13}$ \\
\midrule
$\Delta_{417}^\circ$ & $10$ & $95.0745$ & $20^{12}$ & $1.603 \times 10^9$ & \\
$\Delta_{838}^\circ$ & $10$ & $94.9092$ & $20^{12}$ & $4.461 \times 10^9$ & \\
$\Delta_{782}^\circ$ & $10$ &  $94.9019$ & $20^{12}$ & $3.684 \times 10^9$ & \\
$\Delta_{377}^\circ$, $\Delta_{499}^\circ$, $\Delta_{503}^\circ$ & $10$ & $93.6500$ & $20^{12}$ & $4.461 \times 10^9$ & \\
$\Delta_{1348}^\circ$ & $10$ & $93.7075$ & $20^{12}$ & $4.285 \times 10^9$ & \\
\midrule
$\Delta_{882}^\circ$, $\Delta_{856}^\circ$ & $10$ & $93.6546$ & $20^{12}$ & $3.180 \times 10^9$ & \\
$\Delta_{1340}^\circ$ & $10$ & $92.2989$ & $20^{12}$ & $4.496 \times 10^9$ & \\
$\Delta_{1879}^\circ$ & $10$ & $92.3015$ & $20^{12}$ & $4.461 \times 10^9$ & \\
$\Delta_{1384}^\circ$ & $10$ & $90.8524$ & $20^{12}$ & $7.040 \times 10^9$ & \\
\bottomrule
\end{tabular}
\end{equation}
Recall that these geometries are one-to-one to fine, regular, star triangulations (FRSTs) of reflexive 3-dimensional polytopes. Some of these polytopes are very complicated and the number of triangulations can only be estimated \cite{HT17}. In the last two columns, we have listed the known lower and upper bounds to the number of FRSTs. If the exact number of FRSTs is known, we have not listed an upper bound $\widehat{N}_{\text{FRST}}$ as we hope that this helps to read this table.

Let us analyze these numbers. A natural question is to consider the union of all 33 families of F-theory QSMs and then to ask how likely it is to find no vector-like exotics on the quark-doublet curve. In other words, if we throw a dart at the board of all these QSM families, how likely it is that we hit a configuration without vector-like exotics on the quark-doublet curve? To this end, we focus on the quantity
\begin{align}
P = \frac{\sum \limits_{i \in I}{P^{(i)}_{(3)} \cdot N^{(i)}_{\text{roots}} \cdot N^{(i)}_{\text{FRST}}}}{\sum \limits_{i \in I}{N^{(i)}_{\text{roots}} \cdot N^{(i)}_{\text{FRST}}}} \, ,
\end{align}
where $I$ is the indexing set of the polytopes in the above table. The complicating fact is that we cannot uniquely tell the number of FRSTs for 10 polytopes. Rather, for these polytopes we only know a lower and upper bound to the number of triangulations \cite{HT17}. So what we do, is we minimize this quantity in the domain of the admissible triangulations. This task can easily be completed e.g. with \texttt{Mathematica}\cite{Mathematica} \footnote{One can also work out this result with the \emph{Karush–Kuhn–Tucker} conditions \cite{Kar39, KT51}, which generalize the method of Lagrange multiplier. The key step is to employ the complementary slackness conditions.}. We find that the likeliness for exactly three global sections on the quark doublet curve is at least $93.91\%$. This minimum percentage is attained by maximizing the number of FRSTs for $\Delta^\circ_{798}$ and minimizing the number of all other FRSTs.

In principle one could be tempted to repeat this analysis to also find an upper bound. Such a naive approach would use the sum of the percentage of roots with exactly three sections and that of roots for which we could only bound the number of global sections below by three. The outcome of this computation is then $96.96\%$ -- attained by maximizing the number of FRSTs for $\Delta^\circ_{8}$, $\Delta^\circ_{88}$ and minimizing the number of all other FRSTs. Note however, that we actually want to count global sections of root bundles on the physical quark-doublet curve and not the nodal curves, for which we execute the computations presented in this work. Along the smoothing $C^\bullet_{(\mathbf{3},\mathbf{2})_{1/6}} \to C_{(\mathbf{3},\mathbf{2})_{1/6}}$ we may lose vector-like pairs/global sections. Indeed, we cannot rule out the possibility that all root bundles may drop down to having exactly three global sections after this smoothing. This effect is not taken into account by the naive upper bound $96.96\%$, which must therefore be discarded.

A complementary answer is obtained by estimating the chances for exactly one vector-like exotic on the nodal quark-doublet curve $C^\bullet_{(\mathbf{3},\mathbf{2})_{1/6}}$. It is important to recall that we want to estimate this number on the original smooth matter curves. Due to the deformation process from the nodal to the smooth, irreducible curves, we can merely provide an upper bound from our estimates on the nodal curve. The latter is readily obtained with e.g. \texttt{Mathematica} \cite{Mathematica}, namely the chances for this are at most $5.17\%$. This maximal value is obtained for a maximal number of FRSTs for $\Delta_{798}^\circ$ and minimal number of FRSTs for all other polytopes.

\section{Summary and outlook} \label{sec:SummaryAndOutlook}

The geometries of the Quadrillion F-theory Standard Models (QSMs) \cite{CHLLT19} are defined by toric 3-folds, which arise from full, regular, star triangulations of certain polytopes in the Kreuzer-Skarke list of 3-dimensional reflexive polytopes \cite{KS98}. The vector-like spectra in QSM geometries are given by the cohomologies of certain roots $P_{\mathbf{R}}$ (twists of certain powers of) the canonical bundle on the matter curve $C_{\mathbf{R}}$ \cite{BCDLO21}. These roots are not unique, and it is unclear which ones come from F-theory gauge potentials in the Deligne cohomology. Investigating this matter is currently beyond our abilities, so \cite{BCDLO21, BCL21, BCDO22} analyzed \emph{all} mathematically possible root bundles, including those that could be induced from the $G_4$-flux. They conducted a local and bottom-up analysis, dealing with one matter curve at a time and neglecting any correlations among the vector-like spectra on different matter curves.

Counting root bundles on $C_{\mathbf{R}}$ becomes easier when the matter curve is deformed into a nodal curve. Constructing root bundles on smooth curves is challenging, but on nodal curves, it amounts to using the combinatorics of limit roots \cite{CCC07}. Additionally, for each family $B_3(\Delta^\circ)$ of F-theory QSMs, there is a canonical nodal curve $C_{\mathbf{R}}^\bullet$ that depends solely on $\Delta^\circ$ \cite{BCL21}. Therefore, the vector-like spectra for the entire family $B_3(\Delta^\circ)$ can be estimated with a single calculation by studying the limit roots on this nodal curve $C_{\mathbf{R}}^\bullet$.

After deformation, we turn to counting limit root bundles with a specific number of global sections. This was the main goal of \cite{BCDO22}, which focused on the family $B_3(\Delta_4^\circ)$ of $\mathcal{O}(10^{11})$ F-theory QSMs. The majority of the limit roots were defined over tree-like rational curves, and their line bundle cohomologies were computed using an algorithm that simplified a line bundle on a tree-like nodal curve without changing the number of global sections. The remaining limit roots were defined over circuit-like curves, and their cohomologies were counted manually. Fortunately, there were only a few of them, namely five for the $B_3(\Delta_4^\circ)$ family, because the involved nodal curves had relatively small Betti numbers and high root indices. On four of the curves, the line bundles were found to have exactly three sections. On the remaining curve, labeled as a \emph{jumping circuit} in \cite{BCDO22}, the line bundle was identified as the canonical bundle, which has as many sections as the arithmetic genus, namely four in this case. This is why at least 99.995\% of limit root bundles on the nodal curves $C_{(\mathbf{3}, \mathbf{2})_{1/6}}^\bullet$, $C_{(\mathbf{\overline{3}}, \mathbf{1})_{-2/3}}^\bullet$, and $C_{(\mathbf{1}, \mathbf{1})_{1}}^\bullet$ have exactly three global sections within the F-theory QSMs of the family $B_3(\Delta_4^\circ)$.

This paper seeks to upgrade the results of \cite{BCDO22} in two ways. First, it develops a systematic algorithm that determines line bundle cohomologies on circuit-like curves with rational and elliptic components. Previously, this was done manually in \cite{BCDO22}. We present two techniques that simplify the dual graphs of nodal curves: the removal of subtrees in \cref{subsec:PruningLeaves} and the removal of interior edges in \cref{subsec:RemovingInteriorEdges}. These simplifications result in a finite number of so-called terminal circuits, which are graphs that cannot be further reduced. The terminal graphs were classified in \cref{subsec:TerminalGraphs}, and their line bundle cohomologies were examined case-by-case. The relevant computations can be found in \cref{subsec:TerminalGraphs} as well as \cref{Appendix:Derivations}, \cref{sec:AppendixEllipticCircuitJumps}. Like in \cite{BCDO22}, some circuits were found to be jumping depending on the descent data, and only a lower bound on the number of global sections can be obtained in these cases.

With this established algorithm, this paper seeks to move beyond the family $B_3(\Delta_4^\circ)$ and widen the scope of analysis. The simplification algorithm was applied to 33 distinct QSM families to assess the probability of no vector-like exotics. The improved statistics can be found in \cref{sec:ImplicationsForFTheoryQSMs}, where \cref{tab:Results1} and \cref{tab:Results3} compare our current findings with those obtained using old techniques from \cite{BCDO22}. As expected, this algorithm improved our ability to conclusively determine line bundle cohomology and helped decrease the statistical ignorance from previous papers. For instance, the total ignorance for elliptic circuits within the family $B_3( \Delta_{88}^\circ )$ dropped from 22.7\% to 0.7\%, a massive upgrade. These results are optimal in the sense that pure combinatorics alone will not offer any more improvement. Further progress requires deeper knowledge of the descent data and the line bundle divisor on elliptic curves.

In \cref{subsec:Absenceofvector-likequark-doublets}, we discovered that the probability of having precisely three global sections on the quark doublet curve is at least 93.91\%. Although this value is lower than the previous probability of 99.995\% mentioned in \cite{BCDO22}, there is a crucial distinction. In our analysis, we encompass multiple QSM families, whereas \cite{BCDO22} only examined a single QSM family. Hence, our current estimation of 93.91\% provides a more comprehensive representation of the likelihood of lacking vector-like exotics on the quark-doublet curve across the majority of QSM setups. It is important to emphasize that 93.91\% serves as a lower bound. For certain QSM families, the exact number of elements remains unknown. Stated differently, for some of the 3-dimensional reflexive polytopes in the Kreuzer-Skarke list \cite{KS98} the exact number of fine regular star triangulations is unknown. Only lower and upper bounds for the number of fine regular star triangulations are known \cite{HT17}. We minimized the absence of vector-like exotics within the domain formed from these lower and upper bounds on triangulations. The stated minimum of 93.91\% is attained for minimal number of FRSTs for polytopes with a high probability of lacking vector-like exotics and maximized the number of FRSTs for polytopes with a (relatively) low probability of lacking vector-like exotics. The latter specifically pertains to the $B_3( \Delta_{272}^\circ )$ and $B_3( \Delta_{274}^\circ )$ families.

It should be noted that only 33 families of QSMs were studied, leaving 675 families remaining. While each comes with only few triangulations, the number of roots grows exponentially. The number of roots is given by
\begin{align}
\left(2 \overline{K}_{B_3}^3\right)^{2g} \, , \qquad g = \frac{2 + \overline{K}_{B_3^3}}{2} \colon \text{ the genus of the quark-doublet curve} \, .
\end{align}
Hence, the cases $\overline{K}_{B_3}^3 = 18$ and $\overline{K}_{B_3}^3 = 30$, which have not been covered in this paper, come with $36^{10}$ and $60^{16}$ roots each. These numbers are huge compared to the setups in our analysis. However, since these cases did not our statistically rooted top-down condition, $h^{2,1}(\widehat{Y}_4) \geq g$ where $g$ is the genus of $C_{(\mathbf{3},\mathbf{2})_{1/6}}$ \cite{BCL21}, they were discarded.\footnote{4 setups were discarded even though they satisfied $h^{2,1}(\widehat{Y}_4) \geq g$, but their canonical nodal quark doublet curve $C^\bullet_{(\mathbf{3},\mathbf{2})_{1/6}}$ have an irreducible component of genus strictly larger than $1$.} Nevertheless, based on the number of (root bundle) configurations that they provide and the fact that large $\overline{K}^3$ corresponds to small $h^{1,1}( \widehat{Y}_4 )$, which is needed for "realistic volumes" of the gauge divisors $V(s_3)$, $V(s_9)$ in stretched Kaehler cones \cite{CHLL20}, one should study these setups eventually.

This work also does not address the fundamental question of understanding which root bundles are induced by F-theory gauge potentials in the Deligne cohomology. We refer the reader to \cite{BCDLO21} to see how the $G_4$-flux quantization condition discussed in \cite{Wit97} gives rise to root bundles. One possible scenario is for all physically allowed gauge potentials to induce only a subset of limit roots on one matter curve. Another scenario is for the fluxes to induce a specific root on one matter curve but induce a few roots on another. To study this complex question, one can begin by setting $G_4 = 0$ and studying the origins of the root bundles in Deligne cohomology. In this setting, the root bundles are spin bundles, and so we seek to answer which spin bundles on the matter curve are compatible with the geometry of the F-theory compactification. For instance, an important condition from the physics is the cancellation of the Freed-Witten anomaly \cite{FW99}. A complete answer to all of this requires detailed investigation, similar to the steps outlined in \cite{BHV09}. One could also investigate the relationship between these spin bundles and the appearance of one half of the second Chern class (of the tangent bundle of $\widehat{Y}_4$) in the $G_4$-flux quantization condition \cite{Wit97}. Although our findings state that at least 93.91\% of limit roots have exactly three global sections, we cannot get a definitive answer regarding the absence of vector-like exotics until this matter is resolved. A detailed examination of such top-down conditions is planned for future research.

We eventually want to study the absence of vector-like exotics on the Higgs curve. However, this involves overcoming some computational challenges. Due to the Higgs curve having high genus and a large number of components, we will end up with a rather complicated curve even after simplification. The root bundle constraint is also different, and we will have to analyze a much larger number of roots. Nevertheless, our newfound techniques may uncover some structure that is not apparent yet. A more general extension could focus on other Standard Model setups, such as those considered in \cite{KMW12*1, CKPOR15, LW16, CLLO18, JTT22}, and investigate these constructions with the techniques presented in this work. We reserve these studies for future work.

This paper, and in particular \cref{tab:Results1} and \cref{tab:Results3}, should be seen as a computational approximation to describing the Brill-Noether Theory of root bundles on nodal curves. Such a theory is underdeveloped in the mathematical literature. These arithmetic perspectives may encourage more robust theorems to be made in the future. For instance, one may gain more insight by studying the descent data underlying limit root bundles and its behavior under deformation in \cite{CCC07}. As a common technique used in Brill-Noether theory, the application of limit linear series to limit root bundles is also compelling \cite{EH86, O14}. It would be interesting to compare the phenomena observed in our investigation to classical Brill-Noether theory. In particular, some circuits exhibit jumping behavior, which seems to resemble Brill-Noether jumps on smooth irreducible curves.

To conclude, one potential application in machine learning is to study and detect patterns linking the graphs to the corresponding Brill-Noether approximations. In the past, for many line bundles on nodal curves with relatively simple dual graphs, our computed Brill-Noether numbers featured a high ignorance in the numbers of global sections. Thus, the data was insufficient to learn conclusive results from. Now that we have improved our computational capabilities, we anticipate that most of this ignorance can be removed and, based on this refined data, a much better trained algorithm can be obtained. This algorithm could then either be used to estimate the number of global sections on the Higgs curve or in an attempt to predict a mathematical theorem underlying the Brill-Noether theory of root bundles on nodal curves.

\subsection*{Acknowledgements}

The \texttt{OSCAR} computer algebra system was used for key computations in this work. The reliable computations conducted by the supercomputer \texttt{plesken.mathematik.uni-siegen.de} allowed us to count limit root bundles. This we truly appreciate. M.~B. expresses gratitude towards his colleague Tobias Metzlaff for valuable discussions. The work of M.~B. was supported by the SFB-TRR 195 \emph{Symbolic Tools in Mathematics and their Application} of the German Research Foundation (DFG) and the Forschungsinitiative des Landes Rheinland-Pfalz through \emph{SymbTools – Symbolic Tools in Mathematics and their Application}. The work of M.C.~is supported by DOE Award DE-SC0013528Y and the Simons Foundation Collaboration grant \#724069 on ``Special Holonomy in Geometry, Analysis and Physics''. M.C.~thanks the Slovenian Research Agency \mbox{No. P1-0306} and the Fay R.~and Eugene L.~Langberg Chair for support. R.D. and M.O.~are partially supported by the NSF grant DMS~2001673 and by the Simons Foundation Collaboration grant \#390287 on ``Homological Mirror Symmetry''. M.O. is grateful for the support by the Ph.D. Presidential Fellowship research fund.

\newpage

\newpage 

\appendix

\section{Jumps on terminal graphs}

\subsection{Rational circuits} \label{Appendix:Derivations}

\begin{remark}
Consider a rational circuit $\Gamma$. By definition, this is a terminal graph whose components are all rational. In addition, by our prescription, for each rational component/vertex $V_i$ of $\Gamma$, there is an integer $d_i$ which encodes a line bundle $L_i = \mathcal{O}_{V_i}( d_i ) \cong \mathcal{O}_{\mathbb{P}^1}( d_i )$. Altogether, this defines a family $\mathcal{F}$ of line bundles on the nodal curve $C^\bullet$ whose dual graph is $\Gamma$, such that for every $L \in \mathcal{F}$ and every vertex $V_i$ of $\Gamma$ it holds
\begin{align}
\left. L \right|_{V_i} = L_i = \mathcal{O}_{V_i}( d_i ) \cong \mathcal{O}_{\mathbb{P}^1}( d_i ) \, .
\end{align}
We say that $\Gamma$ is non-jumping if $h^0( C^\bullet, L )$ is the same for all $L \in \mathcal{F}$. Otherwise, we say that $\Gamma$ is jumping.
\end{remark}

\begin{prop}
$G_2$ is jumping iff $0 \leq d \leq 2$:
\begin{align}
\begin{tikzpicture}[scale=0.6, baseline=(current  bounding  box.center)]
\def\s{2.0};
\path[-,out = -30, in = 30, looseness = 3] (-1.9*\s,-0.2*\s) edge (-1.9*\s,+0.2*\s);
\path[-,out = -150, in = 150, looseness = 3] (-2.1*\s,-0.2*\s) edge (-2.1*\s,+0.2*\s);
\node at (-2*\s,0) [stuff_fill_red, scale=0.6]{$d$};
\end{tikzpicture}
\end{align}
\end{prop}

\begin{myproof}
If $d < 0$, then $h^0( G_2, L ) = 0$ for all $L \in \mathcal{F}$. Certainly, $G_2$ is jumping for $d = 0$. Namely, if the descent data is $\lambda_1 = \lambda_2 = 1$, then $h^0( G_2, L ) = 1$. Otherwise, $h^0( G_2, L ) = 0$. The interesting case to discuss is $d > 0$. We first fix notation for the vertex $V_1 = \mathbb{P}^1$. By an $\mathrm{SL}(2, \mathbb{C})$ transformation, we position the nodes connecting the left edge at $[1:0]$ and $[0:1]$. The ones for the right edge are taken as $[1:1]$ and $[p:1]$ where $p \in \mathbb{C} \setminus \{0,1\}$. We can express the most general $\varphi \in H^0( V_1, \left. L \right|_{V_1})$ as follows:
\begin{align}
\varphi = \sum_{i = 0}^{d}{\alpha_i x^i y ^{d-i}} \, .
\end{align}
We impose two gluing conditions:
\begin{align}
\alpha_0 = \lambda_1 \cdot \alpha_d \, , \qquad \left( \lambda_1 + 1 \right) \alpha_d + \sum_{i = 1}^{d-1}{\alpha_i} = \lambda_2 \left( \lambda_1 + p^d \right) \cdot \alpha_d + \lambda_2 \sum_{i = 1}^{d-1}{\alpha_i \cdot p^i} \, .
\end{align}
Equivalently, we can write these conditions as
\begin{align}
\alpha_0 = \lambda_1 \cdot \alpha_d \, , \qquad \alpha_d \cdot \left[ \lambda_1 + 1 - \lambda_1 \lambda_2 - \lambda_2 p^d \right] = \sum_{i = 1}^{d-1}{\alpha_i \cdot \left( \lambda_2 p^i - 1 \right)} \, .
\end{align}
If $d = 1$, then these conditions become
\begin{align}
\alpha_0 = \lambda_1 \cdot \alpha_1 \, , \qquad \alpha_1 \cdot \left[ \lambda_1 + 1 - \lambda_1 \lambda_2 - \lambda_2 p \right] = 0 \, .
\end{align}
Take $\lambda_1 = 2$, $\lambda_2 = 3$ and $p = -1$, then those conditions are satisfied for all $\alpha_1 \in \mathbb{C}$ and $h^0( G_2, L ) = 1$. However, for generic $\lambda_1, \lambda_2 \in \mathbb{C}^\ast$ and $p \in \mathbb{C} \setminus \{0,1\}$, we find $h^0( G_2, L ) = 0$. Hence, $G_2$ is jumping for $d = 1$. Next, take $d = 2$. Then the gluing conditions are
\begin{align}
\alpha_0 = \lambda_1 \cdot \alpha_2 \, , \qquad \alpha_2 \cdot \left[ \lambda_1 + 1 - \lambda_1 \lambda_2 - \lambda_2 p^2 \right] = \alpha_1 \cdot \left( \lambda_2 p - 1 \right) \, .
\end{align}
Let us take $\lambda_1 = \frac{1}{2}$, $\lambda_2 = 2$ and $p = \frac{1}{2}$. Then, these conditions are satisfied for all $\alpha_1, \alpha_2 \in \mathbb{C}$ and so $h^0( G_2, L ) = 2$. However, for generic $\lambda_1, \lambda_2 \in \mathbb{C}^\ast$ and $p \in \mathbb{C} \setminus \{ 0, 1 \}$, we have two non-trivial conditions and so $h^0( G_2, L ) = 1$. Hence, $G_2$ is jumping for $d = 2$. Finally, focus on $d > 2$. Then the gluing conditions are
\begin{align}
\alpha_0 &= \lambda_1 \cdot \alpha_d \, , \\ \alpha_d \cdot \left[ \lambda_1 + 1 - \lambda_1 \lambda_2 - \lambda_2 p^d \right] &= \alpha_1 \left( \lambda_2 p - 1 \right) + \alpha_2 \left( \lambda_2 p^2 - 1 \right) + \sum_{i = 3}^{d-1}{\alpha_i \cdot \left( \lambda_2 p^i - 1 \right)} \, .
\end{align}
Let us assume that $\lambda_2 p - 1 \neq 0$. Then, these conditions are equivalent to
\begin{align}
\alpha_0 &= \lambda_1 \cdot \alpha_d \, , \\ 
\alpha_1 &= \frac{\alpha_d \cdot \left[ \lambda_1 + 1 - \lambda_1 \lambda_2 - \lambda_2 p^d \right] - \alpha_2 \left( \lambda_2 p^2 - 1 \right) - \sum_{i = 3}^{d-1}{\alpha_i \cdot \left( \lambda_2 p^i - 1 \right)}}{\lambda_2 p - 1} \, .
\end{align}
Hence, we then find $h^0( G_2, L ) = d - 1$. Conversely, assume that $\lambda_2 p - 1 = 0$, i.e. $p = \frac{1}{\lambda_2}$. Recall that we must have $p \in \mathbb{C} \setminus \{ 0, 1 \}$, so $\lambda_2 \neq 1$. In particular $\frac{1}{\lambda_2} - 1 \neq 0$. Under these assumptions, the above conditions becomes
\begin{align}
\alpha_0 &= \lambda_1 \cdot \alpha_d \, , \\ 
\alpha_d \cdot \left[ \lambda_1 + 1 - \lambda_1 \lambda_2 - \lambda_2^{-d + 1} \right] &= \alpha_2 \left( \frac{1}{\lambda_2} - 1 \right) + \sum_{i = 3}^{d-1}{\alpha_i \cdot \left( \lambda_2^{-i + 1} - 1 \right)} \, .
\end{align}
Equivalently, we can write
\begin{align}
\alpha_0 &= \lambda_1 \cdot \alpha_d \, , \\ 
\alpha_2 &= \frac{\alpha_d \cdot \left[ \lambda_1 + 1 - \lambda_1 \lambda_2 - \lambda_2^{-d + 1} \right] - \sum_{i = 3}^{d-1}{\alpha_i \cdot \left( \lambda_2^{-i + 1} - 1 \right)}}{\frac{1}{\lambda_2} - 1} \, .
\end{align}
And so, again we find $h^0( G_2, L ) = d - 1$. Hence, $h^0( G_2, L ) = d - 1$ for all $L \in \mathcal{F}$ for $d > 2$, i.e. $G_2$ is non-jumping for $d > 2$.
\end{myproof}

\begin{prop}
$G_3$ is jumping iff $0 \leq d_\alpha, d_\beta \leq 1$:
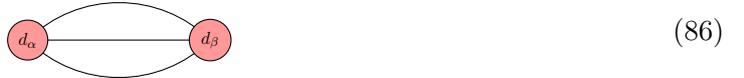
\begin{align}
\begin{tikzpicture}[scale=0.6, baseline=(current  bounding  box.center)]
\def\s{1.0};
\path[-,out = 45, in = 135] (-2*\s,0) edge (+2*\s,0);
\path[-,out = 0, in = 180] (-2*\s,0) edge (+2*\s,0);
\path[-,out = -45, in = -135] (-2*\s,0) edge (+2*\s,0);
\node at (-2*\s,0) [stuff_fill_red, scale=0.6]{$d_\alpha$};
\node at (+2*\s,0) [stuff_fill_red, scale=0.6]{$d_\beta$};
\end{tikzpicture}
\end{align}
\end{prop}

\begin{myproof}
First, if $d_\alpha, d_\beta < 0$, then $h^0( G_3, L ) = 0$ for all $L \in \mathcal{F}$. So $G_3$ is non-jumping for $d_\alpha, d_\beta < 0$.

Before we proceed, let us fix some notation. We use homogeneous coordinates $[x:y]$ for $V_1 = \mathbb{P}^1_\alpha$ and $[u:v]$ for $V_2 = \mathbb{P}^1_\beta$. In terms of these, we can express the most general section $\varphi_\alpha \in H^0( V_1, \left. L \right|_{V_1})$ and $\varphi_\beta \in H^0( V_2, \left. L \right|_{V_2})$ as follows:
\begin{align}
\varphi_\alpha = \sum_{i = 0}^{d_\alpha}{a_i x^i y^{d_\alpha -i}} \, , \qquad \varphi_\beta = \sum_{i = 0}^{d_\beta}{b_i u^i v^{d_\beta -i}} \, .
\end{align}
By an $\mathrm{SL}(2, \mathbb{C})$ transformation, we can assume that the nodes are positioned at $n_1 = [1:0]$, $n_2 = [0:1]$ and $n_3 = [1:1]$. First, let us study situations with $d_\alpha < 0$ and $d_\beta \geq 0$. Then the gluing conditions are as follows:
\begin{align}
0 = \lambda_1 b_{d_\beta} \, , \qquad 0 = \lambda_0 b_{0} \, , \qquad 0 = \sum_{i = 0}^{d_\beta}{b_i} \, .
\end{align}
Consequently, $h^0( G_3, L ) = \max \{ d_\beta - 2, 0\}$ for all $L \in \mathcal{F}$, i.e. $G_3$ is non-jumping. Next, focus on $d_\alpha, d_\beta \geq 0$. Then, the gluing conditions are as follows:
\begin{align}
a_{d_\alpha} = \lambda_1 b_{d_\beta} \, , \qquad a_{0} = \lambda_0 b_{0} \, , \qquad \left( \lambda_1 - 1 \right) b_{d_\beta} + \left( \lambda_0 - 1 \right) b_{0} = \sum_{i = 1}^{d_\beta - 1}{b_i} - \sum_{i = 1}^{d_\alpha - 1}{a_i} \, .
\end{align}
The first two conditions are never trivial. However, the third condition is trivial iff $\lambda_0 = \lambda_1 = 1$ and $0 \leq d_\alpha, d_\beta \leq 1$. Consequently, $G_3$ is jumping iff $0 \leq d_\alpha, d_\beta \leq 1$.
\end{myproof}

\begin{prop}
$G_4$ is jumping iff one of the following applies:
\begin{itemize}
    \item $d_\alpha < 0$ and $d_\beta = 1$,
    \item $d_\alpha = 0$ and $d_\beta \geq 0$,
    \item $d_\alpha = 1$ and $d_\beta < 0$,
    \item $d_\alpha \geq 0$ and $d_\beta = 0$,
    \item $d_\alpha = d_\beta = 1$.
\end{itemize}
\begin{align}
\begin{tikzpicture}[scale=0.6, baseline=(current  bounding  box.center)]
\def\s{1.0};
\path[-,out = -150, in = 150, looseness = 10] (-2*\s,-0.2*\s) edge (-2*\s,+0.2*\s);
\path[-,out = -30, in = 30, looseness = 10] (2*\s,-0.2*\s) edge (2*\s,+0.2*\s);
\path[-,out = 180, in = 0] (-2*\s,0) edge (2*\s,+0);
\node at (-2*\s,0) [stuff_fill_red, scale=0.6]{$d_\alpha$};
\node at (2*\s,0) [stuff_fill_red, scale=0.6]{$d_\beta$};
\end{tikzpicture}
\end{align}
\end{prop}

\begin{myproof}
Certainly, for $d_\alpha, d_\beta < 0$ we have $h^0( G_4, L ) = 0$ for all $L \in \mathcal{F}$. So for both degrees negative, $G_4$ is non-jumping.

We use homogeneous coordinates $[x:y]$ on $V_1 = \mathbb{P}^1_\alpha$ and $[u:v]$ on $V_2 = \mathbb{P}^1_\beta$. Fix the node positions to $[1:0]$, $[0:1]$ and $[1:1]$ by an $\mathrm{SL}(2, \mathbb{C})$ transformation. The node linking $\mathbb{P}^1_\alpha$ and $\mathbb{P}^1_\beta$ has homogeneous coordinates $[1:1]$ on both $V_1$ and $V_2$. With that being said, we can express $\varphi_\alpha \in H^0 \left( V_1, \left. L \right|_{V_1} \right)$ and $\varphi_\beta \in H^0 \left( V_1, \left. L \right|_{V_2} \right)$ as follows:
\begin{align}
\varphi_\alpha = \sum_{i = 0}^{d_\alpha}{a_i x^i y^{d_\alpha -i}} \, , \qquad \varphi_\beta = \sum_{i = 0}^{d_\beta}{b_i u^i v^{d_\beta -i}} \, .
\end{align}
On the self-edges, we enforce descent data $\lambda_\alpha$ and $\lambda_\beta$. We begin by analyzing mixed signs for the line bundle degrees.
\begin{itemize}
 \item $d_\alpha < 0$ and $d_\beta = 0$:\\
Then, the gluing conditions are:
\begin{align}
0 = \lambda_\alpha \cdot 0 \, , \qquad b_0 = \lambda_\beta \cdot b_0 \, , \qquad 0 = b_0 \, .
\end{align}
So $h^0( G_4, L ) = 0$ for all $L \in \mathcal{F}$, i.e. $G_4$ is non-jumping.
\item $d_\alpha < 0$ and $d_\beta = 1$:\\
Then, the gluing conditions are:
\begin{align}
0 = \lambda_\alpha \cdot 0 \, , \qquad b_0 = \lambda_\beta \cdot b_1 \, , \qquad 0 = b_1 \cdot \left( 1 + \lambda_\beta \right) \, .
\end{align}
Hence, if $\lambda_\beta = -1$, then $h^0( G_4, L ) = 1$. However, for $\lambda_\beta \neq -1$, we have $h^0( G_4, L ) = 0$. So $G_4$ is jumping for $d_\alpha < 0$ and $d_\beta = 1$.
\item $d_\alpha < 0$ and $d_\beta > 1$:\\
Then, the gluing conditions are:
\begin{align}
0 = \lambda_\alpha \cdot 0 \, , \qquad b_0 = \lambda_\beta \cdot b_{d_\beta} \, , \qquad b_1 = - b_{d_\beta} \cdot \left( 1 + \lambda_\beta \right) - \sum_{i = 2}^{d_\beta - 1}{b_i} \, .
\end{align}
So in this case, we have $h^0( G_4, L ) = d_\beta - 1$ for all $L \in \mathcal{F}$ and $G_4$ is non-jumping.
\end{itemize}
Next, we focus on situations with $d_\alpha = 0$ and $d_\beta \geq 0$.
\begin{itemize}
 \item $d_\alpha = d_\beta = 0$:\\
Then the gluing conditions are as follows:
\begin{align}
a_0 = \lambda_\alpha a_0 \, , \qquad b_0 = \lambda_\beta b_0 \, , \qquad a_0 = b_0 \, .
\end{align}
For generic $\lambda_\alpha, \lambda_\beta \in \mathbb{C}^\ast$, the only solution to these equations is $a_0 = b_0 = 0$ and hence $h^0( G_4, L ) = 0$. However, if $\lambda_\alpha = \lambda_\beta = 1$, then $a_0 = b_0$ solves these conditions and $h^0( G_4, L ) = 1$. Hence, $G_4$ is jumping for $d_\alpha = d_\beta = 0$.
\item $d_\alpha = 0$ and $d_\beta > 0$:\\
Then, the gluing conditions are as follows:
\begin{align}
a_0 = \lambda_\alpha a_0 \, , \qquad b_0 = \lambda_\beta b_{d_\beta} \, , \qquad b_{d_\beta} \cdot \left( 1 + \lambda_\beta \right) = a_0 - \sum_{i = 1}^{d_\beta - 1}{b_i} \, .
\end{align}
For generic $\lambda_\alpha$ and $\lambda_\beta$, we must have $a_0 = 0$ and $1 + \lambda_\beta \neq 0$. This then means that the only solution to these conditions is then given by
\begin{align}
a_0 = 0 \, , \qquad b_0 = \lambda_\beta b_{d_\beta} \, , \qquad b_{d_\beta} &= - \frac{\sum_{i = 1}^{d_\beta - 1}{b_i}}{1 + \lambda_\beta} \, .
\end{align}
So then $h^0( G_4, L ) = d_\beta - 1$. However, if we assume that $\lambda_\alpha = 1$ and $\lambda_\beta \neq 1$, then any $a_0 \in \mathbb{C}$ solves the first condition. In addition, we find
\begin{align}
b_0 = \lambda_\beta b_{d_\beta} \, , \qquad b_{d_\beta} &= \frac{a_0 - \sum_{i = 1}^{d_\beta - 1}{b_i}}{1 + \lambda_\beta} \, .
\end{align}
So then $h^0( G_4, L ) = d_\beta$ and $G_4$ is jumping for $d_\alpha = 0$ and $d_\beta > 0$.
\end{itemize}
It remains to study situation with $d_\alpha, d_\beta \geq 1$.
\begin{itemize}
\item $d_\alpha = d_\beta = 1$:\\
Then the gluing conditions are
\begin{align}
a_0 = \lambda_\alpha a_1 \, , \qquad b_0 = \lambda_\beta b_1 \, , \qquad a_1 \left( 1 + \lambda_\alpha \right) = b_1 \left( 1 + \lambda_\beta \right) \, .
\end{align}
If $\lambda_\alpha = \lambda_\beta = -1$, then the third condition is trivial and one has $h^0( G_4, L ) = 2$. However, if $\left( \lambda_\alpha,  \lambda_\beta \right) \neq \left( -1,-1 \right)$, then $h^0( G_4, L ) = 1$. So $G_4$ is jumping for $d_\alpha = d_\beta = 1$.
\item $d_\alpha = 1$ and $d_\beta > 1$:\\
Then the gluing conditions are
\begin{align}
a_0 = \lambda_\alpha a_1 \, , \qquad b_0 = \lambda_\beta b_{d_\beta} \, , \qquad a_1 \left( 1 + \lambda_\alpha \right) = b_{d_\beta} \left( 1 + \lambda_\beta \right) + b_1 + \sum_{i = 2}^{d_\beta - 1}{b_i} \, .
\end{align}
Hence, $h^0( G_4, L ) = d_\alpha + d_\beta - 1$ for all $L \in \mathcal{F}$ and $G_4$ is non-jumping.
\item $d_\alpha, d_\beta > 1$:\\
Then the gluing conditions are
\begin{align}
a_0 = \lambda_\alpha a_{d_\alpha} \, , \qquad b_0 = \lambda_\beta b_{d_\beta} \, , \qquad a_{d_\alpha} \left( 1 + \lambda_\alpha \right) + a_1 + \sum_{i = 2}^{d_\alpha - 1}{a_i} = b_{d_\beta} \left( 1 + \lambda_\beta \right) + \sum_{i = 1}^{d_\beta - 1}{b_i} \, .
\end{align}
Hence, $h^0( G_4, L ) = d_\alpha + d_\beta - 1$ for all $L \in \mathcal{F}$ and $G_4$ is non-jumping.
\end{itemize}
This completes the argument.
\end{myproof}

\begin{prop}
$G_5$ is jumping iff $0 \leq d_\alpha, d_\beta \leq 2$:
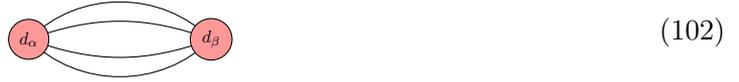
\begin{align}
\begin{tikzpicture}[scale=0.6, baseline=(current  bounding  box.center)]
\def\s{1.0};
\path[-,out = 45, in = 135] (-2*\s,0) edge (+2*\s,0);
\path[-,out = 20, in = 160] (-2*\s,0) edge (+2*\s,0);
\path[-,out = -20, in = -160] (-2*\s,0) edge (+2*\s,0);
\path[-,out = -45, in = -135] (-2*\s,0) edge (+2*\s,0);
\node at (-2*\s,0) [stuff_fill_red, scale=0.6]{$d_\alpha$};
\node at (+2*\s,0) [stuff_fill_red, scale=0.6]{$d_\beta$};
\end{tikzpicture}
\end{align}
\end{prop}

\begin{myproof}
Certainly, for $d_\alpha, d_\beta < 0$, $G_5$ is non-jumping. Before we proceed, let us fix notation. For $V_1 = \mathbb{P}^1_\alpha$ we use the homogeneous coordinates $[x:y]$ and for $V_2 = \mathbb{P}^1_\beta$ we use the homogeneous coordinates $[u:v]$. Thereby, we can write $\varphi_\alpha \in H^0( V_1, \left. L \right|_{V_1} )$ and $\varphi_\beta \in H^0( V_2, \left. L \right|_{V_2} )$ as follows:
\begin{align}
\varphi_\alpha = \sum_{i = 0}^{d_\alpha}{a_i x^i y^{d_\alpha -i}} \, , \qquad \varphi_\beta = \sum_{i = 0}^{d_\beta}{b_i u^i v^{d_\beta -i}} \, .
\end{align}
By an $\mathrm{SL}(2, \mathbb{C})$ transformation, we position the first three nodes at $n_1 = [1:0]$, $n_2 = [0:1]$ and $n_3 = [1:1]$ -- each with descent data $\lambda_1$, $\lambda_0$ and $\lambda_2$. The position of the fourth node is not fixed by an $\mathrm{SL}(2, \mathbb{C})$ transformation. We parametrize the position of this forth node on $V_1 = \mathbb{P}^1_\alpha$ as $[A:1]$ and on $V_2 = \mathbb{P}^1_\beta$ as $[B:1]$, where $A, B \in \mathbb{C} \setminus \{ 0, 1 \}$.

Let us begin by looking at $d_\alpha < 0$ and $d_\beta \geq 0$. Then the gluing conditions are as follows:
\begin{align}
0 = \lambda_1 \cdot b_{d_\beta} \, , \qquad 0 = \lambda_0 \cdot b_0 \, , \qquad 0 = \lambda_2 \cdot \sum_{i = 1}^{d_\beta - 1}{b_i} \, , \qquad 0 = \sum_{i = 1}^{d_\beta - 1}{b_i \cdot B^i} \, .
\end{align}
So then $h^0( G_5, L ) = \max \{ 0, d_\beta - 3 \}$ for all $L \in \mathcal{F}$, i.e. $G_5$ is non-jumping. It remains to discuss $d_\alpha, d_\beta \geq 0$. We do so by discussing several cases:
\begin{itemize}
\item $d_\alpha = d_\beta = 0$:\\
Then the gluing conditions are as follows:
\begin{align}
a_0 = \lambda_1 \cdot b_0 \, , \qquad a_0 = \lambda_0 \cdot b_0 \, , \qquad a_0 = \lambda_2 \cdot b_2 \, , \qquad a_0 = b_0 \, .
\end{align}
So $h^0( G_5, L ) = 1$ iff $\lambda_0 = \lambda_1 = \lambda_2 = 1$ and otherwise $h^0( G_5, L ) = 0$. Hence, $G_5$ is jumping for $d_\alpha = d_\beta = 0$.

\item $d_\alpha = 0$ and $d_\beta = 1$:\\
Then the gluing conditions are as follows:
\begin{align}
a_0 = \lambda_1 \cdot b_1 \, , \qquad a_0 = \lambda_0 \cdot b_0 \, , \qquad a_0 = \lambda_2 \cdot \left( b_0 + b_1 \right) \, , \qquad a_0 = b_0 + b_1 \cdot B \, .
\end{align}
Equivalently, we have
\begin{align}
b_1 = \frac{a_0}{\lambda_1} \, , \quad b_0 = \frac{a_0}{\lambda_0} \, , \quad 0 = a_0 \cdot \left( \frac{\lambda_2}{\lambda_0} + \frac{\lambda_2}{\lambda_1} - 1 \right) \, , \quad 0 = a_0 \cdot \left( \frac{1}{\lambda_0} + \frac{B}{\lambda_1} - 1 \right) \, .
\end{align}
Take $B = 2$, $\lambda_0 = 2$, $\lambda_1 = 4$ and $\lambda_2 = \frac{4}{3}$. Then the last two conditions are trivial and $h^0( G_5, L ) = 1$. However, for generic $B$, $\lambda_0$, $\lambda_1$ and $\lambda_2$, we have $h^0( G_5, L ) = 0$. So $G_5$ is jumping for $d_\alpha = 0$ and $d_\beta = 1$.

\item $d_\alpha = d_\beta = 1$:\\
Then the gluing conditions are as follows:
\begin{align}
a_1 = \lambda_1 \cdot b_1 \, , \quad a_0 = \lambda_0 \cdot b_0 \, , \quad a_0 + a_1 = \lambda_2 \cdot \left( b_0 + b_1 \right) \, , \quad a_0 + a_1 A = b_0 + b_1 B \, .
\end{align}
Equivalently, we have
\begin{align}
b_1 = \frac{a_1}{\lambda_1} \, , \qquad b_0 = \frac{a_0}{\lambda_0} \, , \qquad 0 &= a_0 \cdot \left( \frac{\lambda_2}{\lambda_0} - 1 \right) + a_1 \cdot \left( \frac{\lambda_2 }{\lambda_1} - 1 \right) \, , \\
0 &= a_0 \cdot \left( \frac{1}{\lambda_0} - 1 \right) + a_1 \cdot \left( \frac{B}{\lambda_1} - A \right) \, .
\end{align}
Take $\lambda_0 = \lambda_1 = \lambda_2 = 1$ and $A = B = 2$. Then the last two conditions are trivial and $h^0( G_5, L ) = 2$. But for generic $\lambda_0$, $\lambda_1$, $\lambda_2$, $A$ and $B$, we have $h^0( G_5, L ) < 2$. So $G_5$ is jumping for $d_\alpha = d_\beta = 1$.

\item $d_\alpha = 0$ and $d_\beta = 2$:\\
Then, the gluing conditions become
\begin{align}
a_0 = \lambda_1 \cdot b_2 \, , \qquad a_0 = \lambda_0 \cdot b_0 \, , \qquad b_0 \cdot \left( \lambda_0 - \lambda_2 \right) + b_2 \cdot \left( \lambda_1 - \lambda_2 \right) &= \lambda_2 \cdot b_1 \, , \\
b_0 \left( \lambda_0 - 1 \right) + b_2 \left( \lambda_1 - B^2 \right) &= B \cdot b_1 \, .
\end{align}
Equivalently, we can write
\begin{align}
b_0 = \frac{a_0}{\lambda_0} \, , \qquad b_2 = \frac{a_0}{\lambda_1} \, , \qquad b_1 &= a_0 \cdot \left( \frac{2}{\lambda_2} - \frac{1}{\lambda_0} - \frac{1}{\lambda_1} \right) \, , \\
0 &= a_0 \cdot \left( \frac{2B}{\lambda_2} + \frac{1-B}{\lambda_0} + \frac{B \left( B - 1 \right)}{\lambda_1} - 2 \right) \, .
\end{align}
Take $B = \lambda_2 = 1$. Then these conditions are satisfied for all $\alpha_0 \in \mathbb{C}$ and so $h^0( G_5, L ) = 1$. However, for generic $ \lambda_0, \lambda_1, \lambda_2$ and $B$, the last condition will only be satisfied when $\alpha_0 = 0$ and so, $h^0( G_5, L ) = 0$. Consequently, $G_5$ is jumping for $d_\alpha = 0$ and $d_\beta = 2$.

\item $d_\alpha = 1$ and $d_\beta = 2$:\\
Then, the gluing conditions are
\begin{align}
a_1 = \lambda_1 \cdot b_2 \, , \qquad a_0 = \lambda_0 \cdot b_0 \, , \qquad b_0 \cdot \left( \lambda_0 - \lambda_2 \right) + b_2 \cdot \left( \lambda_1 - \lambda_2 \right) &= \lambda_2 \cdot b_1 \, , \\
b_0 \left( \lambda_0 - 1 \right) + b_2 \left( \lambda_1 A - B^2 \right) &= B \cdot b_1 \, .
\end{align}
Equivalently, we have
\begin{align}
a_1 &= \lambda_1 \cdot b_2 \, , \qquad a_0 = \lambda_0 \cdot b_0 \, , \\
b_1 &= \frac{b_0 \cdot \left( \lambda_0 - \lambda_2 \right) + b_2 \cdot \left( \lambda_1 - \lambda_2 \right)}{\lambda_2} \, , \\
0 &= b_0 \left[ \lambda_2 \left( \lambda_0 - 1 \right) + B \left( \lambda_2 - \lambda_0 \right) \right] + b_2 \left[ A \lambda_1 \lambda_2 + B \left( \lambda_2 - \lambda_1 \right) - B^2 \lambda_2 \right] \, .
\end{align}
Take $\lambda_0 = \lambda_1 = \lambda_2 = 1$, $A = 4$ and $B = 2$. Then the last condition is trivially satisfied for all $b_0, b_2 \in \mathbb{C}$. And so $h^0( G_5, L ) = 2$. However, for generic $\lambda_0$, $\lambda_1$, $\lambda_2$, $A$ and $B$ we find $h^0( G_5, L ) = 1$ and so $G_5$ is jumping for $d_\alpha = 1$ and $d_\beta = 2$.

\item $d_\alpha = d_\beta = 2$:\\
Then, the gluing conditions are
\begin{align}
a_2 = \lambda_1 \cdot b_2 \, , \qquad a_0 = \lambda_0 \cdot b_0 \, , \qquad b_0 \cdot \left( \lambda_0 - \lambda_2 \right) + b_2 \cdot \left( \lambda_1 - \lambda_2 \right) &= \lambda_2 \cdot b_1 - a_1 \, , \\
b_0 \left( \lambda_0 - 1 \right) + b_2 \left( \lambda_1 A^2 - B^2 \right) &= b_1 B - a_1 A \, .
\end{align}
Equivalently, we can write
\begin{align}
a_2 &= \lambda_1 \cdot b_2 \, , \qquad a_0 = \lambda_0 \cdot b_0 \, , \\
a_1 &= \lambda_2 \cdot b_1 - b_0 \cdot \left( \lambda_0 - \lambda_2 \right) - b_2 \cdot \left( \lambda_1 - \lambda_2 \right) \, , \\
0 &= b_0 \cdot \left[ A \left( \lambda_2 - \lambda_0 \right) + \lambda_0 - 1  \right] + b_1 \cdot \left[ A \lambda_2 - B \right] + b_2 \cdot \left[ \lambda_1 A^2 - B^2 + A \left( \lambda_2 - \lambda_1 \right) \right] \, .
\end{align}
Take $\lambda_0 = \lambda_1 = \lambda_2 = 1$ and $A = B = 2$. Then the last condition is trivial and we have $h^0( G_5, L ) = 3$. However, for generic $\lambda_0$, $\lambda_1$, $\lambda_2$, $A$ and $B$, the last condition is non-trivial and we have $h^0( G_5, L ) = 2$. Hence, $G_5$ is jumping for $d_\alpha = d_\beta = 2$.

\item $d_\alpha, d_{\beta} > 2$:\\
Then the four conditions become
\begin{align}
a_{d_\alpha} &= \lambda_1 \cdot b_{d_\beta} \, , \\
a_0 &= \lambda_0 b_0 \, , \\
a_1 &= - b_0 \cdot \left( \lambda_0 - \lambda_2 \right) - b_{d_\beta} \cdot \left( \lambda_1 - \lambda_2 \right) - \sum_{i = 2}^{d_\alpha - 1}{a_i} + \lambda_2 \cdot \sum_{i = 1}^{d_\beta - 1}{b_i} \, , \\
&= - \sum_{i = 2}^{d_\alpha - 1}{a_i} + \Gamma \left( \mathbf{b}, \lambda_0, \lambda_1, \lambda_2 \right) \, , \\
a_1 A + \sum_{i = 2}^{d_\alpha - 1}{a_i A^i} &= \sum_{i = 1}^{d_\beta - 1}{b_i B^i} - b_0 \left( \lambda_0 - 1 \right) + b_{d_\beta} \left( \lambda_1 A^{d_\alpha} - B^{d_\beta} \right) \, .
\end{align}
By plugging the third into the fourth condition we find
\begin{align}
\sum_{i = 2}^{d_\alpha - 1}{a_i \left( A^i - A \right)} &= \sum_{i = 1}^{d_\beta - 1}{b_i B^i} - b_0 \left( \lambda_0 - 1 \right) + b_{d_\beta} \left( \lambda_1 A^{d_\alpha} - B^{d_\beta} \right) - A \cdot \Gamma \left( \mathbf{b}, \lambda_0, \lambda_1, \lambda_2 \right) \, , \\
&= \Gamma_2 \left( A, B, \mathbf{b}, \lambda_0, \lambda_1, \lambda_2 \right)
\end{align}
Since $A \in \mathbb{C} \setminus \left\{0, 1 \right\}$, this is equivalent to
\begin{align}
a_2 = \frac{\Gamma_2 \left( A, B, \mathbf{b}, \lambda_0, \lambda_1, \lambda_2 \right) - \sum_{i = 3}^{d_\alpha - 1}{a_i \left( A^i - A \right)}}{A^2 - A} \, .
\end{align}
Hence, $G_5$ is non-jumping for $2 < d_\alpha, d_\beta$.
\end{itemize}
This completes the argument.
\end{myproof}

\subsection{Elliptic circuits} \label{sec:AppendixEllipticCircuitJumps}

\begin{prop}
$G_6$ is jumping for $d = 0$:
\begin{align}
\begin{tikzpicture}[scale=0.6, baseline=(current  bounding  box.center)]
      \def\s{2.0};
      \node at (-2*\s,0) [stuff_fill_green, scale=0.6]{$d = 0$};
\end{tikzpicture}
\end{align}
\end{prop}

\begin{myproof}
This is well-known for line bundles on elliptic curves.
\end{myproof}

\begin{prop}
The following configuration $G_7$ is jumping for $d \geq 1$:
\begin{align}
\begin{tikzpicture}[scale=0.6, baseline=(current  bounding  box.center)]
      \def\s{2.0};
      \path[-,out = -30, in = 30, looseness = 5] (-2*\s,-0.2*\s) edge (-2*\s,+0.2*\s);
      \node at (-2*\s,0) [stuff_fill_green, scale=0.6]{$d$};
\end{tikzpicture}
\end{align}
\end{prop}

\begin{myproof}
Let us denote the nodes as $p, q \in E$. Assume that $\left. L \right|_E = \mathcal{O}_E \left( (d+2)  \cdot r - p - q \right)$ with $r \in E - \{ p, q \}$. Then it holds
\begin{align}
H^0( E, \left. L \right|_E ) = \{ f \colon E \to \overline{\mathbb{C}} \text{ rational } \, , \, \mathrm{div}(f) \geq - (d+2) \cdot r + p + q \} \, .
\end{align}
Consequently, any $\varphi \in H^0( E, \left. L \right|_E )$ vanishes at the points $p$ and $q$. Therefore, the gluing condition is trivially satisfied and we find $h^0( G_7, L ) = d > d - 1$.
\end{myproof}

\begin{prop}
$G_8$ is jumping for $d \geq 1$:
\begin{align}
\begin{tikzpicture}[scale=0.6, baseline=(current  bounding  box.center)]
      \def\s{2.0};
      \path[-,out = 0, in = 180] (-3*\s,0) edge (-2*\s,0);
      \path[-,out = -30, in = 30, looseness = 5] (-2*\s,-0.2*\s) edge (-2*\s,+0.2*\s);
      \node at (-3*\s,0) [stuff_fill_green, scale=0.6]{$d$};
      \node at (-2*\s,0) [stuff_fill_red, scale=0.6]{$1$};
\end{tikzpicture}
\end{align}
\end{prop}

\begin{myproof}
Choose homogeneous coordinates $[x \colon y]$ for the $\mathbb{P}^1$, descent data $\lambda = 1$ for the self-edge of this rational curve and fix the position of the end points of the node encoded by this self edge to $[1 \colon 0]$ and $[0 \colon 1]$ by an $SL(2, \mathbb{C} )$ transformation. The most general section on this $\mathbb{P}^1$, which satisfies the gluing condition via the self-edge is
\begin{align}
\varphi( [x \colon y ]) = \alpha \lambda x + \alpha y \, , \qquad \alpha \in \mathbb{C} \text{ arbitrary but fixed.}
\end{align}
Next, take the node connecting the $\mathbb{P}^1$ to the elliptic curve E to be $[1 \colon 1]$ and the section on $E$ to be $\psi$. Then the final condition to be imposed is
\begin{align}
\psi( p ) = \alpha \cdot \left( \lambda + 1 \right) \, .
\end{align}
Let us assume that $\lambda = -1$ and that
\begin{align}
\left. L \right|_E = \mathcal{O}_E \left( (d+1) \cdot r - p \right) \, , \qquad r \in E - \{ p \} \, .
\end{align}
Then $\psi(p) = 0$ and the condition $\psi( p ) = \alpha \cdot \left( \lambda + 1 \right)$ is satisfied for all $\alpha \in \mathbb{C}$. Consequently, we then find $h^0( G_8, L ) = d + 1 > \left( d + 2 \right) - 2$. Hence, $G_8$ is jumping for $d \geq 1$.
\end{myproof}

\begin{prop}
$G_9$ is jumping for $d \geq 1$:
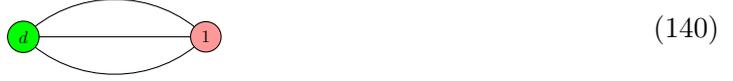
\begin{align}
\begin{tikzpicture}[scale=0.6, baseline=(current  bounding  box.center)]
      \def\s{1.0};
      \path[-,out = 45, in = 135] (-2*\s,0) edge (+2*\s,0);
      \path[-,out = 0, in = 180] (-2*\s,0) edge (+2*\s,0);
      \path[-,out = -45, in = -135] (-2*\s,0) edge (+2*\s,0);
      \node at (-2*\s,0) [stuff_fill_green, scale=0.6]{$d$};
      \node at (+2*\s,0) [stuff_fill_red, scale=0.6]{$1$};
\end{tikzpicture}
\end{align}
\end{prop}

\begin{myproof}
We take coordinates $[x \colon y]$ for the $\mathbb{P}^1$ and place the nodes on this curve at $[1:0]$, $[0:1]$ and $[1:1]$. We label the corresponding points on $E$ as $p_1$, $p_2$ and $p_3$. Hence, the gluing conditions to be imposed are
\begin{align}
\psi(p_1) &= \lambda_1 \alpha \, , \\
\psi(p_2) &= \lambda_2 \beta \, , \\
\psi(p_3) &= \alpha + \beta \, .
\end{align}
Since $\lambda_1, \lambda_2 \in \mathbb{C}^\ast$, we always have
\begin{align}
\frac{\psi(p_1)}{\lambda_1} &= \alpha \, , \qquad \frac{\psi(p_2)}{\lambda_2} = \beta \, .
\end{align}
This uniquely determines the section on the $\mathbb{P}^1$. It remains to satisfy the third condition with the sections on $E$:
\begin{align}
\psi(p_3) &= \frac{\psi(p_1)}{\lambda_1} + \frac{\psi(p_2)}{\lambda_2} \, . \label{equ:Condition}
\end{align}
Let us assume that
\begin{align}
\left. L \right|_E = \mathcal{O}_E \left( (d_\alpha+3) \cdot r - p_1 - p_2 - p_3 \right) \, , \qquad r \in E - \{ p_1, p_2, p_3 \} \, .
\end{align}
Then for every $\psi \in H^0( E, \left. L \right|_E )$ it holds $\psi(p_1) = \psi(p_2) = \psi(p_3) = 0$. Consequently, the condition in \cref{equ:Condition} is trivially satisfied and we $h^0( G_9, L ) = d > \left( d + 2 \right) - 3$.
\end{myproof}

\begin{prop}
$G_{10}$ is jumping for $d \geq 1$:
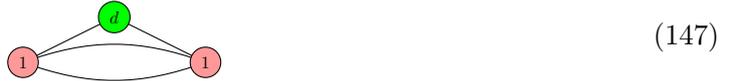
\begin{align}
\begin{tikzpicture}[scale=0.6, baseline=(current  bounding  box.center)]
      \def\s{1.0};
      \draw (0,0) -- (-2*\s,-\s);
      \draw (0,0) -- (2*\s,-\s);
      \path[-,out = 20, in = 160] (-2*\s,-\s) edge (2*\s,-\s);
      \path[-,out = -20, in = -160] (-2*\s,-\s) edge (2*\s,-\s);      
      \node at (0,0) [stuff_fill_green, scale=0.6]{$d$};
      \node at (-2*\s,-\s) [stuff_fill_red, scale=0.6]{$1$};
      \node at (+2*\s,-\s) [stuff_fill_red, scale=0.6]{$1$};
\end{tikzpicture}
\end{align}
\end{prop}

\begin{myproof}
We pick coordinates $[x:y$ and $[u:v]$ for the two $\mathbb{P}^1$s and fix the node positions to $[1:0]$ and $[0:1]$ respectively. Take the descent data to be $\lambda_1$ and $\lambda_2$ along these two nodes. The most general section on the left bottom $\mathbb{P}^1$ be
\begin{align}
\varphi_a = \alpha_1 x + \alpha_2 y \, .
\end{align}
Then the sections on the bottom right $\mathbb{P}_1$ are uniquely fixed to be
\begin{align}
\varphi_b = \lambda_1 \alpha_1 u + \lambda_2 \alpha_2 v \, .
\end{align}
The nodes of the $\mathbb{P}^1$ towards the elliptic curve can be taken to be $[1:1]$. We denote the corresponding points on $E$ as $p$ and $q$. Let the most general section on $E$ be $\psi$. Then we impose the following conditions:
\begin{align}
\psi(p) = \alpha_1 + \alpha_2 \, , \qquad \psi(q) = \lambda_1 \alpha_1 + \lambda_2 \alpha_2 \, .
\end{align}
Equivalently, we have
\begin{align}
\alpha_2 = \psi(p) - \alpha_1 \, , \qquad \psi(q) -  \lambda_2 \psi(p) = \left( \lambda_1 - \lambda_2 \right) \alpha_1 \, .
\end{align}
Let us focus on $\lambda_1 = \lambda_2$ and
\begin{align}
\left. L \right|_E = \mathcal{O}_E \left( (d+2) \cdot r - p - q \right) \, , \qquad r \in E - \{ p, q \} \, .
\end{align}
Then the condition $\psi(q) -  \lambda_2 \psi(p) = \left( \lambda_1 - \lambda_2 \right) \alpha_1$ is trivially satisfied and we find $h^0( G_{10}, L ) = d + 1 > \left( d + 2 + 2 \right) - 4$. This shows that $G_{10}$ is jumping for $d \geq 1$.
\end{myproof}

\bibliography{references-new}{}
\bibliographystyle{JHEP}

\end{document}